\documentclass[draft]{article}
\usepackage{amssymb}
\newcommand{\0}{{\bf 0}}
\newcommand{\z}{{\bf z}}
\newcommand{\y}{{\bf y}}
\newcommand{\x}{{\bf x}}
\newcommand{\w}{{\bf w}}
\newcommand{\e}{{\bf e}}

\newcommand{\A}{{\sf A}}
\newcommand{\Q}{{\bf Q}}

\newcommand{\la}{e}
\renewcommand{\Lambda}{{\cal E}}

\newcommand{\defeq}{\triangleq}
\newcommand{\eq}{\triangleq}

\newcommand{\varliminf}{\mathop{\underline{\lim}}\limits}
\newcommand{\varlimsup}{\mathop{\overline{\lim}}\limits}

\renewcommand{\emptyset}{\varnothing}

\begin{document}

\thispagestyle{empty}

\begin{center}
{\Large\bf Lectures on Designing Screening Experiments~\footnote[1]
{Published in:\quad  Lecture Note Series~10, Feb.~2004,
        Combinatorial and Computational Mathematics Center,
        Pohang  University of Science and Technology (POSTECH),
        Korea Republic, (monograph, pp.~112).}}
\end{center}
\par
\vspace*{1cm}
\par
\begin{center}
{\bf Arkadii G. D'yachkov}
\end{center}

\begin{center}
Moscow State University, Faculty of Mechanics \&
Mathematics,\\
Department of Probability Theory, Moscow, 119992,
Russia.\\
{\small agd-msu@yandex.ru}
\end{center}

\par
\vspace*{1cm}
\par

\section*{Preface}
\begin{center}
\end{center}

Designing Screening Experiments (DSE)
is a class of information - theoretical models for multiple
- access channels (MAC). In Sect.~2-4, we discuss the
combinatorial model of DSE called a disjunct channel
model. This model is the most important for applications
and closely connected with the superimposed code concept.
In Sect.~2, we give a detailed survey of lower and upper
bounds on the rate of superimposed codes. The best known
constructions of superimposed codes are considered in
Sect.~3-4, where we also discuss the development of these
codes (non-adaptive pooling designs) intended for the clone
- library screening problem. In Sect.~5, we obtain lower
and upper bounds on the rate of binary codes for the
combinatorial model of DSE called an adder channel model.
In Sect.~6, we consider the concept of universal decoding
for the probabilistic DSE model called a symmetric model
of~DSE.

\par
\vspace*{1cm}
\par

\section*{Contents}
\begin{center}
\end{center}

\begin{itemize}
\item
[{\bf1}] {\bf Introduction} (p.~1).
\begin{itemize}
\item
[{\bf1.1}] {\sf Statement of problem} (p.~1).
\item
[{\bf1.2}] {\sf List of notations} (p.~3).
\end{itemize}
\item
[{\bf2}] {\bf Disjunct channel model and bounds on the
rate of superimposed codes} (p.~4).
\begin{itemize}

\item
[{\bf2.1}] {\sf Notations, definitions of
superimposed codes and their properties} (p.~4).
\item
[{\bf2.2}] {\sf Upper and lower bounds
on $R(s,L)$ and $R_D(s)$} (p.~6).
\begin{itemize}
\item
[{\bf2.2.1}]  Lower bounds on
$R(s,1)$ and $R_D(s)$ (p.~6).
\item
[{\bf2.2.2}]  Upper bounds on
$R(s,L)$ and $R_D(s)$ (p.~7).
\item
[{\bf2.2.3}]  Lower bound on
$R(s,L)$ (p.~8).
\end{itemize}
\item
[{\bf2.3}] {\sf Kautz-Singleton codes} (p.~13).
\begin{itemize}
\item
[{\bf2.3.1}]  Upper bound on
$R_{KS}(s)$ (p.~14).
\item
[{\bf2.3.2}]  Lower bound on
$R_{KS}(s)$ (p.~14).
\end{itemize}
\item
[{\bf2.4}] {\sf Symmetrical superimposed
 codes} (p.~16).
\end{itemize}

\newpage
\thispagestyle{empty}

\item
[{\bf3}] {\bf Constructions of superimposed codes} (p.~17).
\begin{itemize}
\item
[{\bf3.1}] {\sf Notations and definitions} (p.~17).
\item
[{\bf3.2}] {\sf Application to DNA library screening}
(p.~17).
\item
[{\bf3.3}] {\sf Generalized Kautz-Singleton
codes} (p.~18).
\item
[{\bf3.4}] {\sf Superimposed concatenated codes} (p.~21).
\end{itemize}

\item [{\bf4}]
{\bf Optimal superimposed codes and designs for Renyi
search model} (p.~26).
\begin{itemize}
\item
[{\bf4.1}] {\sf Notations and definitions} (p.~26).
\item
[{\bf4.2}] {\sf Lower bound} (p.~26).
\item
[{\bf4.3}] {\sf Optimal parameters} (p.~27).
\item
[{\bf4.4}] {\sf Homogeneous $q$-ary codes} (p.~29).
\item
[{\bf4.5}] {\sf Proof of Theorem~1} (p.~31).
\item
[{\bf4.6}] {\sf Proof of Theorem~2} (p.~32).
\item
[{\bf4.7}] {\sf Proof of Theorem~3} (p.~34).
\item
[{\bf4.8}] {\sf On $(q,k,3)$-homogeneous $3$-separable
and $3$-hash codes. Proof of Theorem~4} (p.~35).
\begin{itemize}
\item
[{\bf4.8.1}]  Characteristic matrices (p.~35).
\item
[{\bf4.8.2}]  Examples of hash, separable and
hash\&separable codes (p.~36).
\item
[{\bf4.8.3}]  Existence of hash and
hash\&separable codes (p.~37).
\item
[{\bf4.8.4}] Product of
characteristic matrices (p.~38).
\end{itemize}
\item
[{\bf4.9}] {\sf Proof of Theorem~5} (p.~40).
\end{itemize}

\item [{\bf5}] {\bf Adder channel
model and $B_s$-codes} (p.~42).
\begin{itemize}
\item
[{\bf5.1}] {\sf Statement of the
problem and results} (p.~42).
\begin{itemize}
\item
[{\bf5.1.1}]  Upper bounds (p.~42).
\item
[{\bf5.1.2}]  Lower bounds (p.~43).
\end{itemize}
\item
[{\bf5.2}] {\sf Proof of upper bounds
on $R_A(s)$ and $R(s)$} (p.~44).
\begin{itemize}
\item
[{\bf5.2.1}]  Proof of Theorem 1 (p.~44).
\item
[{\bf5.2.2}]  Proof of inequality
$R(2)\le3/5$ (p.~46).
\end{itemize}
\item
[{\bf5.3}] {\sf Proof of Theorem 3} (p.~48).
\end{itemize}
\item
[{\bf6}] {\bf Universal decoding for random design
of screening experiments} (p.~51).
\begin{itemize}
\item
[{\bf6.1}] {\sf Statement of the problem, formulation
and discussion of results} (p.~51).
\item
[{\bf6.2}] {\sf Proof of Theorem 1} (p.~57).
\end{itemize}
\item
[] {\bf References} (p.~62).

\end{itemize}

\newpage
\addtocounter{page}{-2}

\section{Introduction}
\begin{center}
\end{center}

\subsection{Statement of problem}
\begin{center}
\end{center}

Let $1\le s<t$ be fixed integers, $[t]$ be the set of
integers from 1 to $t$. Let $\e\eq(e_1,e_2,\dots,e_s)$,
where $e_i\in[t]$, $1\le e_1<e_2<\dots<e_s\le t$, be an
arbitrary $s$-subset of $[t]$ and, here and below, the
symbol $\eq$ denote the equation by definition. Introduce
${\cal E}(s,t)$ as the collection of all such subsets. Note
that the cardinality (number of elements)
$$
|{\cal E}(s,t)|={t\choose s}=\frac{t!}{s!(t-s)!}.
$$

Suppose that among $t$ {\it factors}, numbered by integers
from 1 to $t$, there are some $s<t$ unknown factors
called {\it significant} factors.  Each collection
of significant factors is identified as an $s$-subset
$\e\in {\cal E}(s,t)$. The problem of {\it screening
experiment design} (DSE) is to {\it find} all significant
factors, i.e. to {\it detect} an unknown subset~$\e$.

To {\it search} $\e$ one can {\it carry out $N$
experiments}. Each experiment is a test of a subset of
$[t]$. These tests could be described by a binary
$N\times t$-matrix $X=\|x_i(u)\|$, $x_i(u)\in\{0;1\}$,
$i=1,2,\dots,N$, $u=1,2,\dots,t$, where
$$
x_i(u)\eq\cases{1, & if the $u$-th factor is included into
the $i$-th test,\cr
0, & otherwise.\cr}
$$
Matrix $X$ is called a {\it code} ({\it design of
experiments}) of {\it length} $N$ and {\it size} $t$.

Fix an arbitrary $i$, $i=1,2,\dots,N$. Let
$\x_i\eq(x_i(1),x_i(2),\dots,x_i(t)$ be the $i$-th row
of~$X$. The $i$-th row $\x_i$ is identified with a
subset of $[t]$, which consists of positions where this row
contains 1's. We say that the $i$-th experiment is
a {\em group} test of this subset.

Let code $X$ be fixed and the symbols
$$
\x(u)\eq(x_1(u),x_2(u),\dots,x_N(u))\in\{0;1\}^N,
\quad u=1,2,\dots,t,
$$
denote the columns ({\it codewords}) of code $X$.
For the given $s$-subset
$\e=(e_1,e_2,\dots,e_s)$ called a {\em message},
consider a non-ordered $s$-{\it collection} of codewords
$$
\x(\e)\eq(\x(e_1),\x(e_2),\dots,\x(e_s)).
$$
We say that $\x(\e)$ {\em encodes} $\e$. Let
$$
\x_i(\e)\eq(x_i(e_1),x_i(e_2),\dots,
x_i(e_s))\in\{0;1\}^s,\quad i=1,2,\dots,N\;,
$$
be the $i$-th row of $s$-collection $\x(\e)$.

Let  $z_i$ be an {\it output} (or {\em result}) of the
$i$-th test and $\z=\z(\e,X)\eq(z_1,z_2,\dots,z_N)$.
To describe the model of such a test output,
we use the terminology of a {\it memoryless multiple-access
channel} (MAC), which has $s$ inputs and one
output~\cite{ck81}\footnote[1]{This terminology was
suggested by I.~Csiszar in 1978.}. Let all $s$ input
alphabets of MAC will be the same and coincide with
$\{0,1\}$. Denote by $Z$ a finite output alphabet. In
Sections~2-5 we will consider the {\it deterministic} model
of MAC. This MAC is defined by the function
$$
z=f(x_1,x_2,\dots,x_s),\quad z\in Z,\quad
x_k\in\{0,1\},\quad k=1,2,\dots,s,
$$
and, by definition,
the result $z_i$ of the $i$-th test is
$$
z_i\eq
f(\x_i(\e))=f(x_i(e_1),x_i(e_2),\dots, x_i(e_s)), \qquad
i=1,2,\dots,N.
$$
The corresponding deterministic model of
DSE is called an $f$--model.  The problem of DSE for the
probabilistic model of MAC will be discussed in Section~6.
\medskip

On the basis of $\z(\e,X)$ an {\it observer} makes a
decision about the unknown $s$-subset~$\e$.  To identify
$\e$ on the basis of $\z(\e,X)$, a code $X$ is assigned the
following definition.
\medskip

{\bf Definition}. We say that a code $X$ is an $(s,N)$-{\it design}
of size~$t$ (or an $(s,t)$-{\it design} of length~$N$)
for the $f$--model, if  all
$\z(\e,X),\;\e\in {\cal E}(s,t)$ are distinct.
\medskip

Let $t_f(s,N)\;(N_f(s,t))$ be the {\it maximal possible size}
of $(s,N)$-design
({\it minimal possible length} of $(s,t)$-design) for $f$--model.
For fixed $s\ge 2$ define the number
$$
R_f(s)\eq\varlimsup_{N\to\infty}
\frac{\log_2t_f(s,N)}{N},
$$
which is called a {\it design rate} of the $f$--model.
Using the terminology of the Shannon coding theory, the number
$R_f(s)$ is called a {\it zero error capacity}
for the $f$-model of DSE.

Let $\sum\limits_{i=1}^sx_i$ denote the arithmetic sum,
i.e.  the number of 1's in the sequence
$x_1,x_2,\dots,x_s$.  We introduce the following {\it
three} combinatorial models of DSE:
\begin{itemize}
\item
A--model ({\em adder} channel model), where the output
alphabet $Z=\{0,1,2,\dots,s\}$ and
$$
f(x_1,x_2,\dots,x_s)=
f_A(x_1,x_2,\dots,x_s)\eq\sum_{i=1}^sx_i;\eqno (1)
$$
\item
D--model ({\em disjunct} channel model), where the output
alphabet $Z=\{0,1\}$ and
$$
f(x_1,x_2,\dots,x_s)=
f_D(x_1,x_2,\dots,x_s)\eq\cases{1, &$\sum_{i=1}^sx_i\ne 0$,\cr
                        0, &$\sum_{i=1}^sx_i=0$,\cr};\eqno (2)
$$
\item
SD--model ({\it symmetrical disjunct} channel model),
where the output alphabet $Z=~\{0,1,*\}$ (the symbol $*$
denotes {\it erasure}) and
$$
f(x_1,x_2,\dots,x_s)=
f_{SD}(x_1,x_2,\dots,x_s)\eq\cases{1, &$\sum_{i=1}^sx_i=n$,\cr
                        0, & $\sum_{i=1}^sx_i=0$,\cr
                        *, & $1\le\sum_{i=1}^sx_i\le n-1$.\cr}\eqno (3)
$$
\end{itemize}

Let $t_A(s,N)$, $N_A(s,t)$, $R_A(s)$ be the parameters of
A--model, $t_D(s,N)$, $N_D(s,t)$, $R_D(s)$ be the
parameters of D--model and $t_{SD}(s,N)$, $N_{SD}(s,t)$,
$R_{SD}(s)$ be the parameters of SD--model.  Obviously, if
$s=2$, then the A--model and SD--model are the same.  Hence
$$
t_A(2,N)=t_{SD}(2,N),\qquad
N_A(2,t)=N_{SD}(2,t),\qquad R_A(2)=R_{SD}(2).
$$

In Sections~2 and~5, we will discuss the properties
of these optimal characteristics. Sections~2-4  focus on
the disjunct channel model. This model is the most
important for applications and closely connected with the
superimposed code concept introduced by Kautz and
Singleton~\cite{ks64}.  In Section~2, we give a detailed
survey of the best lower and upper
bounds~\cite{dr82,dr83,dr89} on the rate of
superimposed codes.  The best known constructions of
superimposed codes~\cite{ks64,dmr00-1,dr01} are considered
in Sections~3-4, where we also discuss the development of
these codes (non-adaptive pooling designs) intended for the
clone - library screening problem~\cite{kbt98,dmr00-2}. In
Section~5, we obtain lower and upper bounds on the rate of
binary codes for the adder channel model. These bounds are
based on a refinement of the results published in
papers~\cite{d75,dr81} and~\cite{rd81,rd83}. In
Section~6, we consider the concept of universal
decoding~\cite{ck81,univ89} for the probabilistic DSE model
called a symmetric model of~DSE.

\subsection{List of notations}
\begin{center}
\end{center}

\begin{itemize}
\item
$[t]=\{1,2,\dots,t\}$---the set of integers from 1 to $t$;
\item
$\lceil b\rceil$---the least integer $\ge b$,\quad
$\lfloor b\rfloor$---the largest integer $\le b$;
\item
$\eq$---equation by definition;
\item
$\vee$---the symbol of Boolean summation,\quad
$|A|$---the number of elements in set $A$;
\item
$\log$---logarithm to the base 2,\quad
$\exp\{a\}=2^a$,\quad
$\ln$---logarithm to the base $e$;
\item
$
X=\|x_i(u)\|,\; x_i(u)\in\{0;1\}\quad i=1,2,\dots,N,\; u=1,2,\dots,t,
$--code (design);
\item
$\x(u)=(x_1(u),x_2(u),\dots,x_N(u)),\;u=1,2,\dots,t,$--codewords;
\item
$\x_i=(x_i(1),x_i(2),\dots,x_i(t)$, $i=1,2,\dots,N$ --
the $i$-th row of code $X$ -- is identified with the $i$-th
group test.
\item
$N$---the code (design) length,\quad
$t$---the code (design) size,\quad
$s$---the code (design) strength;
\item
$(N,R)$-code---the code of length $N$, rate $R$
and  size
$t=\lfloor\exp\{RN\}\rfloor$;
\item
for arbitrary integers $s$ and $t$, we introduce the generalized binomial
coefficients,
$$
{t\choose s}\eq\cases{\frac{t!}{s!(t-s)!}, & if
$0\le s\le t$,\cr
0, & otherwise;\cr}
$$
\item
$\Pr\{E\}$---the probability of event $E$,\quad
$p(x)=\Pr\{X=x\}$---the probability distribution of a discrete random
variable $X$;
\item
${\bf H}(X)\eq-\sum_x p(x)\log p(x)$---the Shannon entropy of $X$;
\item
$h(u)\eq -u\log u-(1-u)\log(1-u)$--binary entropy;
\end{itemize}

\newpage

\centerline{Arkadii G. D'yachkov,\qquad "Lectures on
Designing Screening Experiments"}
\medskip

\section{Disjunct Channel Model and Bounds on\\
the Rate of Superimposed Codes}
\begin{center}
\end{center}

In this section, we consider the connection between the
D--model of DSE and the theory of superimposed codes
 introduced by Kautz and Singleton~\cite{ks64}.

\subsection{Notations, definitions of superimposed
codes\\ and their properties}
\begin{center}
\end{center}

Let $1<s<t$, $1\le L\le t-s$, $N>1$ be integers, and let
$\y(j)=(y_1(j),y_2(j),\dots,\y_N(j)),\;j=1,2,\dots,s$
denote the binary columns of length $N$. The {\it Boolean
sum}
$$
\y=\bigvee_{j=1}^s\y(j)=\;\y(1)\vee\y(2)\vee\cdots\vee\y(s)
$$
of columns $\y(1),\y(2),\dots,\y(s)$ is the binary column
$\y=(y_1,y_2,\dots,y_N)$ with components
$$
y_i=\cases{0, & if $y_i(1)=y_i(2)=\cdots=y_i(s)=0$,\cr
           1, &otherwise.\cr}
$$

Let us say that column $\y$ {\it cover} column $\z$ if
$\y\vee\z=\y$.
\smallskip

Let $X=\|x_i(u)\|,\;i=1,2,\dots,N,\;u=1,2,\dots,t$ be a
binary $N\times t$--matrix (code). Later on the matrix
(code) $X$ is interpreted as a set of $t$ binary columns
(codewords) $\x(1),\x(2),\dots,\x(t)$.  \medskip

{\bf Definition}~\cite{dr83}. An $N\times t$--matrix $X$
is called a {\it list-decoding superimposed code} (LDSC)
{\it of length $N$, size $t$, strength $s$}, and {\it
list-size} $\le L-1$ if the Boolean sum of any $s$-subset
of codewords $X$ can cover not more than $L-1$ codewords
that are not components of the $s$-subset. This code also
will be called an $(s,L,N)$-code of size $t$, or an
$(s,L,t)$-code of length $N$.

Note, that in the most important particular case $L=1$,
the definition is equivalent to the following condition.
The Boolean sum of any $s$-subset of columns $X$ covers
those and only those columns that are the components of
given Boolean sum. Some generalizations of LDSC were
studied in~\cite{dr89,dn93}.

Superimposed $(s,1,N)$-codes were introduced
in~\cite{ks64}.  Applied problems leading to the definition
of $(s,1,N)$-codes and some methods of construction of such
codes are described in~\cite{ks64}. New constructions
and applications developed in papers~\cite{dmr00-1,dmr00-2}
will be given in Section~4.

{\bf Remark}. Each column of the matrix $X$ is identified
with the subset of $[N]$, which consists of positions where
this column contains 1's. Then using the terminology of
sets, the construction of an $(s,L,N)$-code of size $t$ is
equivalent to the following combinatorial problem. A family
of $t$ subsets of the set $[N]$ should be constructed in
which  {\it no union of  $L$ members of the family is
covered by the union of $s$ others}.

Let $t(s,L,N)$ be the maximal possible size of LDSC and
$N(s,L,t)$ be the minimal possible length of LDSC.
For  fixed $s$ and $L$, define the {\it rate} of LDSC
$$
R(s,L)\eq\varlimsup_{N\to\infty}\frac{\log t(s,L,N)}{N}.
$$

For the optimal parameters of D--model, we also use the
notations of {\bf Sect. 1}, i.e., $t_D(s,N)$, $N_D(s,t)$
and $R_D(s)$.  Propositions 1--3 follow easily from
definitions of $(s,N)$-design and LDSC.

{\bf Proposition 1.}
$$
N_D(s,t)\ge\left\lceil\log{t\choose s}\right\rceil,
\quad R_D(s)\le\frac{1}{s}
$$
\medskip

{\bf Proposition 2.} {\it Any $s+1,L,N)$-code is an
$(s,L,N)$-code, and any $(s,L,N)$-code is an
$(s,L+1,N)$-code. Hence},
$$
t(s+1,L,N)\le t(s,L,N)\le t(s,L+1,N).
$$
\medskip

{\bf Proposition 3.} {\it For any $s$-subset of columns of
an $(s,L,t)$-code, there are not more than ${s+L-1\choose
s}$ $s$-subsets of columns, such that the Boolean sums of
columns of these $s$-subsets coincide with the Boolean sum
of columns of a given $s$-subset. Hence,}
$$
N(s,L,t)\ge\left\lceil\log{t\choose s}-\log{s+L-1\choose
s}\right\rceil,\quad R(s,L)\le\frac{1}{s}.
$$

Let $t={N\choose\lceil N/2\rceil}$ and $X$ be an
$N\times t$ matrix whose columns are all distinct and
contain the same number $\lceil N/2\rceil$ of 1's. Then $X$
is a $(1,N)$-design and a $(1,L,N)$-code simultaneously.
Hence, $R_D(1)=R(1,L)=1,\;L=1,2,\dots$.
\medskip

The following Propositions 4--5 are proved easily
{\it by contradiction}.

{\bf Proposition 4}. {\it Any $(s,1,N)$-code is a
$(s,N)$-design, and any $(s,N)$-design is a
$(s-1,2,N)$-code, i.e.},
$$
t(s,1,N)\le t_D(s,N)\le t(s-1,2,N),
\quad R(s,1)\le R_D(s)\le R(s-1,2).
$$

{\bf Proposition 5}. {\it The matrix $X$ simultaneously
satisfies the definitions of the $(s-1,1,N)$-code and the
$(s,N)$-design iff all the Boolean sums composed of not
more than $s$ columns of $X$ are distinct}.

Proposition~5 is very important for the DSE.
It allows us to {\it define}  the rate of the code
satisfying the condition of Proposition~5 as
$\min\{R_D(s);R(s-1,1)\}$.
\medskip

{\bf Proposition 6}.~\cite{dr82} (L.A. Bassalygo, 1975)
$$
N(s,1,t)\ge\min\left\{\frac{(s+1)(s+2)}{2};\;t\right\}
$$
{\it and, therefore, $N(s,1,t)=t$ if $s\ge\sqrt{2t}$. In
other words, for $s\ge\sqrt{2t}$, no $(s,1,t)$-code is
better than the trivial one of length $N=t$, whose matrix
$X$ is diagonal}.

{\bf Proof.} Consider an arbitrary $(s,1,t)$-code $X$.
Let $w_j,\;j=1,2,\dots,t$, called the {\it weight},
denote the number of 1-entries in codeword $\x(j)$
and let $t(w),\; w=1,2,\dots,N,$ denote the number of
codewords of the weight $w$. Evidently,
$$
\sum_{w=1}^N t(w)=t,\quad 0\le t(w)\le t.
$$

{\bf Lemma 1}~\cite{ks64}. {\it If $X$ is an
$(s,1,t)$-code of length $N$, then the number of
its codewords of the weight $\le s$ does non exceed $N$,
i.e}.,
$$
\sum_{w=1}^s t(w)\le N.
$$

{\bf Proof of Lemma 1}. We fix an arbitrary codeword
$\x(j)$ containing $w\le s$ 1's, and consider rows of $X$
that contain 1's of $\x(j)$.  The definition of
$(s,1,t)$-code implies that among these rows there exists
at least one row in which all the remaining elements except
for the element of codeword $\x(j)$, are zero.  To prove
this, it is sufficient to note that otherwise code $X$
would have $s$ codewords, whose Boolean sum covers $\x(j)$.
Using the similar arguments with other codewords of $X$,
whose weight $w\le s$, we obtain Lemma 1.

{\bf Lemma 2}. {\it Let $X$ be a $(s,1,t)$-code of length $N$.
If there exists at least one codeword of weight
$w$, i.e., $t(w)>0$, then}
$$
w\le N-N(s-1,1,t-1).
$$

{\bf Proof of Lemma 2.} Consider codeword $\x(j)$ of
weight $w$. We fix $w$ rows of $X$ containing 1's positions
of $\x(j)$, and we delete them from $X$ together with
$\x(j)$. To obtain Lemma 2, it is sufficient to check that
the remaining matrix ${\tilde X}$ will be
$(s-1,1,t-1)$-code of length $N-w$. This property of matrix
${\tilde X}$ is easily checked by contradiction.

Lemma 2 is proved.

Let $X$ be a $(s,1,t)$-code of length $N\le t-1$. From
Lemma 1 it follows that code $X$ contains at least one
codeword of weight $\ge s+1$, Therefore, Lemma 2  yields
the inequality
$$
N(s,1,t)\ge s+1+N(s-1,1,t-1).
$$
Using the induction on $s$ it follows Proposition~6.
\bigskip

Below, we formulate (without proofs) the generalizations of
of Lemma~1 and Proposition~6 for the case of $(s,L,t)$-code.
The generalization of Lemma~2 is evident.
\smallskip

{\bf Lemma 1$'$}. {\it Let $X$ be a $(s,L,N)$-code of size
$t$. If $L\le s$, then}
$$
\sum_{w=1}^{\lfloor s/L\rfloor} t(w)\le N+L-1.
$$
\smallskip

{\bf Proposition 6$'$}. {\it For} $1\le L\le s<t$,
$$
N(s,L,t)\ge\min\left\{\frac{s(s+1)-L(L-1)}{2L};
t-L+1\right\}.
$$
\smallskip

In the next section,  we give a survey of the best
known upper and lower bounds on $R(s,L)$ and
$R_D(s)$,~$s\ge2$.

\subsection{Upper and lower bounds on $R(s,L)$ and
$R_D(s)$}
\begin{center}
\end{center}

\subsubsection{Lower bounds on $R(s,1)$ and $R_D(s)$}
\begin{center}
\end{center}

Using the random ensemble of {\it constant-weight codes}
of length $N$, size $t$ and weight
$w=\lceil QN\rceil$, $0<Q<1$,
one can prove the following theorem.
\medskip

{\bf Theorem 1.}
$$
R(s,1)\ge\underline{R}(s,1)\eq\frac{A(s)}{s},
$$
{\it where}
$$
A(s)\eq\max_{0<u<1,0<Q<1}\left\{-(1-Q)\log
(1-u^{s})+s\left(Q\log
\frac{u}{Q}+(1-Q)\log\frac{1-u}{1-Q}\right)\right\}.
$$

Theorem 1 is the particular case of the result which was
proved in~\cite{dr89}. If $s\to\infty$, then
$$
\underline{R}(s,1)=\frac{1}{s^2\log e }(1+o(1))=
\frac{0.693}{s^2}(1+o(1)).
$$

The same random ensemble of constant-weight codes also
yields (unpublished) a lower bound on $R_D(s)$.  We denote
this lower bound by $\underline{R}_D(s)$. The definition of
$\underline{R}_D(s)$ (it is omitted here) is similar to
the definition of $\underline{R}(s,1)$. If $s\to\infty$,
then the corresponding asymptotic behavior  has the following form
$$
\underline{R}_D(s)=
\frac{2}{s^2\log e}(1+o(1))=\frac{1.386}{s^2}(1+o(1)).
$$
In {\bf Table 1}, we give some numerical values of
$\underline{R}(s,1)$ and $\underline{R}_D(s)$.
\smallskip
\begin{center}
\begin{tabular}{||c||c|c|c|c|c|c|c||}
\hline
$s$ & 2 & 3 & 4 & 5 & 6 & 7 & 8\\
\hline
\hline
$\underline{R}(s,1)$ & .182 & .079 & .044 & .028 & .019 & .014 & .011\\
\hline
$\overline{R}(s,1)$ & .322 & .199 & .140 & .106 & .083 & .067 & .056\\
\hline
$\underline{R}_D(s)$ & .302 & .142 & .082 & .053 & .037 & .027 &
.021\\
\hline
$\underline{R}(s,2)$ & .236  & .115 & .068 & .046 & .032 & 024 &
.019\\
\hline
$\overline{R}(s,2)$ & 1/2  & 1/3  & 1/4  & 1/5  & .163 & .141 & .117\\
\hline
\end{tabular}
\end{center}
\centerline{{\bf Table 1.}}

\subsubsection{Upper bounds on $R(s,L)$ and $R_D(s)$}

\begin{center}
\end{center}

We remind that $h(u)\eq-u\log u-(1-u)\log(1-u)$ denotes
the binary entropy.  For integers $m=1,2,\dots$, define
$$
f(m,v)\eq h(v/m)-vh(1/m),\, 0<v<1.
$$
Theorem 2 is a generalization of the upper bound on $R(s,1)$
from~\cite{dr82}.
\medskip

{\bf Theorem 2}. (Unpublished). {\it For any fixed $L\ge2$,
the rate
$$
R(s,L)\le\overline{R}(s,L),
$$
where the sequence $\overline{R}(s,L),\,s=2,3,\ldots$, is
defined  by recurrence relations}:
\begin{itemize}
\item
{\it if $s\leq L$, then} $\overline{R}(s,L)=1/s$,
\item
{\it if $s\ge L+1$, then
$\overline{R}(s,L)=\min\{1/s;\,r(s,L)\}$,
and $r(s,L)$ is the unique solution of the equation
$$
r(s,L)=\max_{(*)}f(\lfloor s/L\rfloor,v),
$$
where the maximum is taken over all $v$ satisfying}
$$
0<v<1-\frac{r(s,L)}{\overline{R}(s-1,L)}\eqno (*)
$$
\end{itemize}

The following properties (Theorems 2.1 and 2.2)
of $\overline{R}(s,L)$ take place.
\medskip

{\bf Theorem 2.1}. {\it For any $L\ge 2$, there exists the
integer $s(L)\ge L+1$ such that}
$$
\overline{R}(s,L)=\left\{ \begin{array}{cc}
1/s, & \mbox{if}\; s<s(L),\\
<1/s, & \mbox{if}\; s\ge s(L),\\
\end{array}
\right.
$$
{\it and $s(L)=L\log L(1+o(1))$ for} $L~\to\infty$.
\medskip

Therefore, Theorem~2 improves the upper bound of
Proposition 3 provided that $s\ge s(L)$. The
computations give $s(1)=2$, $s(2)=6$, $s(3)=12$,
$s(4)=20$, $s(5)=25$, $s(6)=36$,\ldots. For $L=2$
and $s=6,7,\dots,13$, the values of $\overline{R}(s,2)$ are
given in {\bf Table 2}.

\smallskip
\begin{center}
\begin{tabular}{|c||c|c|c|c|c|c|c|c|}
\hline
$s$ & 6 & 7 & 8 & 9 & 10 & 11 & 12 & 13\\
\hline
\hline
$\overline{R}(s,2)$ & .163 & .141 & .117 & .102 & .086
&.076 & .066 & .059\\
\hline
\end{tabular}
\end{center}
\centerline{{\bf Table 2.}}
\bigskip

For $s=2,3,\dots,8$, the values of $\overline{R}(s,1)$ and
$\overline{R}(s,2)$ are given in {\bf Table 1}.
\medskip

{\bf Theorem 2.2}. {\it If $L\ge 2$ is fixed and
$s\to~\infty$, then}
$$
\overline{R}(s,L)=\frac{2L\log s}{s^{2}} (1+o(1)).
$$
\smallskip

With the help of Proposition~4, it follows
an upper bound on the design rate $R_D(s)$:
\smallskip

{\bf Corollary}. {\it For $s\ge2$, the design rate
$
R_D(s)\le\overline{R}_D(s)\eq\overline{R}(s-1,2).
$
If $s\to\infty$, then}
$$
\overline{R}_D(s)=\frac{4\log s}{s^{2}} (1+o(1)).
$$

From  the numerical values of {\bf Table 2},
we can conclude $\overline{R}(s-1,2)<1/s$, if $s\ge 11$. Hence,
Theorem~4 implies that for $s\ge 11$, the design rate
$
R_D(s)<1/s.
$
This improves the upper bound of Proposition 1.

\subsubsection{Lower bound on $R(s,L)$}
\begin{center}
\end{center}

To obtain the  random coding (lower) bound on $R(s,L)$,
for $L\ge2$, we  use the ensemble of random codes,
which was suggested in~\cite{nz88} for the particular case
$L=1$.  To formulate our results, we need some notations.
Let $K\ge s$ be an integer, $\xi_1,\xi_2,\dots,\xi_s$ and
$\eta_1,\eta_2,\dots,\eta_L$ be independent identically
distributed random variables with uniform distribution
on the set $[K]$, i.e.,
$$
\Pr\{\xi_i=k\}=
\Pr\{\eta_j=k\}\eq\;\frac{1}{K},\quad i=1,2,\dots,s,
\;j=1,2,\dots,L,\quad k=1,2,\dots,K.
$$
Define the probability
$$
Q_K(s,L)\eq\Pr\left\{\bigcap_{j=1}^L\;
\bigcup_{i=1}^s\;(\eta_j=\xi_i)\right\}
$$
and put
$$
E(s,L)\eq\max_{K\ge s}
\left\{-\frac{\log Q_K(s,L)}{K}\right\}.
$$
\smallskip

{\bf Theorem 3}.~\cite{r90}. {\it For any $s\ge2$ and
$L\ge2$, the rate}
$$
R(s,L)\ge\underline{R}(s,L)\eq\frac{E(s,L)}{s+L-1},
$$
\smallskip

Theorems 3.1 and 3.2 give the properties of the lower
bound $\underline{R}(s,L)$.
\smallskip

{\bf Theorem 3.1}. {\it For fixed $s\ge2$, define
$$
r_s\eq\max_{K\ge s}\left\{\frac{\log(K/s)}{K}\right\}=
\frac{\log(\lceil es\rceil/s)}{\lceil es\rceil},
$$
where we took into account that the maximum is achieved
at $K=\lceil es\rceil$.
The following statements are true}:
\begin{enumerate}
\item
{\it for any $s$ and $L$},
$$
\frac{Lr_s}{s+L-1}\le
\underline{R}(s,L)
\le\frac{Lr_s}{s+L-1}+\frac{\log e}{s+L-1};
$$
\item
{\it for fixed $s\ge2$, there exists}
$$
\underline{R}(s,\infty)
\eq\lim_{L\to\infty}\underline{R}(s,L)=r_s;
$$
\item
{\it for any $s\ge2$, $\underline{R}(s,\infty)\ge\log
e/\lceil e\;s\rceil$ and at $s\to\infty$}
$$
\underline{R}(s,\infty)=\frac{\log e}{e\cdot s}(1+o(1))=
\frac{0.5307}{s}(1+o(1)).
$$
\end {enumerate}
\smallskip

{\bf Theorem 3.2}. {\it If $L\ge1$ is fixed and
$s\to\infty$, then}
$$
\underline{R}(s,L)=\frac{L}{s^2\log e}(1+o(1)).
$$
\smallskip
For the fixed values of $s\ge2$ and $L\ge2$, the bound of
Theorem 3 could be improved~\cite{dra89} (see,
also,~\cite{dr94}) by the random coding method
of~\cite{dr89}.  This method uses the random ensemble of
constant-weight codes. In {\bf Table 1}, for
$s=2,3,\dots,8$, we give the corresponding values of
$\underline{R}(s,1)$ and $\underline{R}(s,2)$.  The
asymptotic lower bounds, obtained in Theorems~3.1-3.2, are
the best known.

\bigskip

{\bf Proof of Theorem 3}.
Let $n\ge1$ be integer and $[K]$ be the set of integers
from 1 to $K$.  Denote by
$Y=\|y_m(j)\|,\;m=1,2,\dots,n;\;j=1,2,\dots,t$ an arbitrary
$(n\times t)$ matrix with elements $y_m(j)\in[K]$. Let
$N=n\cdot K$.  For matrix $Y$, we denote by
$$
X_K=(\x(1),\x(2),\dots,\x(t))
$$
a binary $N\times t$ matrix, whose columns have the following form
$$
\x(j)=(\x^1(j),\x^2(j),\dots,\x^n(j)),\;j=1,2,\dots,t,\quad
$$
$$
\x^m(j)=(x_1^m(j),x_2^m(j),\dots,x_K^m(j)),\;m=1,2,\dots,n,
$$
$$
x_k^m(j)=\cases{1, &if $k=y_m(j)$,\cr
                0, &if $k\ne y_m(j)$.\cr}
$$
Obviously, each column (codeword) $\x(j)$ of matrix (code) $X_K$
contains $n=N/K$ $1$'s and $(N-n)$ $0$'s.

We say that a column $\x(j)$ is ``bad''
or code $X$ if the $\x(j)$ does not
satisfy the definition of LDSC. It follows that among the
rest $t-1$ columns there exist $L-1$ columns
$\x(j_1),\x(j_2),\dots,\x(j_{L-1})$ and $s$ columns
$\x(l_1),\x(l_2),\dots,\x(l_{s})$ for which
$$
\bigvee_{i=1}^s\x(l_i)\quad \mbox{covers} \quad
\x(j)\vee\left[\bigvee_{i=1}^{L-1}\x(j_i)\right].
$$

Let $Y$ be a random matrix whose components be
independent with distribution
$$
\Pr\{y_m(j)=k\}\eq\frac{1}{K},\quad k\in[K].
$$
This ensemble was suggested in~\cite{nz88}.
It is not hard to see that for any column $\x(j)$, the
ensemble probability to be ``bad'' does not exceed
$$
{t-1\choose s+L-1}{s+L-1\choose s}\left[Q_K(s,L)\right]^{N/K}.
$$
Hence, the arguments of the random coding
gives the statement of Theorem 3.
\bigskip

{\bf Proof of Theorem 3.1}. Let $K\ge s$. We need to prove
only Statement 1).  Using the definition of $Q_K(s,L)$ and
the theorem of total probability, we have
$$
Q_K(s,L)\eq\Pr\left\{
\bigcap_{j=1}^L\;\bigcup_{i=1}^s\;(\eta_j=\xi_i)\right\}=
$$
$$
=K^{-s}\sum_{k_1,k_2\dots,k_s}\left[
\Pr\left\{\bigcup_{i=1}^s\;(\eta_1=k_i)\right\}\right]^L,
\leqno 1)
$$
where the summation is taken over all $K^s$ ordered
$s$-collections of integers $(k_1,k_2\dots,k_s)$,
$k_i\in[K]$.

From this formula it follows the evident bound
$Q_K(s,L)\le(s/K)^L$.  Hence,
$$
E(s,L)\eq\max_{K\ge s}\left\{-\frac{\log
Q_K(s,L)}{K}\right\} \ge L\max_{K\ge
s}\left\{\frac{-\log(s/K)}{K}\right\}=r_s
$$
and the lower bound of statement~1)
is proved.

To prove the upper bound,
we introduce the concept of composition for a word
$(k_1,k_2\dots,k_s)$, i.e., the collection of nonnegative
integers $\|n_k\|,\,k=1,2,\dots,K$, where $n_k$ is the number
of positions $i,\,i=1,2,\dots,s$, for which $k_i=k$. We have
$$
0\le n_k\le s,\quad \sum_{k=1}^K n_k=s.
$$
Let the summation in 1) over $(k_1,k_2\dots,k_s)$
be replaced by the summation over $\|n_k\|$, i.e.,
$$
Q_K(s,L)=K^{-s}\sum_{\|n_k\|}\frac{s!}{\prod_{k=1}^K n_k!}
\left(\frac{\sum_{k=1}^K u_k}{K}\right)^L,\leqno 2)
$$
$$
u_k\eq\cases{1, & $n_k>0$,\cr
                0, & $n_k=0$.\cr}
$$

We can rewrite
formula 2) in the form
$$
Q_K(s,L)=K^{-s}\sum_{m=1}^s
B_K(s,m)\left(\frac{m}{K}\right)^L,\leqno 3)
$$
where
$$
B_K(s,m)\eq{K\choose m}
\sum_{(n_1,n_2\dots,n_m)}\frac{s!}{\prod_{i=1}^m n_i!}.\leqno 4)
$$
The summation
is taken over all ordered $m$-collections of
{\it positive} integers $(n_1,\dots,n_m)$, for which
$$
n_1+n_2+\cdots+n_m=s,\quad n_i>0,\;i=1,2,\dots,m.
$$
Obviously,
$$
B_K(s,s)={K\choose s}s!=\prod_{i=0}^{s-1}(K-i).
$$
If we restrict the summation in 3) by the member $m=s$,
then we obtain the lower bound
$$
Q_K(s,L)\ge\left(\frac{s}{K}\right)^L
\frac{\prod_{i=0}^{s-1}(K-i)}{K^s}>
\left(\frac{s}{K}\right)^L\frac{K!}{K^K}>
\left(\frac{s}{K}\right)^L e^{-K}.
$$
The last inequality is the consequence of the Stirling inequality
$K!>K^K\,e^{-K}$.
It follows
$$
-\log Q_K(s,L)\le L\log(K/s)+K\log e.
$$
This inequality yields the upper bound of Statement 1).

Theorem 3.1 is proved.
\medskip

{\bf Remark}. It is known (see, for example~\cite{com}, problem 3.9) that
$$
\sum_{(n_1,n_2\dots,n_m)}\frac{s!}{\prod_{i=1}^m n_i!}=\cases{
\sum_{i=0}^m\left(-1\right)^i{m\choose i}\left(m-i\right)^s, & if $s\ge m$,\cr
                                    0, & if $s<m$.\cr}
$$
Hence, from 3) it follows that
$$
Q_K(s,L)=\Pr\left\{\bigcap_{j=1}^L\;\bigcup_{i=1}^s\;(\eta_j=\xi_i)\right\}=
$$
$$
=K^{-s}\sum_{m=1}^{\min\{K;s\}}{K\choose m}\left(\frac{m}{K}\right)^L
\sum_{i=0}^m\left(-1\right)^i{m\choose i}\left(m-i\right)^s.
$$
\medskip

{\bf Proof of Theorem 3.2}. Let $L\le s\le K$.
For independent random variables $\xi_1,\xi_2,\dots,\xi_s$, we introduce
binary random variables $\alpha_1,\alpha_2,\dots,\alpha_K$, where
$$
\alpha_k\eq\cases{1, & if there exists $i=1,2,\dots,s$ and $\xi_i=k$,\cr
                     0, & otherwise.\cr}
$$
Let the {\it overline} denote the averaging over random variables
$\xi_1,\xi_2,\dots,\xi_s$. It is easy to check that formula 2) can be
rewritten in the form
$$
Q_K(s,L)=K^{-L}\overline{\left(\sum_{k=1}^K\alpha_k\right)^L}=
K^{-L}\sum_{m=1}^L B_K(L,m)q_K(s,m),\leqno 5)
$$
where coefficients $B_K(L,m),\; m=1,2,\dots,L,$ are defined by 4) and
$$
q_K(s,m)\eq\overline{\prod_{i=1}^m\alpha_i},\quad m=1,2,\dots,L.
$$
\medskip

{\bf Lemma}. {\it Let $L\le s\le K$. For $m=1,2,\dots,L$,
the following formula is true}
$$
q_K(s,m)=\sum_{i=0}^m\left(-1\right)^i{m\choose i}\left(\frac{K-i}{K}\right)^s.\leqno 6)
$$

{\bf Proof of Lemma}. Let $m=2,3\dots,L$. We have
$$
q_K(s,m)=\Pr\{\alpha_1=\alpha_2=\cdots=\alpha_m=1\}=
\Pr\left\{\bigcap_{k=1}^m\bigcup_{i=1}^s(\xi_i=k)\right\}=
$$
$$
=\sum_{j=1}^{s-(m-1)}\left(1-\frac{m}{K}\right)^{j-1}\frac{m}{K}
\;\Pr\left\{\bigcap_{k=1}^{m-1}\bigcup_{i=j+1}^s(\xi_i=k)\right\}.
$$
Hence, the following recurrent formula takes place
$$
q_K(s,m)=\sum_{j=1}^{s-(m-1)}\left(1-\frac{m}{K}\right)^{j-1}\;
\frac{m}{K}\;q_K(s-j,m-1).
\leqno 7)
$$

With the help of 7), we can check 6) by induction on $m=1,2,\dots,L$.
If $m=1$, then
$$
q_K(s,1)=\overline{\alpha_1}=\Pr\left\{\bigcup_{i=1}^s(\xi_i=1)\right\}=
1-\left(1-\frac{1}{K}\right)^s,
$$
i.e., for $m=1$, formula 6) is true. Let 6) be true for $q_K(s,m-1)$.
Consider
$$
q_K(s,m)=\sum_{j=1}^{s-(m-1)}\left(\frac{K-m}{K}\right)^{j-1}\;\frac{m}{K}
\left[\sum_{l=0}^{m-1}(-1)^l{m-1\choose l}
\left(\frac{K-l}{K}\right)^{s-j}\right]=
$$
$$
=1-\left(\frac{K-m}{K}\right)^{s-(m-1)}+\sum_{l=1}^{m-1}
\left[(-1)^l\;\frac{m{m-1\choose l}}{K}\left(\frac{K-l}{K}\right)^{s-1}
\left(1-\left(\frac{K-m}{K-l}\right)^{s-(m-1)}\right)\frac{K-l}{m-l}\right]=
$$
$$
=1-\left(\frac{K-m}{K}\right)^{s-(m-1)}+
\sum_{l=1}^{m-1}(-1)^l{m\choose l}\left(\frac{K-l}{K}\right)^s
\left[1-\left(\frac{K-m}{K-l}\right)^{s-(m-1)}\right]=
$$
$$
=\sum_{l=0}^{m-1}(-1)^l{m\choose l}\left(\frac{K-l}{K}\right)^s-
\left(\frac{K-m}{K}\right)^{s-(m-1)}\cdot
\sum_{l=0}^{m-1}(-1)^l{m\choose l}\left(\frac{K-l}{K}\right)^{m-1}.
$$
In order to complete the proof of Lemma, we need to check that
$$
\sum_{l=0}^{m-1}(-1)^l{m\choose l}\left(1-\frac{l}{K}\right)^{m-1}=0,\leqno 8)
$$
provided that $K\ge m$.

It is known (see,~\cite{com}, problem 3.10) that for any
$k=1,2,\dots,m-1$
$$
\sum_{l=0}^{m-1}(-1)^l{m\choose l}l^k=
\sum_{l=0}^{m-1}(-1)^l{m\choose l}(m-l)^k=0.
$$
It follows 8). Lemma is proved.
\medskip

To complete the proof of Theorem 3.2, we fix an arbitrary $L\ge1$.
Consider the following asymptotic conditions
$$
s\to\infty,\quad K\to\infty,\quad s/K=\lambda\le1\;-\;const.\leqno 9)
$$
Let $m=1,2,\dots,L$ be fixed. From 4) and 6) it follows that
$$
\lim_{9)}\frac{B_K(L,m)}{K^L}=\cases{0, & if $m\le L-1$,\cr
                                     1, & if $m=L$,\cr}
$$
$$
\lim_{9)}q_K(s,m)=\left(1-e^{-\lambda}\right)^m.
$$
Therefore, formula 5) yields
$$
\lim_{9)} Q_K(s,L)=\left(1-e^{-\lambda}\right)^L.\leqno 10)
$$
From the definition of $\underline{R}(s,L)$ and 10) it follows that
$$
\lim_{s\to\infty}s^2\underline{R}(s,L)=
L\max_{0\le\lambda\le1}\left\{-\lambda\log\left(1-e^{-\lambda}\right)\right\}
=\frac{L}{\log e},
$$
and the extreme  value of $\lambda=(\log e)^{-1}=\ln 2=0.6931.$

Theorem 3.2 is proved.

\subsection{Kautz-Singleton codes}
\begin{center}
\end{center}

Theorems 1 and 3 of {\bf Sect. 2.2} are only theorems of
existence.  They do not give any method for the
construction of the ``good'' codes. The first question,
arising when one tries to apply Theorem 1, is the
following. How many {\it steps $S$ of computation} one
must make to check, that a given matrix $X$ with
parameters corresponding to the bound of  Theorem~1,
satisfies the definition of an $(s,1,t)$-code?  If one step
is the  computation of a Boolean sum and checking of
covering of the two binary codewords of length $N$, then
the number $S$ evidently has the order of $t^{s+1}$.  For
$t=10^3,\dots,10^4$, and $s=5,\dots,15$, which occur in
applications~\cite{ks64,rd81,rd88}, $S$ becomes
astronomically great
$$
S=10^{18},\dots,10^{64}.
$$

Is it possible to find any simple sufficient condition for
matrix $X$ to be an  $(s,1,t)$-code and the checking of this
condition takes essentially less computation steps? The important
sufficient condition is given in~\cite{ks64} and formulated
below as Theorem 4.
\smallskip

{\bf Theorem 4}~\cite{ks64}. {\it Let $X$ be a
constant-weight code, i.e., $X$ be a binary $N\times t$
matrix, whose columns (codewords) $\x(j)$ have the same
number of $1's$
$$
w=\sum_{i=1}^N x_i(j),\quad j=1,2,\dots,t,
$$
and
$$
\lambda\eq\max_{k\ne j}\sum_{i=1}^N x_i(k)\;x_i(j)
$$
be the maximal correlation of codewords. Then the matrix
$X$ is $(s,1,t)$-code for any $s$, satisfying the
inequality}
$$
s\le\left\lfloor\frac{w-1}{\lambda}\right\rfloor.
$$

The computation of the number $\lceil\frac{w-1}{\lambda}\rceil$
for the matrix $X$ , whose columns have the same weight $w$ takes
$S={t\choose 2}\sim t^2/2$ steps, where one step is the
computation of
$$
\lambda_{k\,j}\eq\sum_{i=1}^N
x_i(k)\;x_i(j).
$$
For the above-mentioned values of
parameters $s$ and $t$, this number has the order $$
S=10^6,\dots,10^8,
$$
which is acceptable from the practical point of view.
The most of known constructions~\cite{ks64,dh93} of
$(s,1,N)$-codes were obtained with the help of Theorem~4.
Below, in this section, we consider the upper and lower
bounds on the optimal parameters of such codes.
\medskip

{\bf Definition}.~\cite{dr83}. Let $1\le\lambda\le w\le N$
be given integers, and let $X$ be a code of size $t$,
length $N$ with parameters $w$ and $\lambda$.  A code $X$
will be called an {\it KS-superimposed code (KSSC) of length
$N$, size $t$ and strength $s$}, if inequality
$s\le\lceil\frac{w-1}{\lambda}\rceil$ holds.  This code
also will be called an $(s,N)$-KS-code of size $t$, or
an $(s,t)$-KS-code of length $N$.

Let $t_{KS}(s,N)$ be the maximal possible size of KSSC and
$N_{KS}(s,t)$ be the minimal possible length of KSSC. For fixed
$s\ge1$, define the {\it rate} of KSSC.
$$
R_{KS}(s)\eq
\varlimsup_{N\to\infty}\frac{\log t_{KS}(s,N)}{N}.
$$

From Theorem 4 it follows that any $(s,N)$-KS-code is also
$(s,1,N)$-code.  Therefore,
$$
N_{KS}(s,t)\ge N(s,1,t),\quad t_{KS}(s,N)\le t(s,1,N),\quad
R_{KS}(s)\le R(s,1).
$$
Thus, the lower bound on $N(s,1,t)$, given by Proposition~6
can be considered as the lower bound on $N_{KS}(s,t)$.
The following Proposition~7 gives an
improved (roughly twice) lower bound on $N_{KS}(s,t)$.

{\bf Proposition 7.}~\cite{dr83}
$$
N_{KS}(s,t)\ge\min\left
\{t;\;\frac{s(s+1)}{1+s/t}\right\}\ge\min\{t;\,s^{2}\}.
$$

{\bf Proof}. Let $X$ be a constant-weight code of length
$N$ and size $t$ with parameters $w$ and $\lambda$. The
well-known Johnson bound~\cite{j62} yields the inequality
$$
N\ge\frac{tw^2}{\lambda(t-1)+w}.
$$
By virtue of Lemma 1, if $X$ is a $(s,1,t)$-code of length
$N\le(t-1)$, then weight $w\ge(s+1)$. Hence, the
above-mentioned inequality  yields the statement of
Proposition 7.
\medskip

\subsubsection{Upper bound on $R_{KS}(s)$}
\begin{center}
\end{center}

We give here (without proof) the best known
upper bound on $R_{KS}(s)$. This bound was obtained
with the help of the best known~\cite{mrrw77} upper bound
on the rate for the constant-weight codes.

{\bf Theorem 5}. {\it For any $s\ge1$}
$$
R_{KS}(s)\le\overline{R}_{KS}(s)
\eq h\left(\frac12-\frac{\sqrt{s(s-1)}}{2s-1}\right).
$$
\medskip

We have $\overline{R}_{KS}(2)=0,187,\;\overline{R}_{KS}(3)=0,081$
and as $s\to\infty$
$$
\overline{R}_{KS}(s)=\frac{\log s}{8s^2}(1+o(1)).
$$

\subsubsection{Lower bound on $R_{KS}(s)$}
\begin{center}
\end{center}

The following Theorem 6 is called a {\em random coding
bound on $R_{KS}(s)$}.
\medskip

{\bf Theorem 6}~\cite{dr83}. {\it For any $s\ge1$
$$
R_{KS}(s)\ge\underline{R}_{KS}(s)
\eq\max_{0<p<s^{-1}}E(s,p),\leqno 1)
$$
where}
$$
E(s,p)=h(p)-ph(s^{-1})-(1-p)
h\left(\frac{p(s-1)}{(1-p)s}\right).\leqno 2)
$$
\medskip

Let the maximum in 1) be achieved at $p=p_s$. The equation
for computation $p_s$ could be written in the form
$$
p=\frac{(1-p)^{2s}(s-1)^{2(s-1)}}{s-(2s-1)p}.\leqno 3)
$$
If $s=1$, then the root of 3) is $p_1=1$. If $s\ge2$, then
3) can be solved numerically by the method of consecutive
approximation.  For $s=2$ and $s=3$, the following values
are obtained:
$$
p_2=0.13846,\;
\underline{R}_{KS}(2)=0,09415;\quad p_3=0.08222,\;
\underline{R}_{KS}(3)=0,03495.
$$
In addition, if $s\to\infty$, then
$$
p_s=\frac{a}{s}(1+o(1)),
$$
where
$a=.203188$ is the unique solution of equation
$a=e^{2(a-1)}$. It follows that if $s\to\infty$, then
the lower bound
$$
\underline{R}_{KS}(s)=\frac{-a\log[ae^{1-a}]}{s^2}(1+o(1))=
\frac{0,23358}{s^2}(1+o(1)).
$$
\medskip

{\bf Proof of Theorem 6}. Let $X=\|x_i(j)\|$ be a fixed $N\times 2t$
matrix, whose columns have weight $w$, $w=1,2,\dots,N$. A column
$\x(j)$ is called ``bad'', if there exists a column $\x(k),\;k\ne j$
that the correlation
$$
\lambda_{kj}\ge\left\lfloor\frac{w-1}{s}\right\rfloor\eq\lambda_0.
$$
Otherwise, the column $\x(j)$ is called ``good''. Denote by
$t_1=t_1(X)$ ($t_2=t_2(X)$) the number of ``good'' (``bad'')
columns of $X$. Obviously, $t_1+t_2=2t$.

Let $X$ be the random $N\times 2t$ matrix, whose $2t$ columns
are selected
with replacement from the set of ${N\choose w}$
constant-weight columns. Introduce the events
$$
B_{kj}=\{\lambda_{kj}\ge\lambda_0\},\quad
B^j=\bigcup_{k=1\,k\ne j}^{2t}B_{kj}.
$$
The probability of event $B_{kj}$ does not depend on $k$ and $j$ and
$$
\Pr\{B_{kj}\}=\frac{\sum_{m=\lambda_0}^w{w\choose m}{N-w\choose w-m}}
{{N\choose w}}\eq q_s(N,w).
$$
Hence, $\Pr\{B^j\}\le 2tq_s(N,w)$ and the expectation
$$
{\bf M}(t_2(X))=\sum_{j=1}^{2t}\Pr\{B^j\}\le 4t^2 q_s(N,w).
$$
With the help of the  arguments of the random
coding method it follows
\medskip

{\bf Lemma}. {\it If $q_s(N,w)<(4t)^{-1}$, then there exists
$(s,N)$-KS-code of size $t$}.
\medskip

Let $p,\;0<p<s^{-1}$ be fixed. From Lemma it follows that the rate of KSSC
$$
R_{KS}(s)\ge\varliminf_{N\to\infty}\frac{-\log q_s(N,\lfloor Np\rfloor)}{N}=
h(p)-ph(s^{-1})-(1-p)h\left(\frac{p(s-1)}{(1-p)s}\right).
$$
To obtain the last equality we applied the well-known asymptotic
behavior  of the binomial coefficients~\cite{g68}.

Theorem 6 is proved.
\newpage

\subsection{Symmetrical superimposed codes}
\begin{center}
\end{center}

Let $\y(j)=(y_1(j),y_2(j),\dots,y_N(j)),\; j=1,2,\dots,s$
be binary columns of length $N$. By definition, the {\it symmetrical}
Boolean sum
$$
\y=\y(1)\diamond\y(2)\diamond\cdots\diamond\y(s)
$$
of columns $\y(1),\y(2),\dots,\y(s)$ is the binary column
$\y=(y_1,y_2,\dots,y_N)$ with components
$$
y_i=\cases{0, & if $y_i(1)=y_i(2)=\cdots=y_i(s)=0$,\cr
           1, & if $y_i(1)=y_i(2)=\cdots=y_i(s)=1$,\cr
           *, & otherwise.\cr}
$$

Let us say that the symmetrical Boolean sum
$\y(1)\diamond\y(2)\diamond\cdots\diamond\y(s)$
{\it cover} binary column $\z$, if  column $\z$
do not change this sum, i.e.,
$$
\y(1)\diamond\y(2)\diamond\cdots\diamond\y(s)=
\y(1)\diamond\y(2)\diamond\cdots\diamond\y(s)\diamond\z.
$$

{\bf Definition}. An $N\times t$--matrix $X$ is called a
{\it list-decoding symmetrical superimposed code} (LDSSC)
{\it of length $N$, size $t$, strength $s$}, and {\it
list-size} $\le L-1$ if the symmetrical Boolean sum of any
$s$-subset of codewords $X$ can cover not more than $L-1$
codewords that are not components of the $s$-subset. This
code also will be called an $(s,L,N)^{\diamond}$-code of
size $t$, or $(s,L,t)^{\diamond}$-code of length $N$.

Let $t_{\diamond}(s,L,N)$ be the maximal possible size of LDSSC
and $N_{\diamond}(s,L,t)$ be the minimal possible length of LDSSC.
For  fixed $s$ and $L$, define
the {\it rate} of LDSSC
$$
R_{\diamond}(s,L)\eq\varlimsup_{N\to\infty}
\frac{\log t_{\diamond}(s,L,N)}{N}.
$$

We apply the notations of {\bf Sect. 1} for the optimal
parameters of SD--model, i.e., $t_{SD}(s,N)$,
$N_{SD}(s,t)$ and $R_{SD}(s)$.  \medskip

Below we give two evident propositions, showing the
connection between codes for symmetrical Boolean sum and
superimposed codes of {\bf Sect 2.1}.

{\bf Proposition 8}. {\it Any superimposed code
is a code for the symmetrical Boolean sum}.  \smallskip

{\bf Proposition 9}.
{\it Let an $N\times t$ matrix
$X=\|x_i(j)\|,\;i=1,2,\dots,N,\;j=1,2,\dots,t$ be a symmetrical
superimposed code of length $N$. Consider $N\times t$ matrix
$X'=\|x'_i(j)\|$ with elements}
$$
x'_i(j)\eq\cases{1, & if $x_i(j)=0$,\cr
               0, & if $x_i(j)=1$.\cr}
$$
{\em Then the $2N\times t$ matrix
composed of the rows of the two matrices $X$ and $X'$ is a
superimposed code}.
\smallskip

From propositions 9 and 10 it follows that
$$
\frac12\,N(s,L,t)\le N_{\diamond}(s,L,t)\le N(s,L,t),\quad
R(s,L)\le R_{\diamond}(s,L)\le2R(s,L).
$$
These inequalities permit to obtain the bounds on the length
and rate of symmetrical superimposed codes using the corresponding
bounds which are presented in~{\bf Sect.~2.2-2.3}.

\newpage
\centerline{Arkadii G. D'yachkov,\qquad "Lectures on
Designing Screening Experiments"}
\medskip

\section{Constructions of Superimposed Codes}
\begin{center}
\end{center}

Kautz-Singleton (1964)~\cite{ks64} suggested a class of
binary superimposed codes which are based on the
$q$-ary Reed-Solomon codes (RS-codes)~\cite{ms83}.
Applying a concatenation of the binary constant-weight
error-correcting codes~\cite{ms83} and the shortened
RS-codes, we obtain new constructions of superimposed
codes.  Tables of their parameters are given. From the
tables it follows that the rate of obtained codes exceeds
the corresponding random coding bound~\cite{dr89}.

\subsection{Notations and definitions}
\begin{center}
\end{center}

Let $1\le s<t,\; N>1$ be integers and $X=\|x_i(u)\|$,
$i=1,2,\dots,N$, $u=1,2,\dots,t,$ be a binary $(N\times
t)$--matrix (code) of {\it size} $t$ and {\it length} $N$
with columns (codewords) $\x(1),\x(2),\dots,\x(t)$, where
$\x(u)=(x_1(u),x_2(u),\dots,x_N(u))$.  Let $\defeq$ denote
the equation by definition.  For code $X$, let $w$ and
$\lambda$ be defined by
$$
w\defeq\min_u\sum_{i=1}^N x_i(u),\quad
\lambda\defeq\max_{u,v}\sum_{i=1}^N x_i(u)x_i(v).
$$
$w$ is the {\it minimal weight of codewords}
and  $\lambda$ is the
{\it maximal correlation of codewords}.

We say that the binary column $\x$ {\it covers} the binary
column $\y$ if the Boolean sum $\x\vee\y=\x$. The code $X$
is called~\cite{ks64,dr82,dr83,dh93} a {\it superimposed}
$(s,N,t)$-code, or {\it $s$-disjunct} code if the
Boolean sum of any $s$-subset of columns of $X$ covers only
those columns of $X$ which are the terms of the given
Boolean sum.

One can consider an arbitrary fixed code $X$ as the
incidence matrix of a $t$-family of subsets of the $N$-set.
In this interpretation, the $s$-{\it disjunct code $X$
one-to-one corresponds to the  family in which no set is
covered by the union of $s$ others}.

Superimposed codes were introduced by Kautz -
Singleton~\cite{ks64} who worked out the constructive
methods and some applications.  Dyachkov - Macula -
Rykov~\cite{dmr00-2} investigated the development of
constructions for superimposed codes (non-adaptive pooling
designs) intended for the clone-library screening problem.
(See Knill, Bruno, Torney~\cite{kbt98}). In Section~3.2, we
give a brief introduction to the problem.

In Sect.~3.3 and~3.4, we study the most important class of
superimposed codes which are based on the $q$-ary
Reed-Solomon codes (RS-codes)~\cite{ms83}. The given class
was invented by Kautz-Singleton~\cite{ks64}. We introduce
some generalizations of the Kautz-Singleton codes and
identify the parameters of the best known superimposed
codes.

\subsection{Application to DNA library screening}
\begin{center}
\end{center}

To  understand what a DNA library is, think of
{\em several copies} of an identical but incredible
long word (of length $\sim10^8$, e.g., a chromosome)
from letters of the  quaternary alphabet
$\{A,C,G,T\}$. Each copy of the word has been cut
in thousands of contiguous pieces (of length
$\sim10^4$, e.g., chromosome fragments).  Take those
pieces and copy their letter strings onto their own
separate small piece of paper. The thousands of little
pieces of paper (i.e., clones) that result essentially
constitute a DNA library.  In other words, each
{\em clone}
represents some contiguous subpiece of a contiguous
superpiece of DNA.  The {\em DNA library}, or the
{\em clone-library} consists of thousands separate
clones.

An unique  and contiguous
sub-subpiece of DNA (of length $\sim10^2$)
is called a {\em sequenced tagged site} (STS). For a fixed
 STS, a clone is called positive (negative) for that STS if
it contains (does not contain) that given STS.

{\bf Example}. Let the following $s=4$
copies of the DNA superpiece be given and
$\{C_1,C_2,C_3,C_4,C_5\}$ be the library of 5
clones.
$$
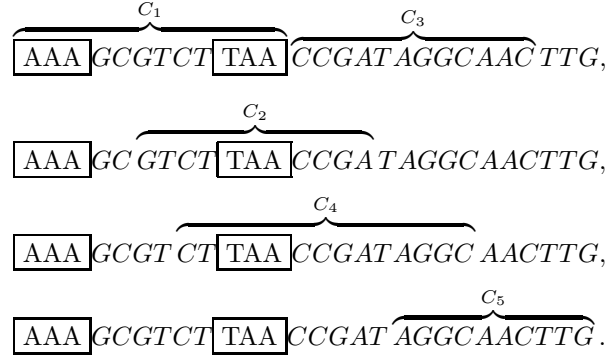
\overbrace{\fbox{AAA}GCGTCT\fbox{TAA}}^{C_1}
\overbrace{CCGATAGGCAAC}^{C_3}TTG,
$$
$$
\fbox{AAA}GC\overbrace{GTCT\fbox{TAA}CCGA}^{C_2}
TAGGCAACTTG,
$$
$$
\fbox{AAA}GCGT\overbrace{CT\fbox{TAA}CCGATAGGC}^{C_4}
AACTTG,
$$
$$
\fbox{AAA}GCGTCT\fbox{TAA}CCGAT
\overbrace{AGGCAACTTG}^{C_5}.
$$
Clones $\{C_1,C_3\}$ could be taken from the same copy of
the DNA superpiece. Clones $\{C_2,C_4\}$ are taken from the
different copies. Let $\mbox{STS}_1=\fbox{AAA}$ and
$\mbox{STS}_2=\fbox{TAA}$.  Then $C_1$ is positive for
$\fbox{AAA}$ and $C_1,\;C_2$ and $C_4$ are positive for
$\fbox{TAA}$. Note that $C_1$ is positive for both
$\fbox{AAA}$ and $\fbox{TAA}$. Clones $C_3,\;C_5$ are
negative  for both $\fbox{AAA}$ and $\fbox{TAA}$.

A {\em pool} is a subset of clones.
Each pool is tested as a group by exposing
that entire group to a chemical probe
(e.g. polymerase chain reaction) which can detect
 a given STS. A pool is
called {\em positive for the STS} if the probe
indicates that some member of that group contains the given
STS. In other words, if the tests are error-free, then a
pool is positive for an STS if that pool contains at least
one clone that contains the given STS.
\medskip

 Let $1\le s<t,\; N>1$ be integers.  Mathematically,
{\em clone-library  screening} for positive clones is
modeled by  searching a $t$-set of objects
(clone-library) for a particular $p$-subset,
$p\le s$, called a {\em subset of positive clones}.  A
non-adaptive pooling design is a
series of $N$ apriori
group tests that can often be carried out
simultaneously. Every parallel pooling design
is non-adaptive.
The pooling design can be described by a
binary $N\times t$-matrix $X=\|x_i(u)\|$, $i=1,2,\dots,N$,
$u=1,2,\dots,t$, where an element $x_i(u)=1$ if the $u$-th
clone is in the $i$-th pool and $x_i(u)=0$,
otherwise.
{\em A pool  outcome} (result of the group test)
is said to be positive if one of the pool's clones is
positive, negative otherwise.  Using this binary
$N$-sequence of outcomes, an investigator {\em has to
identify} the $p$-subset, $p\le s$, of positive clones.

Let $p\le s$ be the number of positive clones in a
clone-library of size $t$. To identify
an unknown $p$-subset of
positive clones, we apply the pooling design $X$ which is
the superimposed $(s,N,t)$-code. Obviously, the binary
$N$-sequence $\y$ of the pool outcomes is the Boolean sum
of the unknown $p$-subset of columns of~$X$. The definition
of $(s,N,t)$-code means that the
unknown $p$-subset is represented by all columns
which are below $\y$.
Thus, we need to carry out $\le t$ successive
comparisons of the Boolean sum $\y$ with codewords of $X$.
Hence, the identification complexity of $(s,N,t)$-code does
not exceed $t$.

\subsection{Generalized Kautz-Singleton codes}
\begin{center}
\end{center}

Let ${\cal P}$ be the set of all primes or
prime powers $\ge2$, i.e.,
$$
{\cal P}\defeq
\{2,3,4,5,7,8,9,11,13,16,17,19,23,25,27,29,31,32,37,\dots\}.
$$
Let $q_0\in{\cal P}$ and $2\le k\le q_0+1$ be fixed
integers for which there exists the $q_0$-ary
Reed-Solomon code (RS-code)
$B$ of size $q^k_0$, length $(q_0+1)$ and
distance
$d=q_0-k+2=q_0+1-(k-1)$~\cite{ms83}. We will identify
the code $B$ with an
$\left((q_0+1)\times q_0^k\right)$--matrix whose
columns, (i.e., $(q_0+1)$-sequences from the alphabet
$\{0,1,2,\dots,q_0-1\}$) are the codewords
of $B$. Therefore, the maximal
possible number of positions (rows) where its two
codewords (columns) can coincide, called a {\em
coincidence} of code $B$, is equal to $k-1$.

Fix an arbitrary integer $r=0,1,2,\dots,k-1$ and introduce
the {\it shortened} RS-code
$\tilde B$ of size $t=q_0^{k-r}$, length $q_0+1-r$ that
has the same distance $d=q_0-k+2$. Code $\tilde B$
is obtained by the {\it shortening} of the {\it subcode} of
$B$ which contains $0's$ in the first $r$ positions (rows)
of $B$.  Obviously, the coincidence of code $\tilde B$
 is equal to
$$
\lambda_0\defeq (q_0+1-r)-d=q_0+1-r-(q_0-k+2)=k-r-1.
\eqno (1)
$$

Consider the following standard transformation of the
$q_0$-ary code $\tilde B$ into the binary constant-weight
code $X$ of size $q_0^{k-r}$,
length $(q_0+1-r)q_0$ and weight
$w=q_0+1-r$. Each symbol of the $q_0$-ary alphabet
$\{0,1,2,\dots,q_0-1\}$ is substituted for the corresponding
binary column of the length $q_0$ and the weight $1$,
namely:
$$
0\Leftrightarrow\underbrace{(1,0,0,\dots,0)}_{q_0},\quad
1\Leftrightarrow\underbrace{(0,1,0,\dots,0)}_{q_0},\quad
\dots,\quad
q_0-1\Leftrightarrow\underbrace{(0,0,0,\dots,1)}_{q_0}.
$$
From (1) it follows that for binary code $X$,
the maximal correlation of codewords is
$\lambda=\lambda_0=k-r-1$.

Let $X$ be a binary code with parameters $w$ and
$\lambda$. Kautz-Singleton~\cite{ks64} suggested
the following evident sufficient condition
of the $s$-disjunct property:
$$
s\lambda\le w-1. \eqno (2)
$$
Hence, by virtue of (1) and (2), code $X$
is the $s$-disjunct code if
$$
s(k-r-1)=s\lambda_0\le w-1=q_0-r. \eqno (3)
$$
For the particular case $r=0$, this construction of
$s$-disjunct codes was suggested in~\cite{ks64}.
\medskip

Denote by $\lceil b\rceil$  the least integer $\ge b$.
Let  $m\ge1$  and $2\le s<2^m$
be arbitrary fixed integers.
Define
$$
{\cal P}(m,s)\defeq\left\{q\,:q\in{\cal P},\;
s\left(\left\lceil\frac{m}{\log_2q}\right\rceil
-1\right)\le q\right\}. \eqno (4)
$$
Consider a binary code $X$
identified by parameters $0\le r\le k-1\le q_0\;$,
$q_0\in{\cal P}(m,s)$.  The sufficient condition (3) for
$s$-disjunct property of $X$ could be written in the form
$
k-r\le\frac{q_0+1-k}{s-1}+1.
$
Hence, if
$$
\left\lceil\frac{m}{\log_2q_0}\right\rceil
\le k-r\le\frac{q_0+1-k}{s-1}+1,
\eqno (5)
$$
then code $X$ has $s$-disjunct property and its size
$t=q_0^{k-r}\ge 2^m$.

For fixed value $q_0\in{\cal P}(m,s)$, denote by
$$
N(q_0,s,m)=\min_{(5)}
\left\{q_0(q_0+1-r)\right\}, \eqno (6)
$$
the {\it minimal possible length of $s$-disjunct codes of
size $\ge 2^m$ which are based on
the $q_0$-ary shortened RS-codes}.
The minimum in (6) is taken over all parameters
$0\le r\le k-1\le q_0$ for which (5) is true.
\medskip

{\bf Lemma}. {\it The minimum in $(6)$ is achieved at
$$
k\,=\, q_0+s-(s-1)
\left\lceil\frac{m}{\log_2q_0}\right\rceil,
\eqno (7)
$$
$$
r\,=\, k-
\left\lceil\frac{m}{\log_2q_0}\right\rceil=
q_0-s\left(\left\lceil\frac{m}{\log_2q_0}\right\rceil-
1\right)\ge0 \eqno (8)
$$
and the optimal length is given by the formula
$$
N(q_0,s,m)=q_0\left[
s\left(\left\lceil\frac{m}{\log_2q_0}\right\rceil
-1\right)+1\right]. \eqno (9)
$$
}
\medskip

{\bf Proof}.  Fix an arbitrary $q_0\in{\cal P}(m,s)$.
Let the integer $k$ be defined by (7).
One can easily check that $k$
is the root of equation
$
\left\lceil\frac{m}{\log_2q_0}\right\rceil
=\frac{q_0+1-k}{s-1}+1.
$
Hence, the given value of $k$ is the maximal possible
integer satisfying (5). From the left-hand side of (5)
it follows
$
r\,\le\, k-
\left\lceil\frac{m}{\log_2q_0}\right\rceil.
$
It means that the maximal possible value of $r$
is given by (8). Therefore, the minimum in (6) is
achieved at $r$ defined by (8) and the corresponding
minimal code length is equal to the right-hand side (9).

Lemma is proved.
\medskip

We can summarize as follows.

{\bf Proposition 1}. {\it Fix an arbitrary $q_0\in{\cal
P}(m,s)$, where the subset
${\cal P}(m,s)\subset{\cal P}$ is defined by $(4)$.
For the binary code $X$ based on the $q_0$-ary
shortened RS-code with parameters $(7)-(8)$, we have}:
\begin{enumerate}
\item
$X$ {\it is
the $s$-disjunct constant-weight code of size $t$, length
$N$, weight $w$ and the maximal correlation of codewords
$\lambda$, where}
$$
\lambda=\lambda_0=k-r-1=\left\lceil
\frac{m}{\log_2q_0}\right\rceil-1,\quad
w=q_0+1-r=
s\lambda_0+1,
$$
$$
t=q_0^{k-r}=q_0^{\lambda_0+1}=
q_0^{\left\lceil\frac{m}{\log_2q_0}\right\rceil}
\ge2^m,
\qquad N=q_0(s\lambda_0+1);
$$
\item
{\it the length $N$ of code $X$ coincides with the minimal
possible length $N(q_0,s,m)$ defined by} (9).
\end{enumerate}

Denote by $N(s,m)$ the minimal
possible length of $s$-disjunct codes of size $t\ge 2^m$
which are based on the shortened  RS-codes, i.e.,
$$
N(s,m)=\min_{{\cal P}(s,m)}N(q,s,m)=
\min_{{\cal P}(s,m)}\left\{q\left[
s\left(\left\lceil\frac{m}{\log_2q}\right\rceil
-1\right)+1\right]\right\},
$$
$$
{\cal P}(m,s)=\left\{q\,:q\in{\cal P},\;
s\left(\left\lceil\frac{m}{\log_2q}\right\rceil
-1\right)\le q\right\}.
$$

In  Table 1, we give numerical values of  $N=N(s,m)$, $q_0$
and $\lambda_0$, when $s=2,3,\dots,8$, $m=5,6,\dots,30$ and
the code size $t=q_0^{\lambda_0+1}$ satisfies the
inequalities $2^m\le t<2^{m+1}$.  The optimal parameters
are identified as follows:
$$
q_0\ge s\lambda_0,\quad r=q_0-s\lambda_0,\quad
k=r+\lambda_0+1, \quad
\lambda_0=\left\lceil
\frac{m}{\log_2q_0}\right\rceil-1,
$$
$$
\eqno (10)
$$
$$
w=n_0=s\lambda_0+1, \quad N=q_0(s\lambda_0+1), \quad
t=q_0^{\lambda_0+1},\quad 2^m\le t<2^{m+1}.
$$
Table 1 shows the solutions of (10), i.e., $\lambda_0$
and $q_0$, yielding the the  minimal length $N=N(s,m)$ if
for the given integers $s$ and $m$, these solutions  exist.
The optimal solutions were calculated with the help of a
{\em computer program}.

\medskip

{\bf Example}. For the case $s=3,\,m=10$, Table 1
gives $q_0=11$, $\lambda_0=2$, $N=77$. It means that there
exists $3$-disjunct constant-weight code with parameters
$$
\lambda=\lambda_0=2,\quad
w=s\lambda_0+1=7,\quad
t=q_0^{\lambda_0+1}=11^3=1331,\quad
N=11\cdot7=77.
$$
This code is obtained from shortened RS-code with
parameters $q_0=11$, $k=7$ and $r=4$.
\medskip

{\bf Remark 1}.  By the boldface type,
we marked two examples of
the superimposed code parameters which were known
from~\cite{ks64,dr83}.
\medskip

{\bf Remark 2}. In Sect.~4, based on paper~\cite{dr01},
we give the detailed investigation of superimposed
codes with parameter $\lambda=1$.
\medskip

{\bf Discussion}.\\
Table 1 contains the values $\underline{R}(s)$
(lower bound on the rate $R(s)$ of the optimal code),
the values $\overline{R}(s)$
(upper bound on the rate $R(s)$ of the optimal code)
and the values of the rate
for several obtained codes, namely: the values of
fraction $\frac{m}{N},\;m=12,\,20,\,25,\,29$. The
comparison yields the following conclusions.
\begin{itemize}
\item
If $s=2$ and $m\le15$, then the values $\frac{m}{N}$
exceed the random coding rate
$\underline{R}(2)=.182$.
\item
If $s\ge3$ and $m\le30$, then the values $\frac{m}{N}$
exceed the random coding rate
$\underline{R}(s)$.
\end{itemize}


\subsubsection*{Table 1. Parameters of  constant-weight
(s,N,t)-codes of strength $s,\;2\le s\le 8$,\\
weight $w$,  length $N$, size $t=q_0^{\lambda_0+1}$,
$2^m\le t<2^{m+1}$, $5\le m\le30$, based on the\\
$q_0$-ary shortened Reed-Solomon codes.}

\begin{center}
\begin{tabular}{|c||c|c|c|c|c|c|c|}
\hline
$s$ & 2 & 3 & 4 & 5 & 6 & 7 & 8\\
\hline
$\underline{R}(s)$  &$.182$
&$.079$ &$.044$
&$.028$ &$.019$
&$.014$ &$.011$\\
\hline
$\overline{R}(s)$ &$.322$ &
$.199$ &$.140$ &
$.106$ &$.083$ &
$.067$ &$.056$\\
\hline
$m$ & $q_0,\,\lambda_0,\,N$ &
$q_0,\,\lambda_0,\,N$ & $q_0,\,\lambda_0,\,N$ &
$q_0,\,\lambda_0,\,N$ & $q_0,\,\lambda_0,\,N$ &
$q_0,\,\lambda_0,\,N$ & $q_0,\,\lambda_0,\,N$\\
\hline
$5$ & $-$ & $7,\,1,\,28$ & $7,\,1,\,35$
& $7,\,1,\,42$ & $7,\,1,\,49$ & $-$ & $-$ \\
\hline
$6$ & $4,\,2,\,20$ &
$8,\,1,\,32$ & $8,\,1,\,40$ & $8,\,1,\,48$
& $8,\,1,\,56$ & $9,\,1,\,72$ & $11,\,1,\,99$ \\
\hline
$7$ & $-$ &
$-$ & $13,\,1,\,65$ &$13,\,1,\,78$&
$13,\,1,\,91$&$13,\,1,\,104$&$13,\,1,\,117$\\
\hline
$8$ & $7,\,2,\,35$ &
$7,\,2,\,49$ & $-$ &
$16,\,1,\,96$ & $16,\,1,\,112$&$16,\,1,\,128$
&$16,\,1,\,144$\\
\hline
$9$ & $8,\,2,\,40$ &
$8,\,2,\,56$ & $8,\,2,\,72$ &
$-$ & $23,\,1,\,161$ &
$23,\,1,\,184$ & $23,\,1,\,207$\\
\hline
$10$ & $-$ &
$11,\,2,\,77$ & $11,\,2,\,99$ &
$11,\,2,\,121$ & $-$ &
$-$ & $-$\\
\hline
$11$ & $7,\,3,\,49$ &
$-$ & $13,\,2,\,117$ &
$13,\,2,\,143$ & $13,\,2,\,169$ &
$-$ & $-$\\
\hline
$12$ & $8,\,3,\,56$ &
$9,\,3,\,90$ & $16,\,2,\,144$ &
$16,\,2,\,176$ & $16,\,2,\,208$ &
$16,\,2,\,240$ & {\bf16,\,2,\,272}\\
$\frac{12}{N}$ &.214 &.133 &.083 &.068 &.058 &.050 &
.044\\
\hline
$13$ & $-$ &
$11,\,3,\,110$ & $-$ &
$23,\,2,\,253$ & $23,\,2,\,299$ &
$23,\,2,\,345$ & $23,\,2,\,391$\\
\hline
$14$ & $-$ &
$13,\,3,\,130$ & $13,\,3,\,169$ &
$-$ & $27,\,2,\,351$ &
$27,\,2,\,405$ & $27,\,2,\,459$\\
\hline
$15$ & $8,\,4,\,72$ &
$-$ & $-$ &
$-$ & $-$ &
$32,\,2,\,480$ & $32,\,2,\,544$\\
\hline
$16$ & $-$ &
$16,\,3,\,160$ & $16,\,3,\,208$ &
$16,\,3,\,256$ & $19,\,3,\,361$ &
$-$ & $-$\\
\hline
$17$ & $11,\,4,\,99$ &
$-$ & $-$ &
$-$ & $-$ &
$-$ & $-$\\
\hline
$18$ & $13,\,4,\,117$ &
$13,\,4,\,169$ & $-$ &
$23,\,3,\,368$ & $23,\,3,\,437$ &
$23,\,3,\,506$ & $25,\,3,\,625$\\
\hline
$19$ & $-$ &
$-$ & $-$ &
$27,\,3,\,432$ & $27,\,3,\,513$ &
$27,\,3,\,594$ & $27,\,3,\,675$\\
\hline
$20$ & $11,\,5,\,121$ &
$16,\,4,\,208$ & {\bf16,\,4,\,272} &
$-$ & $32,\,3,\,608$ &
$32,\,3,\,704$ & $32,\,3,\,800$\\
$\frac{20}{N}$ &.165 &.096 &.074 &-&.034 &.028 &
.025\\
\hline
$21$ & $-$ &
$-$ & $19,\,4,\,323$ &
$-$ & $-$ &
$-$ & $41,\,3,\,1025$\\
\hline
$22$ & $13,\,5,\,143$ &
$-$ & $23,\,4,\,391$ &
$23,\,4,\,483$ & $-$ &
$-$ & $-$\\
\hline
$23$ & $-$ &
$-$ & $25,\,4,\,425$ &
$25,\,4,\,525$ & $25,\,4,\,625$ &
$-$ & $-$\\
\hline
$24$ & $-$ &
$16,\,5,\,256$ & $-$ &
$27,\,4,\,609$ & $29,\,4,\,725$ &
$29,\,4,\,841$ & $-$\\
\hline
$25$ & $13,\,6,\,169$ &
$19,\,5,\,304$ & $-$ &
$-$ & $32,\,4,\,800$ &
$32,\,4,\,928$ & $32,\,4,\,1056$\\
$\frac{25}{N}$ &.148 &.082 &-&-&.031 &.027 &
.024\\
\hline
$26$ & $-$ &
$-$ & $-$ &
$-$ & $37,\,4,\,925$ &
$37,\,4,\,1073$ & $37,\,4,\,1221$\\
\hline
$27$ & $-$ &
$-$ & $23,\,5,\,483$ &
$-$ & $-$ &
$43,\,4,\,1247$ & $43,\,4,\,1419$\\
\hline
$28$ & $16,\,6,\,208$ &
$-$ & $27,\,5,\,702$ &
$25,\,5,\,650$ & $-$ &
$-$ & $49,\,4,\,1617$\\
\hline
$29$ & $-$ &
$19,\,6,\,361$ & $29,\,5,\,609$ &
$29,\,5,\,754$ & $31,\,5,\,961$ &
$-$ & $-$\\
$\frac{29}{N}$ &$-$&.080 &.048 & .038 & .030 &$-$ &
$-$\\
\hline
$30$ & $-$ &
$-$ & $-$ &
$32,\,5,\,832$ & $32,\,5,\,992$ &
$-$ & $-$\\
\hline
\hline
\end{tabular}
\end{center}

\subsection{Superimposed concatenated codes}
\begin{center}
\end{center}

A further extension of the Kautz-Singleton
superimposed $(s,N,t)$-codes is based on the following
{\it
concatenated codes} which were suggested in~\cite{ks64}
and also were discussed in~\cite{z83} and~\cite{agm92}.

Consider the $q_0$-ary shortened Reed-Solomon
code with parameters (10) where $q_0$ is a prime power.
Let $q_0$-ary  {\it symbols}  of this code {\it be
substituted, i.e., be coded, for the binary codewords of a
known constant-weight $s$-disjunct code of size $q'\ge
q_0$, length $q\le q_0$} and weight $w'<q$.  Denote this
binary superimposed code as an $(s,q,q')$-code.  {\it We
will apply  $(s,q,q')$-codes which are standard
constant-weight $(n,D,w')$-codes of size $A(n,D,w')=q'$,
length $n=q$, weight $w'=s\lambda'+1$, distance
$D=2(w'-\lambda')$ and the maximal correlation~$\lambda'$.}
It is easy to check that the result of Sect. 2 can be
generalized as follows.
\medskip

{\bf Proposition 2}. {\it The given substitution
yields the concatenated code which is the binary
constant-weight superimposed
$(s,\,qn_0,\,q_0^{\lambda_0+1})$-code of
weight $w=w'n_0$}.

In Sect.~3.3, we used only the {\it
trivial} substitution, where $q_0=q=q'$ and $w'=1$.
\medskip

{\bf Remark 3}. If we apply the trivial substitution
$q_0=q=q'$, i.e.,  $w'=1$, then we obtain
the {\it concatenated code} which {\it is a standard
constant-weight $(N,\,D,\,w)$-code of size $t$}, where
$$
N=q_0n_0,\quad w=n_0,\quad t=q_0^{k-r},
\quad D=2(n-\lambda_0)=2(q_0-k+2)=2d.
$$

{\bf Remark 4}. If $q<q_0$ and $w'>1$, then one knows only
the weight $w=w'n_0$ of the constant-weight concatenated
code.  We cannot identify its distance $D$ and the maximal
correlation $\lambda$.
\medskip

\subsubsection*{\bf Table 2. Parameters
for $2$- and $3$-disjunct codes
based on $(N,\,D,\,w)$-codes}
\begin{center}
\begin{tabular}{|c|c|c|c|c||c|c|c|c|c|}
\hline
\multicolumn{5}{|c||}
{Parameters of $(2,N,t)$-codes}&
\multicolumn{5}{|c|}
{Parameters of $(3,N,t)$-codes}\\
$\quad N\;$ & $\quad D\;$ & $\quad w\;$ & $\;\lambda\;$
&  $\quad t\quad$ &
$\quad N\;$ & $\quad D\;$ & $\quad w\;$ & $\;\lambda\;$
&  $\quad t\quad$ \\
\hline
9& 4 &3  &1   &12  &  &  &    &  &\\
10 & 4 & 3 & 1   &13    &  &  &  &&\\
11 & 4 & 3 & 1  & 17   &  &  &  &  &\\
12 & 4 & 3 & 1  & 20   &  &  &  &  &\\
13 & 4 & 3 & 1 &  26   &  &  &  &  &\\
14 & 6 & 5 & 2 &  28   &  &  &  &  &\\
{\bf15} & 6 & 5 & 2 &  {\bf42}   &  &  &  &  &\\
{\bf16} & 6 & 5 & 2 &  {\bf48} & 16 & 6 & 4 & 1 & 20\\
17 & 6 & 5 & 2 &  68 & 17 & - & - & - & -\\
18 & - & - & - &  - & 18 & 6 & 4 & 1 & 22\\
19 & 6 & 5 & 2 & 76 & 19 & 6 & 4 & 1 & 25\\
20 & 6 & 5 & 2 & 84 & 20 & 6 & 4 & 1 & 30\\
21 & 8 & 7 & 3 & 120 & 21 & 6 & 4 & 1 & 31\\
22 & 8 & 7 & 3 & 176 & 22 & 6 & 4 & 1 & 37\\
23 & 8 & 7 & 3 & 253 & 23 & 6 & 4 & 1 & 40\\
24 & - & - & - & - & 24 & 6 & 4 & 1 & 42\\
25 & 8 & 7 & 3 & 254 & 25 & 6 & 4 & 1 & 50\\
26 & 6 & 5 & 2 & 260 & 26 & 6 & 4 & 1 & 52\\
27 & 8 & 7 & 3 & 278 & 27 & 6 & 4 & 1 & 54\\
28 & 8 & 7 & 3 & 296 & 28 & 6 & 4 & 1 & 63\\
29 & 8 & 7 & 3 & 300 & 29 & 6 & 4 & 1 &65 \\
30 & 8 & 7 & 3 & 327 & 30 & 6 & 4 & 1 &67 \\
31 & 8 & 7 & 3 & 362 & 31 & 6 & 4 & 1 &76 \\
32 & 8 & 7 & 3 & 403 & 32 & 6 & 4 & 1 &80 \\
33 & 8 & 7 & 3 & 442 & 33 & 6 & 4 & 1 &82 \\
34 & 8 & 7 & 3 & 494 & 34 & 6 & 4 & 1 &92 \\
35 & 8 & 7 & 3 & 555 & 35 &  6 & 4 & 1 &96 \\
36 & 8 & 7 & 3 & 622 & 36 &  6 & 4 & 1 &99 \\
37 & 8 & 7 & 3 & 696 & 37  &  6 & 4 & 1 &111 \\
38 & 8 & 7 & 3 & 785 & 38  &  6 & 4 & 1 &114 \\
39 & 8 & 7 & 3 & 869 & 39  &  6 & 4 & 1 &117 \\
40 & 8 & 7 & 3 & 965 & 40  &  6 & 4 & 1 &130 \\
41 & 8 & 7 & 3 & 1095 & 41  & 6 & 4 & 1 &133 \\
42 & 8 & 7 & 3 & 1206 & 42  &  6 & 4 & 1& 136 \\
43 & 8 & 7 & 3 & 1344 & 43  &  6 & 4 & 1& 149 \\
44 & 8 & 7 & 3 & 1471 & 44  &  6 & 4 & 1 &154 \\
{\bf45} & ? & 15 & ? &$42^2=${\bf1764} &
45  & 6 & 4 & 1 &157 \\
46 & 8 & 7 & 3 & 1795 & 46 & 6 & 4 & 1 &171 \\
47 & 8 & 7 & 3 & 1976 & 47  & 6  & 4 & 1 &176 \\
{\bf48}& ? & 15 & ? &$48^2=${\bf2304} & 48 & 6 &4 &1&180\\
{\bf49}&{\bf8}&{\bf7}&{\bf3}&{\bf2401}&{\bf49}&10
&7&2&{\bf343}\\
\hline
\end{tabular}
\end{center}

Let $D=2,4,6,\dots$, $D\le n$ and $w\le n$ be arbitrary
integers. Denote by $A(n,D,w)$  the {\it
maximal size} of the corresponding constant-weight code
which {\it is known up to now}.  The tables of  $A(n,D,w)$
called Standard Tables (ST) are available~\cite{bsss90}
and :

\centerline{\sf
http://www.research.att.com/$\sim$
njas/codes/Andw/index.html}
\medskip

 Tables 2--4 give the
numerical values of the {\it best known parameters for
superimposed concatenated $(s,N,t)$-codes, when $s=2$ and
$s=3$}.

\subsubsection*{Table 3. Parameters of constant-weight
concatenated $(2,N,t)$-codes\\ of weight $w$,
length $N$ and size $t$,
$2^m\le t<2^{m+1},\;10\le m\le32$}
\begin{center}
\begin{tabular}{|c||c|c|c||c|c|c|c|c|c||c|c|}
\hline
&\multicolumn{3}{|c||}{$(2,q,q')$-code}&
\multicolumn{6}{|c||}{$q_0$-ary shortened RS-code}&
\multicolumn{2}{|c|}{$(2,N,t)$-code}\\
\hline
$m$ & $q$ & $q'$ & $w'$ & $q_0$ & $k$ & $r$ &
$ t=q_0^{k-r}$ & $n_0$ & $\lambda_0$ &
$N=qn_0$ & $w$\\
\hline
\hline
10 & 15 & 42 & 5 & 41 & 41 & 39 & $41^2$
& 3 & 1 & 45 &15\\
10 & 15 & 42 & 5
&\multicolumn{6}{|c||}{Latin square $42\times42=1764$}
& 45 &15\\
\hline
11 & 16 & 48 & 5 & 47 & 47 & 45 & $47^2$
& 3 & 1 & 48 &15\\
11 & 16 & 48 & 5
&\multicolumn{6}{|c||}{Latin square $48\times48=2304$}
& 48 &15\\
11 & 7 & 7 & 1 & 7 & 5 & 1 & $7^4$&
7 & 3 & 49 &7\\
\hline
12 & 17 & 68 & 5 & 67 & 67 & 65 & $67^2$
& 3 & 1 & 51 & 15\\
12 & 17 & 68 & 5
&\multicolumn{6}{|c||}{Latin square $68\times68=4624$}
& 51 & 15\\
12 & 19 & 76 & 5 & 73 & 73 & 71 &
$73^2$ & 3 & 1 & 57 & 15\\
12 & 19 & 76 & 5
&\multicolumn{6}{|c||}{Latin square $76\times76=5776$}
& 57 & 15\\
12 & 20 & 84 & 5 & 83 &
83 & 81 & $83^2$ & 3 & 1 & 60 & 15\\
12 & 20 & 84 & 5
&\multicolumn{6}{|c||}{Latin square $84\times84=7056$}
& 60 & 15\\
\hline
13 & 21 & 120 & 7
&\multicolumn{6}{|c||}{Latin square $120\times120=14400$}
& 63 & 21\\
13 & 9 & 12 & 3 & 11 & 9 & 5 & $11^4$ & 7 & 3
& 63 & 21\\
13 & 13 & 26 & 3 & 25 & 24 & 21 & $25^3$ & 5 &
2 & 65 & 15\\
\hline
15 & 23 & 253 & 7 & 251 & 251 & 249
& $251^2$ & 3 & 1 & 69 & 21\\
15 & 23 & 253 & 7
&\multicolumn{6}{|c||}{Latin square $253\times253=64009$}
& 69 & 21\\
\hline
16 & 11 & 17 &
3 & 17 & 15 & 11 & $17^4$ & 7 & 3 & 77 & 21\\
\hline
17 & 9 & 12 & 3 & 11 & 8 & 3 &
$11^5$ & 9 & 4 & 81 & 27\\
17 & 17 & 68 & 5 & 67 & 66 & 63 &
$67^3$ & 5 & 2 & 85 & 25\\
\hline
18 & 10 & 13 & 3 & 13 & 10 & 5 &
$13^5$ & 9 & 4 & 90 & 27\\
18 & 13 & 26 & 3 & 25 & 23 & 19 &
$25^4$ & 7 & 3 & 91 & 21\\
\hline
19 & 9 & 12 & 3 & 11 & 7 & 1 &
$11^6$ & 11 & 5 & 99 & 33\\
\hline
21 & 12 & 20 & 3 & 19 & 16 & 11 &
$19^5$ & 9 & 4 & 108 & 27\\
\hline
22 & 10 & 13 & 3 & 13 & 9 & 3 &
$13^6$ & 11 & 5 & 110 & 33\\
\hline
23 & 23 & 253 & 7 & 251 & 250 & 247 &
$251^3$ & 5 & 2 & 115 & 35\\
\hline
24 & 11 & 17 & 3 & 17 & 13 & 7 &
$17^6$ & 11 & 5 & 121 & 33\\
\hline
25 & 10 & 13 & 3 & 13 & 8 & 1 &
$13^7$ & 11 & 6 & 130 & 39\\
\hline
28 & 11 & 17 & 3 & 17 & 12 & 5 &
$17^7$ & 13 & 6 & 143 & 39\\
\hline
29 & 12 & 19 & 3 & 19 & 12 & 5 &
$19^7$ & 13 & 6 & 156 & 39\\
\hline
31 & 23 & 253 & 7 & 251 & 249 & 245 &
$251^4$ & 7 & 3 & 161 & 49\\
\hline
32 & 11 & 17 & 3 & 17 & 11 & 3 &
$17^8$ & 15 & 7 & 165 & 45\\
\hline
\hline
\end{tabular}
\end{center}
\medskip


{\bf Description of Table 2}.\\
We tabulate the values of size $t>N$ for
$(s,N,t)$-superimposed codes of strength $s=2,\,3$ and
length $N=9,10,\dots,47$. The symbol $t=A(N,D,w)$ denotes
the code size, i.e., the number of codewords (columns), $N$
is the code length, $w=s\lambda+1$ is the code weight and
$D=2(w-\lambda)$ is the code distance. Table~2 is
calculated as follows.  Let $N=9,10,\dots,47$ be fixed.
In ST, we are looking for the pair $(D,\,w)$ such that
$$
w=s\lambda+1,\quad D=2(w-\lambda)
\quad\Longleftrightarrow\quad
w(s-1)=\frac{D}{2}\,s-1
$$
and the {\it value of size $A(N,D,w)$ is maximal}.
Table~2
gives these maximal values if $s=2,\,3$. For $s=2$,
$N=45,48$ and $s=3$, $N=49$, we use the boldface type
because the corresponding codes are concatenated. We will
repeat them in  Table~3 and Table~4. For $s=2,3$
and $N=49$, the  concatenated codes are the
standard constant-weight codes because they are obtained by
the trivial substitution.  These codes {\it are new} for
the Standard Tables and should be included in ST.
\smallskip

{\bf Description of Tables 3 and 4}.\\
We tabulate the parameters of the best
known concatenated
$(s,\,N,\,t)$- codes of strength $s=2,\,3$ and
length $N\ge45$. For values $N$ of
the form $N=qn_0$, {\it we give the parameters of the
$(s,q,q')$-code and the shortened RS-code yielding the
maximal possible size $t$ of the form $t=q_0^{k-r}$}. To
calculate the parameters of the shortened RS-code, we apply
the particular cases of formulas (10) corresponding to
$s=2$ or $s=3$. The groups of $t$ values for which $2^m\le
t<2^{m+1}$, $m=8,9,\dots$, {\it are separated}.
The Reed-Solomon codes are a known class of {\it
maximum-distance separable} codes
(MDS-codes)~\cite{s64,ms83} which could be applied for
our concatenated construction.  In Table~3, we give the
parameters of several concatenated codes, which could be
obtained with the help of the Latin squares $q_0\times
q_0$, when $q_0\ge2$ is an arbitrary integer.  We
remind~\cite{s64,ms83} that any Latin square $q_0\times
q_0$ yields the $q_0$-ary MDS-code of size $q_0^2$, length
$n_0=3$ and distance $d=2$, i.e., $\lambda_0=1$.
\smallskip

\subsubsection*{Table 4. Parameters of constant-weight
concatenated $(3,N,t)$-codes\\ of weight $w$, length $N$
and size $t$, $2^m\le t<2^{m+1},\; m=8,9,\dots,62$}

\begin{center}
\begin{tabular}{|c||c|c|c||c|c|c|c|c|c||c|c|}
\hline
&\multicolumn{3}{|c||}{$(3,q,q')$-code}&
\multicolumn{6}{|c||}{$q_0$-ary shortened RS-code}&
\multicolumn{2}{|c|}{$(3,N,t)$-code}\\
\hline
$m$ & $q$ & $q'$ & $w'$ & $q_0$ & $k$ & $r$ &
$t=q_0^{k-r}$ & $n_0$ & $\lambda_0$ &
$N=qn_0$ & $w$\\
\hline
\hline
8 & 7 & 7 & 1 & 7 & 4 & 1 &
$7^3$ & 7 & 2 & 49 & 7\\
\hline
9 & 8 & 8 & 1 & 8 & 5 & 2 &
$8^3$& 7 & 2 & 56 & 7\\
\hline
10 & 11 & 11 & 1 & 11 & 8 & 5 &
$11^3$& 7 & 2 & 77 & 7\\
\hline
12 & 9 & 9 & 1 & 9 & 6 & 3 &
$9^4$& 10 & 3 & 90 & 10\\
\hline
13 & 11 & 11 & 1 & 11 & 6 & 2 &
$11^4$& 10 & 3 & 110 & 10\\
\hline
14 & 13 & 13 & 1 & 13 & 8 & 4 &
$13^4$& 10 & 3 & 130 & 10\\
\hline
15 & 22 & 37 & 4 & 37 & 34 & 31 &
$37^3$& 7 & 2 & 154 & 28\\
\hline
16 & 16 & 16 & 1 & 16 & 11 & 7 &
$16^4$& 10 & 3 & 160 & 10\\
\hline
18 & 13 & 13 & 1 & 13 & 6 & 1 &
$13^5$& 13 & 4 & 169 & 13\\
\hline
19 & 20 & 30 & 4 & 29 & 24 & 20 &
$29^4$& 10 & 3 & 200 & 40\\
\hline
21 & 16 & 20 & 4 & 19 & 12 & 7 &
$19^5$& 13 & 4 & 208 & 52\\
\hline
23 & 19 & 25 & 4 & 25 & 18 & 13 &
$25^5$& 13 & 4 & 247 & 52\\
\hline
25 & 16 & 20 & 4 & 19 & 10 & 4 &
$19^6$& 16 & 5 & 256 & 64\\
\hline
26 & 22 & 37 & 4 & 37 & 30 & 25 &
$37^5$& 13 & 4 & 286 & 52\\
\hline
29 & 16 & 20 & 4 & 19 & 8 & 1 &
$19^7$& 19 & 6 & 304 & 76\\
\hline
31 & 22 & 37 & 4 & 37 & 28 & 22 &
$37^6$& 16 & 5 & 352 & 64\\
\hline
37 & 19 & 25 & 4 & 25 & 12 & 4 &
$25^8$& 22 & 7 & 418 & 88\\
\hline
41 & 19 & 25 & 4 & 25 & 10 & 1 &
$25^9$& 25 & 8 & 475 & 100\\
\hline
46 & 22 & 37 & 4 & 37 & 22 & 14 &
$37^9$& 25 & 8 & 550 & 100\\
\hline
52 & 22 & 37 & 4 & 37 & 20 & 10 &
$37^{10}$& 28 & 9 & 616 & 112\\
\hline
62 & 22 & 37 & 4 & 37 & 16 & 4 &
$37^{12}$& 34 & 11 & 748 & 136\\
\hline
\hline
\end{tabular}
\end{center}

{\bf Discussion}.\\
The preliminary conclusions could be
formulated as follows.
\begin{itemize}
\item
For fixed integers $s=2,3,\dots$ and $m\ge1$, we
constructed the family ${\cal F}={\cal F}(s,\,m)$ of binary
concatenated $s$-disjunct constant-weight codes of size
$t,\,2^m\le t<2^{m+1},$ based on the $q_0$-ary shortened
RS-codes ($q_0$ satisfies the inequality in (10)) and
the standard binary constant-weight codes.
\item
For $s=2$,  Table 3 shows the {\it stability of the
code rate}, i.e., $\frac{m}{N}\approx0.2>0.182$,
$m=11,12,\dots20$. It means that for $s=2$, the {\it code
rate of the family ${\cal F}$ exceeds the random coding
rate} $\underline{R}(2)=0.182$~\cite{dr89}.

\item
For $s=3$ and $m=15,16,\dots21$, the  stable code
rate  $\frac{m}{N}\approx0.1$ also exceeds the
corresponding random coding rate
$\underline{R}(3)=0.079$~\cite{dr89}.
\item
It seems to us that the calculations of the {\it best}
parameters of ${\cal F}={\cal F}(3,\,m)\;$,
$m\ge8$, are not completed and some values from Table 4
could be improved.
\end{itemize}

{\bf Example}. For the length $N=45=3\cdot15$, the
$(s=2,\,N=45,\,t=42^2=1764)$-code is the concatenated code
based on the binary constant-weight $(n=15,D=6,w=5)$-code
of size $A(15,6,5)=42$ and the Latin square
$42\times42=1764$.

\newpage
\centerline{Arkadii G. D'yachkov,\qquad
"Lectures on Designing Screening Experiments"}
\medskip

\section{Optimal Superimposed Codes and Designs\\
for Renyi's Search Model}
\begin{center}
\end{center}

In 1965, Renyi~\cite{renyi65}
suggested a combinatorial group testing model, in which
the size of a testing group was restricted. In this
model, Renyi considered the search of one defective element
(significant factor) from the finite set of elements
(factors). The corresponding optimal search designs were
obtained by Katona~\cite{katona66}.  In this section,
we study Renyi's search model of several significant
factors. This problem is closely related to the concept of
binary superimposed codes, which were introduced by
Kautz-Singleton~\cite{ks64} and were investigated by
Dyachkov-Rykov\cite{dr83}. Our goal is to prove a lower
bound on the search length and to construct the optimal
superimposed codes and search designs. The results of
Sect.~4 will be published in paper~\cite{dr01}.

\subsection{Notations and definitions}
\begin{center}
\end{center}

Let $1\le s<t,\;1\le k<t,\; N>1$ be integers and
$X=\|x_i(u)\|,\;i=1,2,\dots,N,\;u=1,2,\dots,t,$ be a binary
$(N\times t)$--matrix (code) with columns (codewords)
$\x(1),\x(2),\dots,\x(t)$ and rows
$\x_1,\x_2,\dots,\x_N$,
where $\x(u)=(x_1(u),x_2(u),\dots,x_N(u))$ and
$\x_i=(x_i(1),x_i(2),\dots,x_i(t))$
Let
$$
w=\min_u\sum_{i=1}^N x_i(u),\quad
k=\max_i\sum_{u=1}^t x_i(u),\quad
\lambda=\max_{u,v}\sum_{i=1}^N x_i(u)x_i(v)
$$
be the {\it minimal weight of codewords},
the {\it maximal weight of rows} and the
{\it maximal dot product of codewords}.

We say that the binary column $\x$ {\it covers} the binary
column $\y$ if the Boolean sum $\x\vee\y=\x$. The code $X$
is called~\cite{ks64,dr83} a {\it superimposed
$(s,t)$-code} if the Boolean sum of any $s$-subset of
columns of $X$ covers those and only those columns of $X$
which are the terms of the given Boolean sum.
The code $X$
is called~\cite{ks64,dr83} a {\it superimposed
$(s,t)$-design} if all Boolean sums
composed of not more than $s$ columns of $X$ are distinct.
\medskip

{\bf Definition 1}.  An $(N\times t)$--matrix $X$ is called
a superimposed $(s,t,k)$-code (design) of {\it
length $N$, size $t$, strength} $s$ and {\it constraint}
$k$ if code $X$ is a superimposed $(s,t)$-code (design)
whose the maximal row weight is equal to $k$.

The above-mentioned constraint $k$ was introduced by
Renyi~\cite{renyi65} and was studied
by~Katona~\cite{katona66} for the search designs.

\subsection{Lower bound}
\begin{center}
\end{center}

{\bf Proposition 1}.  {\it Let $t>k\ge s\ge 2$
and $N>1$ be integers}.
\begin{enumerate}
\item
{\it For any superimposed $(s-1,t,k)$-code
$((s,t,k)-design)$ $X$ of length $N$, the following
inequality takes place}
$$
N\ge\left\lceil\frac{st}{k}\right\rceil\,.\eqno (1)
$$
\item
{\it If  $k\ge s+1$, $st=kN$ and there exists
the optimal superimposed
$(s-1,t,k)$-code  $X$ of length $N=st/k$, then}
\begin{enumerate}
\item
{\it code $X$ is a constant weight code of weight $w=s$,
for any $i=1,2,\dots,N$, the weight of row $\x_i$ is equal
to $k$ and the maximal dot product $\lambda=1$};
\item
{\it  the following inequality is true}
$$
k^2-\frac{k(k-1)}{s}\le t. \eqno(2)
$$
\end{enumerate}
\end{enumerate}

{\bf Proof}. {\bf 1.} It is known~\cite{dr83} that code
$X$ is a superimposed $(s,t,k)$-design if and only if $X$
is superimposed $(s-1,t,k)$-code and all ${t\choose s}$
Boolean sums composed of $s$ columns of $X$ are distinct.
Hence, we need to prove inequality~(1)  for superimposed
$(s-1,t,k)$-codes only.  Let $s\ge2,\;1\le k<t$ be fixed
integers.  Consider an arbitrary superimposed
$(s-1,t,k)$-code $X$ of length $N$. Let $n,0\le n\le t$, be
the number of codewords of $X$ having a weight $\le s-1$.
From definition of superimposed $(s-1,t)$-code it follows
(see,~\cite{ks64}) that $n\le N$ and, for each codeword of
weight $\le s-1$, there exists a row in which all the
remaining elements, except for the element of this
codeword, are 0's.  We delete these $n$ rows from $X$
together with $n$ codewords of weight $\le s-1$. Consider
the remaining $(N-n)\times (t-n)$ matrix $X'$. Obviously,
each column of $X'$ has a weight $\ge s$ and each its row
contains $\le k$ 1's. Since $k\ge s$, we have
$$
s(t-n)\le k(N-n),\quad
ts\le kN-n(k-s)\le kN. \eqno(3)
$$
Statement 1 is proved.

{\bf 2.} Let  $k\ge s+1$, $st=kN$ and $X$  be the
optimal superimposed $(s-1,t,k)$-code of length
$N=st/k$.
\begin{itemize}
\item
Since $k\ge s+1$, inequality (3) has signs of equalities if
and only if $X$ is the constant weight code of weight $w=s$
and for any $i=1,2,\dots,N$, the weight of row $\x_i$ is
equal to $k$.  By contradiction, using the constant weight
property $w=s$ one can easily check that the maximal dot
product $\lambda=1$.\\
Statements (2a) is proved.
\item
To prove Statement (2b), we apply  the the well-known
Johnson inequality
$$
t{w\choose \lambda+1}\le{N\choose
\lambda+1}
$$
which is true for any constant weight code $X$ of length
$N$, size $t$, weight $w$ and the maximal dot product
$\lambda$. In our case, $\lambda=1$, $w=s$, $tw=kN$ and
$N=st/k$. This gives
$$
tw(w-1)\le N(N-1),\quad k(s-1)\le N-1=\frac{st}{k}-1,\quad
k^2(s-1)+k\le st,\quad k^2-\frac{k(k-1)}{s}\le t.
$$
Proposition~1 is proved.
\end{itemize}

Denote by $N(s,t,k),\;(\tilde{N}(s,t,k))$ the minimal
possible length of superimposed $(s,t,k)$-code
($(s,t,k)$-design). From Proposition 1 it follows:
\begin{itemize}
\item
if  $k\ge s+1$, then
$$
\tilde{N}(s,t,k)\ge N(s-1,t,k)\ge
\left\lceil\frac{st}{k}\right\rceil.
$$
\item
if $k\le s$, then $N(s-1,t,k)=\tilde{N}(s,t,k)=t.$
\end{itemize}

\subsection{Optimal parameters}
\begin{center}
\end{center}

Let $s\ge 2$ and $k\ge s+1$ be fixed integers. Denote by
$q\ge 2$  an arbitrary integer.  We shall consider the {\it
optimal} superimposed $(s-1,kq,k)$-codes and {\it optimal}
superimposed $(s,kq,k)$-designs of length $N=sq$ whose
parameters satisfy (1) with the sign of equality. By
virtue of (2)
\begin{itemize}
\item
if $q\ge k-\frac{k-1}{s}$, then there exists a  possibility
to find the optimal superimposed $(s-1,kq,k)$-code of
length $N=sq$;
\item
if $q< k-\frac{k-1}{s}$, then lower bound (1) is not
achieved and the interesting {\it open problem} is
{\it how to obtain a new nontrivial lower bound on
$N(s-1,t,k)$ provided that}
$$
k^2-\frac{k(k-1)}{s}> t.
$$
\end{itemize}
\medskip

Some constructions of superimposed $(2,kq,k)$-designs of
length $N=2q$ and superimposed $(2,kq,k)$-codes of length
$N=3q$ were obtained in~\cite{ks64}. By virtue of
Proposition~1, they are optimal. We give here the
parameters of these designs and codes. The following
statements are true:
{\begin{itemize}
\item if $k-1\ge2$ is a prime
power and $q=k^2-k+1$, then there exists an superimposed
$(2,kq,k)$-design of length $N=2q$,
\item
for pair $(k=3,\;q=5)$ and pair $(k=7,\;q=25)$,
there exists an superimposed
$(2,kq,k)$-design of length $N=2q$.
\item
if $k\ge4$ and $q=k-1$, or $q=k$, then there exists
an superimposed $(2,kq,k)$-code of length $N=3q$.
\end{itemize}
}

Our aim -- to prove Theorems~1--4.
\medskip

{\bf Theorem 1}. {\it Let $s=2$ and $k\ge 3$ be integers.
Then
\begin{itemize}
\item[$1)$]
for any integers $q\ge k\ge 3$ there exists an optimal
superimposed $(1,kq,k)$-code of length $N=2q$, i.e.,
$
N(1,kq,k)=2q,\; q\ge k;
$
\item[$2)$]
for any integer $q\ge 2^k-1$ there exists an optimal
superimposed $(2,kq,k)$-design of length $N=2q$, i.e.,
$
{\tilde N}(2,kq,k)=2q,\; q\ge 2^k-1.
$
\end{itemize}}

{\bf Theorem 2}.  {\it Let $s\ge 3$, $k\ge s+1$ be fixed
integers and $q=k^{s-1}$. Then there exists an optimal
superimposed $(s,kq,k)$-design $X$ of length $N=sq$, i.e},
$
\tilde{N}(s,k^{s},k)=sk^{s-1}.
$
\medskip

{\bf Theorem 3}.  {\it Let $k=4,5,\dots,$ be a fixed
integer. For any integer $q\ge k+1$, there exists an
optimal superimposed $(2,kq,k)$-code of length
$N=3q$, i.e.},
$
N(2,kq,k)=3q,\; q\ge k+1.
$
\medskip

{\bf Remark}. Let $s\ge3$. For the case of superimposed
$(s,kq,k)$-codes, Theorem 3 is generalized (the proof is
omitted) as follows.
{\it Let $p_i$, $i=1,2\dots I$, be arbitrary prime
numbers and
$r_i$, $i=1,2\dots I$, be arbitrary integers. If
$$
q=p_1^{r_1}p_2^{r_2}\cdots p_I^{r_I},\qquad
3\le s\le\min_{i}\left\{p_i^{r_i}\right\}-1,
$$
then for any $k$, $s+1\le k\le q+1$, the
optimal length $N(s,kq,k)=(s+1)q$}.
\medskip

The following theorem supplements Theorem 2 if $s=3$ and
$k=4$.
\smallskip

{\bf Theorem 4}. {\it If $k=4$  and $q\ge12$,
then there exists an
optimal superimposed $(3,kq,k)$-design of length
$N=3q$, i.e.},
$$
\tilde{N}(3,4q,4)=3q,\quad q\ge 12.
$$
\bigskip

To prove Theorems 1--4, we  apply concatenated
codes using a class of homogeneous $q$-nary codes of size
$t=kq$. The description of cascade construction,
definitions and  properties of homogeneous $q$-nary codes
will be given  Sect.~4.4.
The proofs of Theorems~1-4 will be given in
Sect.~4.5--4.8.
\bigskip

The following theorem yields a different family
of optimal superimposed $(s,t,k)$--codes.
It will be proved in Sect.~4.9.
\smallskip

{\bf Theorem 5}. {\it Let $s\ge1,\;k\ge s+2$ be fixed
integers. Then there exists an $(s,t,k)$--code of size
$t={k+s\choose s+1}$ and length
$$
N=\frac{(s+1)t}{k}=\frac{(s+1){k+s\choose s+1}}{k}=
{k+s\choose s},
$$
i.e., the optimal length}
$$
N\left(s,{k+s\choose s+1},k\right)=
{k+s\choose s}.
$$

For Theorem 5, the optimal code constructions were invented
in~\cite{m96}.

\subsection{Homogeneous $q$-nary codes}
\begin{center}
\end{center}

Let $q\ge s\ge 1,\;k\ge2,\;k\le t\le kq,\;J\ge 2$
be integers,
$A_q=\{a_1,a_2,\dots,a_q\}$ be an arbitrary $q$-ary
alphabet and
$
B=\|b_j(u)\|,\;j=1,2,\dots,J,\; u=1,2,\dots,t,
$
be an $q$-nary ($b_j(u)\in A_q$)
$(J\times t)$-matrix (code) with $t$ columns (codewords)
and $J$ rows
$$
{\bf b}(u)=(b_1(u),b_2(u),\dots,b_{J}(u)),
\; u=1,2,\dots,t,\quad
{\bf b}_j=(b_j(1),b_j(2),\dots,b_j(t)),
\; j=1,2,\dots,J.
$$
Denote the number of $a$-entries in the
$j$-th row ${\bf b}_j$ by $n_j(a)$, where
$a\in A_q,\;j=1,2,\dots,J$. We suppose that for any
$j=1,2,\dots,J$ and any $a\in A_q$, the value $n_j(a)\le k$.
\medskip

{\bf Definition 2}.  Let $t=kq$. Code $B$ is called an
{\it $(q,k,J)$-homogeneous}  code if for any
$j=1,2,\dots,J$ and any $a\in A_q$, the number $n_j(a)=k$.
\medskip

{\bf Definition 3}. Code $B$ will be called an {\it
$s$-disjunct} if for any codeword  ${\bf b}(u)$ and any
$s$--subset of codewords $\{{\bf b}(u_1),{\bf
b}(u_2),\dots,{\bf b}(u_s)\}$, there exists a coordinate
$j=1,2,\dots,J$ for which $b_j(u)\ne
b_j(u_i),\;i=1,2,\dots,s$.
\medskip

For two codewords ${\bf b}(u),\,{\bf b}(v),\,u\ne v$,
define the $q$-ary Hamming distance
$$
D({\bf b}(u);{\bf b}(v))=
\sum_{j=1}^J\chi(b_j(u);b_j(v)),
$$
$$
\chi(a;b)=\cases{1, & if $a\ne b$,\cr
                      0, & if $a=b$.\cr}
$$
Let
$
D=D(B)=\min_{u\ne v}D({\bf b}(u);{\bf b}(v))\le J
$
be the Hamming distance of code $B$.
By contradiction, one can easily  prove the following
statement which gives the analog of
the well-known Kautz-Singleton~\cite{ks64} condition.
\medskip

{\bf Proposition 2}. {\it If
$
s(J-D(B))\le J-1,
$
then code $B$ is $s$-disjunct. In addition,
$(q,k,s)$-homogeneous code $B$ is $(s-1)$-disjunct code if
and only if $D(B)=s-1$.}
\medskip

Let $n\le t$ be a fixed  integer
and
$
\e=\{e_1,e_2,\dots,e_{n}\},
\;1\le e_1<e_2<\cdots<e_{n}\le t
$
be an arbitrary $n$-subset of the set $[t]=\{1,2,\dots,t\}$.
For a given code $B$ and
any $j=1,2,\dots,J$, denote by
$
{\A}_j(\e,B)\subseteq A_q
$
--the {\it set of all pairwise distinct elements of the
sequence} $b_j(e_1),b_j(e_2),\dots,b_j(e_{n})$.
The set ${\A}_j(\e,B)$ is called the $j$-th,
$j=1,2,\dots,J$, {\it coordinate} set of subset
$\e\subseteq[t]$ over code $B$.
For its
{\it cardinality} $|{\A}_j(\e,B)|$, we have
$$
1\le|{\A}_j(\e,B)|\le\min\{n,\,q\}.
$$

{\bf Definition 4}. Let $s\ge1,\;n\le s,\;m\le s$ be
arbitrary integers. Code $B$ is called
an $s$-{\it separable}
code if for any two distinct subsets
$$
\e=\{e_1,e_2,\dots,e_{n}\},
\;1\le e_1<e_2<\cdots< e_n\le t,
\quad
\e'=\{e'_1,e'_2,\dots,e'_{m}\},
\;1\le e'_1<e'_2<\cdots e'_{m}\le t,
$$
of the
set $[t]$, there exists $j=1,2,\dots,J$, for which the
corresponding coordinate sets are distinct, i.e.,
${\A}_j(\e,B)\ne{\A}_j(\e',B)$.
In other words, for an arbitrary $n$-subset
$
\e=\{e_1,e_2,\dots,e_{n}\},
$
of the set $[t]$,
there exists the  {\it possibility to identify}
this $n$-subset $\e=\{e_1,e_2,\dots,e_{n}\}$
(or the corresponding $n$-subset
of codewords
$\{{\bf b}(e_1),\;{\bf b}(e_2),\dots,{\bf b}(e_{n})\}$
of code $B$) on the basis of sets:
$$
{\A}_1(\e,B),{\A}_2(\e,B),\dots,{\A}_J(\e,B),\quad
\A_j(\e,B)\subseteq A_q.
$$

{\bf Remark}. In Definitions 3 and 4, we used the
terminology of~\cite{dh93}.

One can easily prove (by contradiction)
the following ordering among these properties:
\begin{center}
$s$-disjunct $\;\Longrightarrow\;$
$s$-separable $\;\Longrightarrow\;$
$(s-1)$-disjunct.
\end{center}

{\bf Definition 5}. Code $B$ is called
an {\it $s$-hash}~\cite{fk84} if for
an arbitrary $s$-subset
$$
\e=\{e_1, e_2,\dots, e_{s}\},\quad
1\le e_1<e_2<\cdots<e_{s}\le t,
$$
of the set $[t]$, there exists a
coordinate $j=1,2,\dots J$, where
the cardinality $|{\A}_j(\e,B)|=s$, i.e.,
the elements
$b_j(e_1), b_j(e_2),\dots, b_j(e_s)$
are {\it all different}.
\medskip

Obviously, the following {\it ordering} takes place:
$s$-hash $\Longrightarrow$ $(s-1)$-disjunct.
\medskip

{\bf Definition 6}. Code $B$ is called
an {\it $s$-hash$\&$separable} if
it has both of these properties.
\medskip

Let $q$-nary alphabet $A_q=[q]=\{1,2,\dots,q\}$.
To illustrate Definitions 2--6 and the proof of Theorem 1,
we give two examples of disjunct and separable codes.
\medskip

{\bf Example 1}.
Let $k=q=2,3,\dots$ be fixed integers.
The evident $(k,k,2)$-homogeneous $1$-disjunct code
$B$ of distance $D=1$ has the following $t=k^2$ columns
(codewords):
$$
B=\left(\begin{array}{cccccccccc}
111&\dots&1&222&\dots&2&\dots&kkk&\dots&k\\
123&\dots&k&123&\dots&k&\dots&123&\dots&k
\end{array}\right).
$$
\smallskip

{\bf Example 2}. For $k=3,\;q=7$, the $(7,3,2)$-homogeneous
$2$-hash\&separable code
$B$ of distance $D=1$ has $kq=21$ codewords:
$$
B=\left(\begin{array}{ccccccc}
111&222&333&444&555&666&777\\
124&235&346&457&156&267&137
\end{array}\right).
$$
\smallskip

The idea of the following two examples of
$(q,k,3)$-homogeneous
$3$-hash\&separable codes will be used to
prove Theorem 2.

{\bf Example 3}. For $k=3,\;q=9$, the $(9,3,3)$-homogeneous
$3$-hash\&separable code $B$ of distance $D=2$
has $kq=27$ columns (codewords):
$$
B=\left(\begin{array}{ccccccccccc}
111&222&333&|&444&555&666&|&777&888&999\\
123&123&123&|&456&456&456&|&789&789&789\\
123&456&789&|&123&456&789&|&123&456&789
\end{array}\right).
$$
Code $B$ contains $k=3$ groups of codewords.
In the first and second rows, we use the construction idea
which could be called an {\it alphabet separating between
groups}.
\medskip

{\bf Remark}.
Obviously, $3$-separable code
$B$ from example 3 is not
$3$-disjunct code. Hence, in general,
the ordering
$s$-separable$\Longrightarrow$$s$-disjunct
is not true.
\medskip

{\bf Example 4}. For $k=4,\;q=16$, the
$(16,4,3)$-homogeneous $3$-hash\&separable code
$B$ of distance $D=2$ has $kq=64$ columns (codewords)
which are divided into $k=4$ groups:
\begin{itemize}
\item
the first $16$ codewords take the form
$$
\begin{array}{cccccccccc}
1111&2222&3&3&3&3&4&4&4&4\\
1234&1234&1&2&3&4&1&2&3&4\\
1234&5678&9&10&11&12&13&14&15&16,
\end{array}
$$
\item
the construction of the last $48$ codewords of $B$
applies the same method of alphabet separating:
\begin{itemize}
\item
for $j=1,2$ and $u=16m+l,\;m=1,2,3,\;l=1,2,\dots,16$,
the element\\ $b_j(u)=b_j(16m+l)=b_j(l)+4m$,
\item
for $j=3$ and $u=16m+l,\;m=1,2,3,\;l=1,2,\dots,16$,
the element\\
$b_3(u)=b_3(16m+l)=b_3(l)=l$.
\end{itemize}
\end{itemize}
\medskip

Let $q$-nary alphabet $A_q=[q]=\{1,2,\dots,q\}$.
For code $B$, we denote by
$$
X_B=(\x(1),\x(2),\dots,\x(t)),\quad  k\le t\le kq,
$$
a binary $Jq\times t$ matrix (code),
whose columns (codewords) have the form
$$
\x(u)=(\x^1(u),\x^2(u),\dots,\x^{s}(u)),\quad
u=1,2,\dots,t,
$$
$$
\x^j(u)=(x_1^j(u),x_2^j(u),\dots,x_q^j(u)),\quad
j=1,2,\dots,J,
$$
$$
x_l^j(u)=\cases{1, & if $l=b_j(u)$,\cr 0,
                & if $l\ne b_j(u),\;l=1,2,\dots,q.$\cr}
$$
In other words, a symbol $b\in[q]$ of $q$-ary matrix $B$
is replaced by the binary $q$-sequence in which all elements
are $0$'s, except for the element with number $b$.
Obviously, each codeword $\x(u)$ of (code) $X_B$
contains $J$ $1$'s and $(Jq-J)$ $0$'s and
each row $\x_i$ of
code $X_B$ contains $\le k$ $1$'s.
For $(q,k,J)$-homogeneous code $B$, each row $\x_i$ of
code $X_B$ contains $k$ 1's and $(kq-k)$ 0'. In addition,
the stated below Proposition~3 follows easily by
Definitions~2--4 and Propositions~1--2.
\medskip

{\bf Proposition 3}. {\it Let $q>k\ge s+1$ and $B$ be a
$(q,k,s)$-homogeneous code. The following
two statements are true.
\begin{itemize}
\item
If  $B$ is a $(s-1)$-disjunct code  $X_B$, then $X_B$ will
be the optimal superimposed $(s-1,kq,k)$-code of length
$N=sq$.
\item
If  $B$ is a $s$-separable code, then $X_B$ will be the
optimal superimposed $(s,kq,k)$-design of length $N=sq$.
\end{itemize}}

Hence, to prove Theorems~1--4, it is
sufficient to construct the corresponding
$(q,k,s)$-homoge-\\neous codes. In particularly, the
constructive method of examples 3 and 4 yields Theorem~2
for the case $s=3$, i.e.,
$
\tilde{N}(3,k^{3},k)=3k^{2},\;k=4,5,\dots
$.

\subsection{Proof of Theorem 1}
\begin{center}
\end{center}

Let  $s=2,\,q\ge k$, $q$-nary
alphabet $A_q=[q]=\{1,2,\dots,q\}$ and
$B=({\bf b}(1),{\bf b}(2),\dots,{\bf b}(kq))$
be an arbitrary $(q,k,2)$-homogeneous code
$1$-disjunct code, i.e., $B$ has pairwise distinct
codewords
${\bf b}(u)=(b_1(u),b_2(u)),\,u=1,2,\dots,kq$.
Following~\cite{ks64}, we  introduce  the {\it
binary characteristic $(q\times q)$-matrix}
$C=\|c_i(j)\|,\,i=1,2,\dots,q,\,j=1,2,\dots,q$, where
$$
c_i(j)=\cases{1, & if there exists codeword
${\bf b}(u)=(i,j)$,\cr
0, & otherwise.\cr}
$$

{\bf Example 5}.
For $(7,3,2)$-homogeneous code $B$ of Example 2,
the characteristic $(7\times 7)$-matrix is
$$
C=\pmatrix{
1&1&0&1&0&0&0\cr
0&1&1&0&1&0&0\cr
0&0&1&1&0&1&0\cr
0&0&0&1&1&0&1\cr
1&0&0&0&1&1&0\cr
0&1&0&0&0&1&1\cr
1&0&1&0&0&0&1\cr}.
$$

Obviously, the $1$-disjunct code B is a
$(q,k,2)$-homogeneous code if and only if the {\it weight
of each row} and the {\it weight of each column} of $C$ are
equal to $k$.  It is not difficult to understand that this
condition is true for any {\it circulant} matrix. The
circulant matrix $C$ is defined as follows:
\begin{itemize}
\item
the first row ${\bf c}_1=(c_1(1),c_1(2),\dots,c_1(q))$ of
circulant matrix $C$ is an arbitrary binary sequence of
length $q$ and weight $k\le q$,
\item
the $m$-th $m=2,3,\dots,q$ row ${\bf
c}_m=((c_m(1),c_m(2),\dots,c_m(q))$ of $C$ is the {\it
cyclic shift} of the  $(m-1)$-th row, i.e.,
$$
c_m(j)=\cases{c_{m-1}(q), & if $j=1$,\cr
c_{m-1}(j-1), & if $j=2,3,\dots,q$.\cr}
$$
\end{itemize}
The first statement of Theorem 1 is proved.

To prove the second statement of Theorem 1, we apply
the evident necessary and sufficient condition of
$2$-separable property which is given in~\cite{ks64}:
{\it no two 1's in $C$ must occupy the same pair of rows
and columns as two other 1's; that is, no row of $C$ can
contain a pair of 1's in the same two positions as another
row}.

It is easy to check that the circulant
matrix $C$ of  Example~5 satisfies this condition.
Let $q\ge 2^k$. As the simple generalization, we consider
the circulant matrix $C$ whose first row ${\bf
c}_1=(c_1(1),c_1(2),\dots,c_1(q))$ is defined as follows $$
c_1(j)=\cases{1, & if $j=2^{n-1},\,n=1,2,\dots,k$,\cr
0, & otherwise.\cr}
$$
Theorem 1 is proved.

\subsection{Proof of Theorem 2}
\begin{center}
\end{center}

The following Proposition~4
gives the recurrent  construction method of $(s+1)$-
separable codes with the help of $s$-separable codes.
\medskip

{\bf Proposition 4}. {\it If there exists an
$(q,k,s)$-homogeneous $s$-separable code
$B(q,k,s)$ with elements from $A_q=[q]$,
then there exists  the
$(kq,k,s+1)$-homogeneous $(s+1)$-separable
code $B(kq,k,s+1)$ with elements from $A_{kq}=[kq]$}.

{\bf Proof}. Let
$$
B(q,k,s)=\|b_j^s(u)\|,\;b_j^s(u)\in[q],\;
j=1,2,\dots,s,\;u=1,2,\dots,kq
$$
be an arbitrary
$(q,k,s)$-homogeneous $s$-separable code.
Consider the following two-step recurrent construction
(cf. examples 3 and 4) for
$(kq,k,s+1)$-homogeneous code $$
B(kq,k,s+1)=\|b_j^{s+1}(u)\|,\;b_j^{s+1}(u)\in[kq],\;
j=1,2,\dots,s+1,\;u=1,2,\dots,k^2q.
$$
\begin{itemize}
\item
The first $kq$ codewords of $B(kq,k,s+1)$
have the form
$$
\begin{array}{cccc}
b_1^s(1)&b_1^s(2)&\dots&b_1^s(kq)\\
b_2^s(1)&b_2^s(2)&\dots&b_2^s(kq)\\
\dots&\dots&\dots&\dots\\
b_s^s(1)&b_s^s(2)&\dots&b_s^s(kq)\\
1&2&\dots&kq,
\end{array}
$$
i.e., for $u=1,2,\dots,kq$, the
element $b_j^{s+1}(u)=b_j^{s}(u)$, if
$j=1,2,\dots,s$, and
$b_{s+1}^{s+1}(u)=u$.
\item
If the number
$u=kqm+l,\;m=1,2,\dots,k-1,\;l=1,2,\dots,kq$, then
\begin{itemize}
\item
for $j=1,2,\dots s$,
the element
$b_j^{s+1}(u)=b_j^{s+1}(kqm+l)=b_j^{s+1}(l)+qm$,
\item
for $j=s+1$,
the element
$b_{s+1}^{s+1}(u)=b_{s+1}^{s+1}(kqm+l)=
b_{s+1}^{s+1}(l)=l$.
\end{itemize}
\end{itemize}
Note that $t=k^2q$ codewords of $B(kq,k,s+1)$
(or the set $[k^2q]$) could be
divided into $k$ groups of the equal
cardinality $kq$ where the $m$-th group $G_m(q,k,s)$,
$m=1,2,\dots,k$,
has the form
$$
G_m(q,k,s)=\pmatrix{
b_1^s(1)+(m-1)q&b_1^s(2)+(m-1)q&\dots&b_1^s(kq)+(m-1)q\cr
b_2^s(1)+(m-1)q&b_2^s(2)+(m-1)q&\dots&b_2^s(kq)+(m-1)q\cr
\dots&\dots&\dots&\dots\cr
b_s^s(1)+(m-1)q&b_s^s(2)+(m-1)q&\dots&b_s^s(kq)+(m-1)q\cr
1&2&\dots&kq\cr}.
$$
Let
$$
A_{kq}^{(m)}=\{(m-1)kq+1,(m-1)kq+2,\dots,mkq\},\;|A_{kq}^{(m)}|=kq,
\;\bigcup_{m=1}^k A_{kq}^{(m)}=[k^2q],
$$
be the set of numbers of codewords which belong to
$G_m(q,k,s)$.
Consider the $s\times kq$ matrix $B_m(q,k,s)$ composed of
the first $s$ rows of $G_m(q,k,s)$, $m=1,2,\dots,k$.
Obviously, $B_m(q,k,s)$ is
$(q,k,s)$-homogeneous code.
In addition, all
elements of $B_m(q,k,s)$ belong to the alphabet
$$
A_q^{(m)}=\{(m-1)q+1,(m-1)q+2,\dots,mq\},\;|A_q^{(m)}|=q,
\;\bigcup_{m=1}^k A_q^{(m)}=[kq],
$$
and, hence, they
do not may occur in $B_n(q,k,s)$, if $n\ne
m,\;n=1,2,\dots,k$. On account of the $s$-separable
property of $B(q,k,s)$, it follows the $s$-separable
property of $B_m(q,k,s),\;m=1,2,\dots,k$.

To prove the $(s+1)$-separable property of $B(kq,k,s+1)$,
we consider an
arbitrary $(s+1)$-subset of the set $[k^2q]$:
$\;
\e=\{e_1,e_2,\dots,\e_{s+1}\}, \;
1\le e_1<e_2<\cdots<e_{s+1}\le k^2q.
$
Let
$$
{\A}_1(\e,B),{\A}_2(\e,B),\dots,{\A}_s(\e,B),
{\A}_{s+1}(\e,B)
$$
be the corresponding subsets of the set $[kq]$ and
$$
\e=\sum_{m=1}^k\e_m,
\quad \e_m=\e\bigcap A_{kq}^{(m)} .
$$
The above-mentioned property of groups
$G_m(q,k,s)$, $m=1,2,\dots,k$ implies that
for any $j=1,2,\dots,s$, the set ${\A}_j(\e,B)$
could be written in the form
$$
{\A}_j(\e,B)=\sum_{m=1}^k{\A}_j(\e_m,B),
$$
where ${\A}_j(\e_m,B)\subseteq A_q^{(m)}$.
Hence, for any fixed $j=1,2,\dots,s$, all nonempty sets
${\A}_j(\e_m,B)$, $m=1,2,\dots,k,$
could be identified on the basis of the set
${\A}_j(\e,B)$.

We have two possibilities.
\begin{itemize}
\item
There exists the unique value $m=1,2,\dots,k$
such that $\e_m=\e,\;|\e_m|=s+1$.
It follows that for any $j=1,2,\dots,s$,
the set ${\A}_j(\e_m,B)\ne\emptyset$ and,
for any $n\ne m$, the set ${\A}_j(\e_n,B)=\emptyset$.
Hence, one can identify
the set $\e$ on the basis of the set
${\A}_{s+1}(\e,B)$.
\item
For any $m=1,2,\dots,k$, the cardinality $|\e_m|\le s$.
Accounting the $s$-separating property of $B_m(q,k,s)$,
the set $\e_m$ could be identified
on the basis of $s$ subsets
$
{\A}_j(\e_m,B),
\; j=1,2,\dots,s
$.
It follows the possibility to identify
$\e=\sum_{m=1}^k\e_m$.
\end{itemize}
Proposition 4 is proved.
\medskip

Let an $(k,k,2)$-homogeneous $1$-separable code $B(k,k,2)$
be the code from  example~1. Consider the
corresponding $(k^2,k,3)$-homogeneous code $B(k^2,k,3)$
obtained from  $B(k,k,2)$ on the basis of Proposition~4.
For $k=3,4$, constructions of $B(k^2,k,3)$ are given in
examples~3 and~4.  To prove Theorem~2,
it is sufficient to establish the $3$-separable property
of code $B=B(k^2,k,3)$ for $k=4,5,\dots$.

We shall use symbols which were introduced to prove
Proposition~4. Note that $t=k^3$ codewords of $B$ could be
divided (in increasing order) into $k$ groups
$G_m(k,k,2),\;m=1,2,\dots,k$ of the equal cardinality
$k^2$. Consider the $2\times k^2$ matrix $B_m(k,k,2)$
composed of the first $2$ rows of $G_m(k,k,2)$,
$m=1,2,\dots,k$.  Obviously, $B_m(k,k,2)$ is the
$(k,k,2)$-homogeneous $1$-separable code.
In addition, all
elements of $B_m(k,k,s)$ belong to the alphabet
$$
A_k^{(m)}=\{(m-1)k+1,(m-1)k+2,\dots,mk\},\;|A_k^{(m)}|=k,
\;\bigcup_{m=1}^k A_k^{(m)}=[k^2],
$$
and, hence, they
do not may occur in $B_n(k,k,2)$,
if $n\ne m,\;n=1,2,\dots,k$.

Let $\e=\{e_1,e_2,e_3\},\;1\le e_1<e_2<e_3\le k^3$
be an arbitrary fixed $3$-subset of the set $[k^3]$ and
$\{{\bf b}(e_1),\;{\bf b}(e_2),\;{\bf b}(e_3)\}$
be the corresponding triple of codewords
of code $B$.
To identify the codewords ${\bf b}(e_i),\;i=1,2,3$, using
the properties of $B_m(k,k,2),\;m=1,2,\dots,k$,
mentioned above, it suffices to analyze the following
three cases.
\begin{itemize}
\item
There are known three numbers
$1\le m_1<m_2<m_3\le k$ such that the codeword
${\bf b}(e_i),\;i=1,2,3$ belongs to
the group $G_{m_i}(k,k,2)$. In this case,
${\bf b}(e_i)$ could be identified on the basis of
$1$-separable property of  $B_{m_i}(k,k,2)$.

\item
There is known the number $m=1,2,\dots,k$ such that
all three codewords
${\bf b}(e_1),\;{\bf b}(e_2),\;{\bf b}(e_3)$
belong to the group $G_m(k,k,2)$.
In this case, the triple
$\{{\bf b}(e_1),\;{\bf b}(e_2),\;{\bf b}(e_3)\}$,
can be identified on the basis of the set
$\A_3(\e,B)$ whose cardinality $|\A_3(\e,B)|=3$.
\item
There are known two numbers $1\le m<n\le k$ such that
(without loss of generality) codeword ${\bf b}(e_1)$
belongs to the group $G_m(k,k,2)$ and two other codewords
${\bf b}(e_2)$ and ${\bf b}(e_3)$ belong to the
group $G_n(k,k,2)$. In this case, we have the following
three-step identification:
\begin{itemize}
\item
the codeword ${\bf b}(e_1)=(b_1(e_1),b_2(e_1),b_3(e_1))$
is identified on the basis of
$1$-separable property of $B_m(k,k,2)$,
\item
the set $\{b_3(e_2),b_3(e_3)\}$
evidently identified on the basis of symbol $b_3(e_1)$
and the set $\A_3(\e,B)$,
\item
codewords ${\bf b}(e_2)$ and ${\bf b}(e_3)$
are identified on the basis of
the set $\{b_3(e_2),b_3(e_3)\}$.
\end{itemize}
\end{itemize}
Theorem 2 is proved.

\subsection{Proof of Theorem 3}
\begin{center}
\end{center}

Let  $q\ge k+1,\;k\ge 4$, and $q$-ary
alphabet $A_q=[q]=\{1,2,\dots,q\}$.
We need to construct
$(q,k,3)$-homogeneous code $B$ of distance $D(B)=2$.
Consider the construction of $(q,k,3)$-homogeneous
code  $B=\|b_j(u)\|$ whose rows
$
{\bf b}_j=(b_j(1),b_j(2),\dots,b_j(kq)),\;j=1,2,3,
$
are defined as follows:
\begin{enumerate}
\item
for $j=1$, the first row
$
{\bf b}_1=({\bf b}_1^{(1)},{\bf b}_1^{(2)},
\dots,{\bf b}_1^{(q)}),\;
{\bf b}_1^{(m)}=\underbrace{(m,m,\dots,m)}_{k},
\;m=1,2,\dots,q;
$
\item
for $j=2$, the second row
$
{\bf b}_2=({\bf b}_2^{(1)},{\bf b}_2^{(2)},
\dots,{\bf b}_2^{(k)}),\;
{\bf b}_2^{(m)}=(1,2,\dots,q),\;m=1,2,\dots,k;
$
\item
for $j=3$, the third row
$
{\bf b}_3=({\bf b}_3^{(1)},{\bf b}_3^{(2)},
\dots,{\bf b}_3^{(k)}),
$
where  the subsequence
${\bf b}_3^{(m)}$ of length $q$ is the {\it
$(m-1)$-step cyclic shift} of the sequence
$(1,2,\dots,q)$:
$$
{\bf b}_3^{(m)}=\cases{(1,2,\dots,q), & if
$m=1$,\cr
m,m+1,\dots,q-1,q,1,2,\dots,m-1, &
if $m=2,3,\dots,k.$\cr}
$$
\end{enumerate}
Obviously, this construction guarantees the distance
$D(B)=2$. From Proposition 2 it follows $2$-disjunct
property of $B$. \\
Theorem 3 is proved.

{\bf Example 6}. As an illustration, we yield the
$(6,4,3)$-homogeneous $2$-disjunct code $B$ with $kq=24$
codewords
$$
B=\left(\begin{array}{cccc}
111122&223333&444455&556666\\
123456&123456&123456&123456\\
123456&234561&345612&456123
\end{array}\right).
$$

{\bf Remark}. For $2$-disjunct code
$B$ of Example 6, it is easy to check the
following properties.
\begin{itemize}
\item
The $3$-subsets $\e=\{2,8,13\}$ and
$\e'=\{2,7,13\}$ of the set $[24]$ have the
same coordinate sets, namely:  $\A_1=\{1,2,4\}$,
$\A_2=\{1,2\}$ and $\A_3=\{2,3\}$. From
this it follows that the code $B$ is not
$3$-separable code, i.e., in general, the ordering
$(s-1)$-disjunct$\Longrightarrow$$s$-separable
is not true.
\item
The $3$-subset $\e=\{1,2,7\}$
of the set $[24]$ has the equal
coordinate sets $\A_1=\A_2=\A_3=\{1,2\}$
of cardinality $2$. Hence, the code $B$
is not $3$~-~hash code, i.e., in general, the ordering
$(s-1)$-disjunct$\Longrightarrow$$s$-hash
is not true.
\end{itemize}

\subsection{On $(q,k,3)$-homogeneous $3$-separable
and $3$-hash codes.\\ Proof of Theorem 4}
\begin{center}
\end{center}

\subsubsection{Characteristic matrices}
\begin{center}
\end{center}

Consider an arbitrary $(q,k,3)$-homogeneous
$2$-disjunct code $B$. From
Proposition 2 it follows that we can introduce
{\it characteristic} $(q\times q)$-matrix
$\;C=\|c_i(j)\|,\,i=1,2,\dots,q,\,j=1,2,\dots,q,$
with elements from alphabet
$A_{q+1}=\{*,\,[q]\}=\{*,1,2,\dots,q\}$,
where
$$
c_i(j)=\cases{a, & if there exists codeword
${\bf b}(u)=(a,i,j)$,\cr
*, & otherwise.\cr}
$$
We shall say that code {\it $B$ is identified} by the
(characteristic) matrix $C$ which will be called
{\it $C(q,k)$-matrix}.
\medskip

{\bf Example 7.}
For $k=4,\;q=6$, the $(6,4,3)$-homogeneous
$2$-disjunct code $B$ of Example~6 is identified by
$C(q,k)$-matrix
$$
C=\pmatrix{1&2&4&5&*&*\cr
           *&1&2&4&5&*\cr
           *&*&1&3&4&6\cr
           6&*&*&1&3&4\cr
           5&6&*&*&2&3\cr
           3&5&6&*&*&2\cr}.
$$

The evident characterization of $C(q,k)$-
matrix is given by Proposition~5.

{\bf Proposition 5}. {\it The
matrix $C$ is $C(q,k)$-matrix if and only if $C$
has the following properties:
\begin{itemize}
\item
for any $a\in[q]$, the number of $a$-entries in $C$ is
equal to $k$,
\item
for any row (column) of $C$, the number
of $*$-entries in the row (column) is equal to $q-k$,
\item
for any $a\in[q]$ and any row (column) of $C$, the number
of $a$-entries in the row (column) does not exceed $1$.
\end{itemize}}

{\bf Remark}. If {\Large$q=k$}, then $C(q,q)$-matrix
is called the {\it Latin square}.
\medskip

Characteristic matrix $C$ of hash, separable and
hash\&separable code will be called $C_H(q,k)$-matrix,
$C_S(q,k)$-matrix and $C_{HS}(q,k)$-matrix.
\medskip

One can easily check the following characterization of
$C_{H}(q,k)$-matrix.

{\bf Proposition 6}. {\it Matrix $C$ is $C_{H}(q,k)$-matrix
if and only if $C$ has the  properties of Proposition 5 and
the following two equivalent conditions take place:
\begin{itemize}
\item
if for $i\ne m$ and $j\ne n$, the element
$c_i(j)=c_m(n)=a\ne *$, then $c_i(n)=c_m(j)=*$.
\item
if for $i\ne m$, $j\ne n$ and $a\ne b$, code $B$ contains
codewords $(a,i,j)$ and $(a,m,n)$, then $B$ does not contain
the word $(b,m,j)$.
\end{itemize}}

The evident characterization of $C_{HS}(q,k)$-matrix
is given by Proposition~7.
\medskip

{\bf Proposition 7}. {\it Let $a,\,b$ and $c$ be arbitrary
pairwise distinct elements of $[q]$. Matrix $C$ is
$C_{HS}(q,k)$-matrix if and only if $C$ has  properties of
Propositions~5 and~6 and the following property is true.
Matrix $C$ does not contain any $(3\times 3)$-submatrix of
the form:
$$
\pmatrix{ *&a&c\cr a&*&b\cr
c&b&*\cr}, \pmatrix{ c&a&*\cr b&*&a\cr *&b&c\cr}, \pmatrix{
*&c&a\cr a&b&*\cr c&*&b\cr},
$$
$$
\pmatrix{ a&*&c\cr
*&a&b\cr
b&c&*\cr},
\pmatrix{
a&c&*\cr
*&b&a\cr
b&*&c\cr},
\pmatrix{
c&*&a\cr
b&a&*\cr
*&c&b\cr}.
$$
These prohibited matrices are  the permutations of the same
three columns}.
\medskip

{\bf Remark}. The characterization of $C_{S}(q,k)$-matrix
has a tedious form and it is omitted here. Below, we give
the examples of $C_{S}(q,k)$-matrices which are not
$C_{HS}(q,k)$-matrices.

\subsubsection{Examples of hash, separable
and hash\&separable codes}
\begin{center}
\end{center}

Let an integer $k\ge2$ be fixed. How to find the minimal
possible integer $q_k\ge k$ such that there exists
$C_{S}(q_k,k)$-matrix, $C_{H}(q_k,k)$-matrix or
$C_{HS}(q_k,k)$-matrix? From Examples 3 and 4 it follows
that one can put $q_k=k^2$. For $k=2,3,4$, the following
Examples~8-10 improve this result and yield $C_{S}(q_k,k)$,
$C_{H}(q_k,k)$ and $C_{HS}(q_k,k)$-matrices for which
$q_k<k^2$.  If $k=2,3,4$ and $q_k<q<k^2$, then the
corresponding characteristic matrices could be
given also.
\medskip

{\bf Example 8.} For  $k=2$, $C_{S}(q_2,2)$-matrix,
$C_{H}(q_2,2)$-matrix and $C_{HS}(q_2,2)$-matrix
are\footnote[1]{Here and below,
for $3$-hash codes,
we mark the pairs of "bad" triples
which break the $3$-separable property.}
$$
\pmatrix{
1&2&*\cr
*&1&3\cr
3&*&2\cr},\quad
\pmatrix{
*&\dot3&\ddot1\cr
\dot1&\ddot2&*\cr
\ddot3&*&\dot2\cr},\quad
\pmatrix{
1&*&3&*\cr
*&1&*&3\cr
4&*&2&*\cr
*&4&*&2\cr}
$$
The first matrix ($q_2=3$) identifies the separable (not
hash) code. The second matrix ($q_2=3$) identifies the hash
(not separable) code.  The third matrix ($q_2=4$) is the
particular case of Proposition~4.
\medskip

{\bf Example 9.} Let $k=3$.
For hash code $q_3=6$ and for hash\&separable code $q_3=7$.
The corresponding characteristic matrices are
$$
\pmatrix{
*&*&\dot1&\ddot2&3&*\cr
*&\ddot1&*&\dot5&*&3\cr
1&*&*&*&5&4\cr
*&\dot2&\ddot5&*&*&6\cr
2&*&4&*&6&*\cr
3&4&*&6&*&*\cr},
\pmatrix{*&*&1&2&3&*&*\cr
           *&1&*&5&7&*&*\cr
           1&*&*&*&*&7&3\cr
           *&2&*&*&*&5&4\cr
           2&*&7&*&*&*&6\cr
           *&3&*&4&*&6&*\cr
           5&*&4&*&6&*&*\cr}.
$$

{\bf Example 10.}  Let $k=4$.  For hash
codes, $q_4=8$ and for hash\&separable codes, $q_4=13$.
The corresponding characteristic  $C_{H}(8,4)$ and
$C_{HS}(13,4)$-matrices are
$$
\pmatrix{
*&*&*&1&*&\dot2&3&\ddot5\cr
*&*&1&*&\ddot2&*&4&\dot6\cr
*&1&*&*&3&4&*&7\cr
1&*&*&*&\dot5&\ddot6&7&*\cr
*&2&3&4&*&*&*&8\cr
2&*&5&6&*&*&8&*\cr
3&5&*&7&*&8&*&*\cr
4&6&7&*&8&*&*&*\cr},
$$
$$
\pmatrix{*&*&*&*&*&*&4&2&3&*&8&*&*\cr
           *&*&*&*&*&*&*&7&13&5&1&*&*\cr
           *&*&*&*&*&*&*&9&10&*&*&1&8\cr
           *&*&*&*&*&*&6&12&11&*&*&5&*\cr
           *&*&*&*&*&*&*&*&*&10&11&13&3\cr
           *&*&*&*&*&*&*&*&*&9&12&7&2\cr
           1&*&*&13&*&*&*&*&*&6&*&*&4\cr
           2&7&9&12&*&*&*&*&*&*&*&*&*\cr
           3&6&11&10&*&*&*&*&*&*&*&*&*\cr
           *&5&*&*&11&9&13&*&*&*&*&*&*\cr
           8&4&*&*&10&12&*&*&*&*&*&*&*\cr
           *&*&4&5&6&7&*&*&*&*&*&*&*\cr
           *&*&8&*&3&2&1&*&*&*&*&*&*\cr}.
$$

{\bf Open problems}.
{\bf 1}. Is it possible to construct a $C_{HS}(q,4)$-matrix
if $q<13$?  {\bf 2}. Is it possible to construct
$C_{HS}(q,k)$-matrices, if $k\ge5$ and $q<k^2$?

\subsubsection{Existence of hash and hash\&separable
codes}
\begin{center}
\end{center}

The following obvious Proposition~8 can be used to construct
the new characteristic matrices using the known
ones.
\medskip

{\bf Proposition 8}. (S.M. Yekhanin, 1998).
{\it Let
$v=1,2$ and
there exist $C_{H}(q_v,k)$-matrix
$$
C^v=\|c_i^v(j)\|,\;
i,j\in[q_v],\quad
c_i^v(j)\in\{*,\,[q_v]\}.
$$
Let
$
{\tilde C}^2=\|{\tilde c}^2_i(j)\|$
be  the matrix whose element
$$
{\tilde c}^2_i(j)=\cases{q_1+c^2_i(j), & if
$c_i^2(j)\ne *$ ,\cr
*, & otherwise.\cr}
$$
Then matrix
$$
C=\pmatrix{
C^1&*\cr
*&{\tilde C}^2\cr}
$$
is a $C_{H}(q_1+q_2,k)$-matrix.
The similar statement is also true for
characteristic matrices of hash$\&$separable
codes}.
\bigskip

With the help of the computer checking, we constructed the
finite collection of "non-regular"
$C_{H}(q,4)$-matrices, $q\ge8$,
and $C_{HS}(q,4)$-matrices, $q\ge13$.
Taking into account Proposition 8, we obtain
\medskip

{\bf Proposition 9}.
 1. {\it If  $q\ge8$, then there exists
$C_{H}(q,4)$-matrix}.
2. {\it If  $q\ge13$, then there exists
$C_{HS}(q,4)$-matrix}.
\medskip

The following statement is a generalization of the
hash\&separable construction of Examples~3 and~4.
\medskip

{\bf Proposition 10}. {\it If $q\ge k^2$, then there exists
$(q,k,3)$-homogeneous $3$-hash code}.
\medskip

{\bf Proof}. Let $k=2,3,\dots$ and $q\ge k^2$. Consider the
following construction of $(q,k,3)$-homogeneous code
$B=\|b_j(u)\|$ whose rows
$$
{\bf b}_j=(b_j(1),b_j(2),\dots,b_j(kq)),\;j=1,2,3,
$$
are defined as follows:
\begin{enumerate}
\item
for $j=1$, the first row
$$
{\bf b}_1=({\bf b}_1^{(1)},{\bf b}_1^{(2)},
\dots,{\bf b}_1^{(q)}),\quad
{\bf b}_1^{(m)}=\underbrace{(m,m,\dots,m)}_{k},
\quad m=1,2,\dots,q;
$$
\item
for $j=2$, the second row
$$
{\bf b}_2=({\bf b}_2^{(1)},{\bf b}_2^{(2)},
\dots,{\bf b}_2^{(k)}),\quad
{\bf b}_2^{(m)}=(1,2,\dots,q),
\quad m=1,2,\dots,k;
$$
\item
for $j=3$, the third row
$
{\bf b}_3=({\bf b}_3^{(1)},{\bf b}_3^{(2)},
\dots,{\bf b}_3^{(k)}),
$
where  the subsequence
${\bf b}_3^{(m)},\,m=1,2,\dots,k$
of length $q$ is
the {\it
$k(m-1)$-step cyclic shift} of the sequence
$(1,2,\dots,q)$:
$$
{\bf b}_3^{(m)}=\cases{(1,2,\dots,q), & if $m=1$,\cr
(k(m-1)+1,k(m-1)+2,\dots,q-1,q,1,2,\dots,k(m-1), &
if $m=2,3,\dots,k.$\cr}
$$
\end{enumerate}

As an illustration,
we yield the $(11,3,3)$-homogeneous code
$$
\pmatrix{
111&222&333&444&555&666&777&888&999&aaa&bbb\cr
123&456&789&ab1&234&567&89a&b12&345&678&9ab\cr
123&456&789&ab4&567&89a&b12&378&9ab&123&456\cr},
$$
where, for convenience of notations, we put
$a=10,\,b=11$.

If $q\ge k^2$, then
this construction of
$(q,k,3)$-homogeneous code $B$
has an evident property of
{\it alphabet separation}, which
could be formulated as follows.
{\it Let the symbol $\oplus$ denote modulo $kq$
addition and $u=1,2,\dots,kq$ be an arbitrary fixed
integer. Then  $q$-ary  elements of the $k$-subsequence
$$
b_3(u),b_3(u\oplus1),b_3(u\oplus2),
\dots,b_3(u\oplus (k-1))
$$
do not may occur in the $k$-subsequence}
$$
b_3(u\oplus q),b_3(u\oplus (q+1)),b_3(u\oplus (q+2)),
\dots,b_3(u\oplus (q+k-1)).
$$
By virtue of the second condition of Proposition 6, it
implies $3$-hash property of code $B$.\\
Proposition 10 is proved.
\bigskip

{\bf Conjecture}.
The construction of Proposition 10 yields
hash\&separable codes.

\subsubsection{Product of characteristic matrices}
\begin{center}
\end{center}

In this section, we consider a construction  of homogeneous
codes, which makes possible to obtain the new (more
complicated) codes using the known ones.

Let $v=1,2$ and
$
C^v=\|c_i^v(j)\|$,
$i,j\in[q_v]$,
$c_i^v(j)\in\{*,\,[q_v]\},
$
be $C(q_v,k_v)$-matrix of code $B_v$. Denote by
$$
C=C^1\diamond C^2=\|c_r(u)\|,\;r,u\in[q_1q_2],\;
c_r(u)\in\{*,\,[q_1q_2]\}
$$
the {\it product of characteristic matrices of code $B_1$
and code $B_2$}. Matrix $C$ is defined as follows:  for
arbitrary $i,j\in[q_1]$ and $l,m\in[q_2]$, put
$$
r=q_2(i-1)+l,\quad
u=q_2(j-1)+m,
$$
$$
c_r(u)=\cases{q_2(c_i^1(j)-1)+c_l^2(m), & if
$c_i^1(j)\ne *$ and $c_l^2(m)\ne *$,\cr
*, & otherwise.\cr}
$$

{\bf Example 11.} Let $k_1=k_2=2,\;q_1=q_2=3$, and
$$
C_H(q_1,k_1)=C_H(q_2,k_2)=
\pmatrix{*&1&2\cr 1&*&3\cr
             2&3&*\cr}.
$$
$$
C_H(q_1q_2,k_1k_2)=C_H(q_1,k_1)
\diamond C_H(q_2,k_2)=
$$
$$
=\pmatrix{
*&*&*&*&1&2&*&4&5\cr
*&*&*&1&*&3&4&*&6\cr
*&*&*&2&3&*&5&6&*\cr
*&1&2&*&*&*&*&7&8\cr
1&*&3&*&*&*&7&*&9\cr
2&3&*&*&*&*&8&9&*\cr
*&4&5&*&7&8&*&*&*\cr
4&*&6&7&*&9&*&*&*\cr
5&6&*&8&9&*&*&*&*\cr}.
$$
Such product of matrices remains the hash property.

{\bf Example 12.}
Let $k_1=k_2=2,\;q_1=q_2=3$, and
$$
C_S(q_1,k_1)=C_S(q_2,k_2)=
\pmatrix{
1&2&*\cr
*&1&3\cr
3&*&2\cr}.
$$
The product of matrices
$$
C(q_1q_2,k_1k_2)=C_S(q_1,k_1)
\diamond C_S(q_2,k_2)=
$$
$$
=\pmatrix{
1&\grave2&*&\acute4&5&*&*&*&*\cr
*&\acute1&3&*&\grave4&6&*&*&*\cr
3&*&2&6&*&5&*&*&*\cr
*&*&*&\grave1&\acute2&*&\bar7&\hat8&*\cr
*&*&*&*&1&3&*&7&9\cr
*&*&*&3&*&2&\dot9&*&\ddot8\cr
7&\bar8&*&*&*&*&\hat4&5&*\cr
*&\hat7&\ddot9&*&*&*&*&\bar4&\dot6\cr
9&*&\dot8&*&*&*&\ddot6&*&5\cr}
$$
does not remain the separable properties of factors.
In the figure, we have {\it marked} three pairs of
"bad" triples, namely:
$$
\{(\grave1,\grave2,\grave4)\;(\acute1,\acute2,\acute4)\},
\quad
\{(\bar4,\bar7,\bar8)\;(\hat4,\hat7,\hat8)\},\quad
\{(\dot6,\dot8,\dot9)\;(\ddot6,\ddot8,\ddot9)\}.
$$

This example shows the reason why the separable property of
the product of two separable matrices is not true. To
guarantee the separable property of the product of two
separable matrices, {\it at least one of two factors
should have hash$\&$separable property}. The following
Proposition takes place.
\newpage

{\bf Proposition 11.}
{\bf 1}. {\it The product of $C_H(q_1,k_1)$-matrix
and $C_H(q_2,k_2)$-matrix is $C_H(q_1q_2,k_1k_2)$-matrix.
{\bf 2}. The product of $C_S(q_1,k_1)$-matrix
and $C_{HS}(q_2,k_2)$-matrix
is $C_S(q_1q_2,k_1k_2)$-matrix. In addition,
if the product of two separable matrices has the separable
property, then at least one of these factors should have
the hash$\&$separable property}. (S.M. Yekhanin, 1998).
\bigskip

To explain the second statement of Proposition 11, we give
the following example.
\medskip

{\bf Example 13.}
Let $k_1=k_2=2,\;q_1=3,\;q_2=4$, and
$$
C_S(q_1,k_1)=
\pmatrix{
1&2&*\cr
*&1&3\cr
3&*&2\cr},\quad
C_{HS}(q_2,k_2)=
\pmatrix{
1&*&3&*\cr
*&1&*&3\cr
4&*&2&*\cr
*&4&*&2\cr}.
$$
The product
$
C_S(q_1,k_1)\diamond C_{HS}(q_2,k_2)
$
has the form
$$
\pmatrix{
1&*&3&*&5&*&7&*&*&*&*&*\cr
*&1&*&3&*&5&*&7&*&*&*&*\cr
4&*&2&*&8&*&6&*&*&*&*&*\cr
*&4&*&2&*&8&*&6&*&*&*&*\cr
*&*&*&*&1&*&3&*&9&*&11&*\cr
*&*&*&*&*&1&*&3&*&9&*&11\cr
*&*&*&*&4&*&2&*&12&*&10&*\cr
*&*&*&*&*&4&*&2&*&12&*&10\cr
9&*&11&*&*&*&*&*&5&*&7&*\cr
*&9&*&11&*&*&*&*&*&5&*&7\cr
12&*&10&*&*&*&*&*&8&*&6&*\cr
*&12&*&10&*&*&*&*&*&8&*&6\cr}
$$
which illustrates its separable property.

The changed order product
$ C_{HS}(q_2,k_2)\diamond C_S(q_1,k_1)$
also remains the separable property and has the form
$$
\pmatrix{
1&2&*&*&*&*&7&8&*&*&*&*\cr
*&1&3&*&*&*&*&7&9&*&*&*\cr
3&*&2&*&*&*&9&*&8&*&*&*\cr
*&*&*&1&2&*&*&*&*&7&8&*\cr
*&*&*&*&1&3&*&*&*&*&7&9\cr
*&*&*&3&*&2&*&*&*&9&*&8\cr
10&11&*&*&*&*&4&5&*&*&*&*\cr
*&10&12&*&*&*&*&4&6&*&*&*\cr
12&*&11&*&*&*&6&*&5&*&*&*\cr
*&*&*&10&11&*&*&*&*&4&5&*\cr
*&*&*&*&10&12&*&*&*&*&4&6\cr
*&*&*&12&*&11&*&*&*&6&*&5\cr}.
$$
\bigskip

From Propositions 9, 11 and Example 13 it follows the
statement of Theorem~4.

\subsection{Proof of Theorem 5}
\begin{center}
\end{center}

Let $s\ge2,\;l\ge1$ be fixed integers and
$n>2s+l$ be an arbitrary integer.
Let $[n]$ be the set of integers
from 1 to $n$ and ${\cal E}(s,n)$ be the collection of all
${n\choose s}$ $s$-subsets of $[n]$.
Following~\cite{m96},
we define the binary code $X=\|x_B(A)\|$,
$B\in{\cal E}(s,n),\;A\in{\cal E}(s+l,n)$,
of size $t={n\choose s+l}$
and length $N={n\choose s}$, whose element $x_B(A)=1$
if and only if $B\subset A$.
One can easily understand that $X$ is
the constant weight code with parameters:
$$
t={n\choose s+l},\quad N={n\choose s},\quad
k={n-s\choose l},\quad w={s+l\choose s},\quad
\lambda={s+l-1\choose s},
$$
where $t$--code size, $N$--code length, $w$--weight of
columns (codewords), $k$--weight of rows and $\lambda$--the
maximal dot product of codewords. In addition, let
$A_0,A_1,\dots,A_s,\;A_i\in{\cal E}(s+l,n)$
be an
arbitrary $(s+1)$-collection of pairwise different
$(s+l)$-subsets of $[n]$.
Since $A_0\ne A_i$, for any
$i=1,2\dots,s$, there exists an element $a_i\in A_0$ and
$a_i\notin A_i$. Hence, there exists a $s$-subset $B\subset
A_0$ and for any $i=1,2,\dots,s$, $B\not\subset A_i$. It
follows that $X$ is a superimposed $(s,t,k)$--code.  For the
particular case $l=1$, these properties yield Theorem~5.

\newpage

\centerline{Arkadii G. D'yachkov,\qquad "Lectures on
Designing Screening Experiments"}
\medskip

\section{Adder channel model and $B_s$-codes}
\begin{center}
\end{center}

\subsection{Statement of the problem and results}
\begin{center}
\end{center}

For the optimal parameters of A--model, we use the
notations of  Sect.~1, i.e., $t_A(s,N)$, $N_A(s,t)$
and $R_A(s)$.  Let $1\le u_1\le u_2\le\dots\le u_s\le t$.
In what follows, the sum of columns
$\sum\limits_{k=1}^s\x(u_k)$ is defined as a column of
length $N$ whose $i$-th component, $i=1,2,\dots,N$, is
equal to the arithmetic sum $\sum\limits_{k=1}^s x_i(u_k)$.
\medskip

{\bf Definition}. Matrix $X$ is called a $B_s$-{\it code}
of length $N$ and size $t$, if all
${t+s-1\choose s}$ sums of its columns
$\sum\limits_{k=1}^s\x(u_k)$,
where $1\le u_1\le u_2\le\dots\le u_s\le t$, are distinct.
\medskip

Denote by $N(s,t),\;(t(s,N))$ the minimal (maximal) possible number of
rows (columns) of $B_s$-code. For fixed $s\ge 2$  define the
number
$$
R(s)\eq\varlimsup_{N\to\infty}
\frac{\log t(s,N)}{N},
$$
which is called a rate of $B_s$-code.

Obviously, if $s=2$, then the definition of $B_2$-code and
the definition of $(2,N)$-design are equivalent. Hence,
$N(2,t)=N_A(2,t),\;t(2,N)=t_A(2,N),\;
R(2)=R_A(2)$, while for $s\ge 3$
$$
t_A(s,N)\ge t(s,N),\quad N_A(s,t)\le N(s,t),\quad
R_A(s)\ge R(s).
$$

The concept of $B_s$-code (called a $B_s$-sequence
in~\cite{l72}) was motivated by the concept of
$B_s$-sequence, introduced by Erdos in~\cite{e41}. In this
section we give a survey of the known upper and lower
bounds on  $R_A(s)$ and $R(s)$.

\subsubsection{Upper bounds}
\begin{center}
\end{center}

The upper bounds are given as Theorems 1 and 2.
To formulate Theorem 1, we introduce some notations.
Let
$$
b_s(k,p)\eq{s\choose k}p^k(1-p)^{s-k},\;0\le p\le 1,
\;0\le k\le s\eqno (1)
$$
be binomial probabilities and
$$
H_s(p)\eq-\sum_{k=0}^s b_s(k,p)\log b_s(k,p)\eqno (2)
$$
be the Shannon entropy of the binomial distribution with
parameters $(s,p)$.
\medskip

{\bf Theorem 1}. {\it For any $s\ge 2$ the rate
$$
R(s)\le R_A(s)\le H_s/s,\eqno (3)
$$
where
$$
H_s\eq H_s(1/2)=-\sum_{k=0}^s{s\choose k}2^{-s}
\log\left[{s\choose k}2^{-s}\right].\eqno (4)
$$
}

Theorem 1 is called an {\it entropy bound} and its proof
will be given in Sect.~5.2.
\medskip

{\bf Remark}. Let $p,\; 0\le p\le 1$ be fixed and
$s\to\infty$. Using the Moivre-Laplace local limit
theorem, one can  prove~\cite{d96} that
$$
H_s(p)=\frac12\log s+\frac12\log[2\pi ep(1-p)]+O(1/s),
$$
and for the particular case $p=1/2$
$$
H_s(1/2)=\frac12\log s+\frac12\log\frac{\pi e}{2}+O(1/s^2).
$$
It follows that for $s\gg 1$ upper bound (3)
can be written as
$$
R(s)\le R_A(s)\le\frac{\log s}{2s}+
\frac{1}{2s}\log\frac{\pi e}{2}+ O(1/s^3).\eqno (5)
$$
\medskip

{\bf Theorem 2}.~\cite{dr81,rd83} {\it For any
$s=1,2,\dots$ the rate
$$
R(2s)\le[sH_s^{-1}+sh_s^{-1}]^{-1},\eqno (6)
$$
and for any $s=2,3,\dots$ the rate
$$
R(2s-1)\le[sH_s^{-1}+(s-1)h_s^{-1}]^{-1},\eqno (7)
$$
where $H_s$ is defined by $(4)$ and}
$$
h_s\eq\log(s+1)+s(s+1)^{-1}.
$$
\smallskip

For $B_2$-codes, Theorem 2 gives $R(2)\le3/5$. For the
first time this result was obtained in~\cite{l72}. Direct
calculations show that for $B_{s}$-codes, where
$s=2,3,\dots,10$, bounds (6) and (7) are better than
entropy bound (3).  Bounds (6) and (7) demonstrate the
possibility of improving the entropy bound only for
$B_s$-codes.  Analogous improvement of the entropy bound
for $(s,N)$-designs, when $s\ge3$, remains an open
question.  In Sect.~5.2, we give the proof of Theorem 2
only for the simplest case~$s=2$.
\medskip

{\bf Remark.} For the particular case of $B_4$-codes, bound
(6) was slightly improved in~\cite{rd81}.

\subsubsection{Lower bounds}
\begin{center}
\end{center}

It was shown in~\cite{l75} that the Bose theorem from
additive number theory~\cite{bc62} yields $B_s$-codes with
parameters $N=ks,\;t=2^k$, where $k=1,2,\dots$. This result
gives the lower bound
$$
R_A(s)\ge R(s)\ge1/s.\eqno (8)
$$

The following theorem will be proved in Sect.~5.3 by
the random coding method.
\medskip

{\bf Theorem 3.}~\cite{dr81,p87}. {\it For any $s=2,3,\dots$
$$
R_A(s)\ge R(s)\ge\frac{{\tilde H}_s}{2s-1},\eqno (9)
$$
where}
$$
{\tilde H}_s\eq\log\frac{2^{2s}}{{2s\choose s}}=
\log\frac{(2s)!!}{(2s-1)!!}.\eqno (10)
$$
\medskip

It is easy to see that bound (9)-(10) improves the Bose
bound (8) for $s\ge3$. If $s=2$, then the Bose bound is
better than (9)-(10).
\medskip

{\bf Remark.} Let $s\to\infty$. With the help of the Stirling
formula
$s!\sim s^s\,e^{-s}\sqrt{2\pi\,s}$, one can prove that
$$
{\tilde H}_s=\frac12\log\pi s+\frac{1}{8s\ln2}+O(s^{-2}).
$$
It follows that for $s\gg1$ the rate lower
bound (9) could be written as
$$
R_A(s)\ge R(s)\ge\frac{\log s}{4s}+\frac{1}{4s}\log\pi+o(s^{-1}).\eqno (11)
$$
Hence, as  $s\to\infty$, the ratio of upper bound (5) to
lower bound (11) tends to $2$.
\medskip

\subsection{Proof of upper bounds on $R_A(s)$ and $R(s)$}
\begin{center}
\end{center}
\subsubsection{Proof of Theorem 1}
\begin{center}
\end{center}

{\bf Lemma.}~\cite{m78}. {\it Entropy  $(2)$ takes its
maximal possible value at $p=1/2$, i.e.,}
$$
\max_{0\le p\le 1}H_s(p)=H_s(1/2)=H_s.
$$

{\bf Proof of Lemma}. We will use
logarithms to the base $e$. Denote by the symbols $f'(p)$
and $f''(p)$ the first and the second derivatives of
$f(p)$ with respect to $p$.
Taking into account (1) and the definition of
generalized binomial coefficients (see Sect.~1.3),
one can easily  check that
$$
b'_s(k,p)=sb_{s-1}(k-1,p)-sb_{s-1}(k,p).\leqno 1)
$$
In addition, it is evident that
$$
\sum_{k=0}^s b'_s(k,p)\equiv 0,\qquad 0\le p\le 1.\leqno 2)
$$
It follows from 1) and 2) that
$$
H'_s(p)=-\sum_{k=1}^s sb_{s-1}(k-1,p)\ln b_s(k,p)+
\sum_{k=0}^{s-1}sb_{s-1}(k,p)\ln b_s(k,p).\leqno 3)
$$
The formula 3) could be written as
$$
H'_s(p)=s\sum_{k=0}^{s-1}b_{s-1}(k,p)\ln\frac{(k+1)(1-p)}{(s-k)p}.\leqno 4)
$$
Put
$$
A_s(p)\eq\sum_{k=0}^s b_s(k,p)\ln\frac{k+1}{p}.\leqno 5)
$$
With the help of 5) we can rewrite 4) in the form
$$
H'_s(p)=s[A_{s-1}(p)-A_{s-1}(1-p)].\leqno 6)
$$
Therefore, $H'(1/2)=0$. Further, we want to prove that the
function $H_s(p)$ is convex~$\cap$. It is sufficient to
prove Lemma. The equality~6) means that  the second
derivative $H''_s(p)=s[A'_{s-1}(p)+A'_{s-1}(1-p)]$.
Hence, we need to prove that $A'_{s-1}(p)<0$ for any
$p,\;0<p<1$.  The equalities~1), 2) and 5) yield
$$
A'_s(p)=-\frac1p+\sum_{k=0}^s\ln(k+1)b'_s(k,p)=
$$
$$
=-\,\frac1p\,+\,s\left[\sum_{k=1}^s b_{s-1}(k-1,p)\ln(k+1)-
\sum_{k=0}^{s-1} b_{s-1}(k,p)\ln(k+1)\right]=
-\,\frac1p+s\sum_{k=0}^{s-1} b_{s-1}(k,p)
\ln\frac{k+2}{k+1}=
$$
$$
=-\,\frac1p+\frac1p\,\sum_{k=0}^{s-1}
\frac{s!(1-p)^{s-k-1}p^{k+1}}{(k+1)!(s-k-1)!}(k+1)\,
\ln\,\frac{k+2}{k+1}\,=\,
\frac1p\,\left[-1+\sum_{k=0}^{s-1}
b_s(k+1,p)\ln\left(\frac{k+2}{k+1}\right)^{k+1}\right].
$$
Taking into account the well-known inequality
$\left(\frac{k+2}{k+1}\right)^{k+1}=
\left(1+\frac{1}{k+1}\right)^{k+1}\,<\,e$,
we have
$$
A'_s(p)<\frac1p\,\left[-1+\sum_{k=1}^s b_s(k,p)\right]<0.
$$

Lemma is proved.
\medskip

{\bf Proof of Theorem 1}. Let $X$ be an arbitrary
$(s,t)$-design of length $N$. Introduce the random variable
(message)
$$
\e=(e_1,e_2,\dots,e_s),\qquad e_i\in[t],\qquad
1\le e_1< e_2<\cdots<e_s\le t,
$$
which takes equiprobable values in the set
$\Lambda(s,t)$. The Shannon entropy of $\e$ is
${\bf H}(\e)=\log{t\choose s}$. For $X$ and $\e$,
consider the random variable
$\z=\z(\e,X)=(z_1,z_2,\dots,z_N)$.
The definition of an
$(s,t)$-design $X$ implies that the Shannon entropy
of $\z$~is
$$
{\bf H}(\z)={\bf H}(\z(\e,X))={\bf H}(\e)=\log{t\choose s}.
$$
Therefore, the well-known subadditivity
property~\cite{g68,ck81} of the Shannon entropy gives
$$
\log{t\choose s}\le\sum_{i=1}^N {\bf H}(z_i).\leqno 1)
$$
Let $r_i,\;i=1,2,\dots,N$ denote the number of 1's in the $i$-th
row of the matrix $X$. It is easy to see that the random variable
$z_i$ has hypergeometric probability distribution
$$
\Pr\{z_i=k\}=\frac{{r_i\choose k}{t-r_i\choose s-k}}
{{t\choose s}},\quad \max\{0,s+r_i-t\}\le k\le\min\{s,r_i\}.
$$
Hence, inequality 1) implies that the design length
$$
N\ge\log{t\choose s}/\max_{r_i}{\bf H}(z_i).\leqno 2)
$$
Let $s\ge 2$ and $p,\;0<p<1$ be fixed, $t\to\infty$ and
$r_i\sim pt$.
One can easily check that
the entropy of hypergeometric distribution
${\bf H}(z_i)\sim H_s(p)$.
Lemma and inequality 2) yield inequality (3).

Theorem 1 is proved.

\subsubsection{Proof of inequality $R(2)\le 3/5$}
\begin{center}
\end{center}

Consider an arbitrary $(2,N)$-design
$X=(\x(1),\x(2),\dots,\x(t))$ of size $t$. Let $n,\;1\le
n\le N-1$ be fixed.  Denote by
$$
\z=(z_1,z_2,\dots,z_n)\in\{0,1\}^n
\quad (\y=(y_1,y_2,\dots,y_{N-n})\in\{0,1\}^{N-n})
$$
an arbitrary column of length $n\;(N-n)$ with components
from the alphabet $\{0,1\}$. Denote by
$\mu(\z,\y)$ the number of columns of $X$, which have the
form $(\z,\y)$. Obviously $\mu(\z,\y)\in\{0,1\}$. Let
$$
\mu(\z,\cdot)\eq\sum_{\y}\mu(\z,\y)
$$
be the number of columns, the first $n$ components
of which coincide with $\z$. We have
$$
 \sum_{\z}\sum_{\y}\mu(\z,\y)=
\sum_{\z}\mu(\z,\cdot)=t.\leqno 1)
$$

For an $(N-n)$-ary column $\w=(w_1,w_2,\dots,w_{N-n})\in\{-1,0,1\}^{N-n}$
and $(2,N)$-design $X$, define the number
$$
m(\w)\eq\sum_{\z}\sum_{\y}\mu(\z,\w+\y)\mu(\z,\y).\leqno 2)
$$
\medskip

{\bf Lemma}. {\it Let $X$ be $(2,N)$-design. Then the following two statements
are true. $1)$ If $\w=\0$, then $m(\0)=t$. $2)$ If $\w\ne\0$, then}
$m(\w)\in\{0,1\}$.
\medskip

{\bf Proof of Lemma.} The first statement follows from 1) and 2). The
second statement is proved by contradiction. Let there exists
$\w\ne\0$ for which $m(\w)\ge2$. By symbols $u,u',v,v'$ we denote
arbitrary elements of the set $[t]$.
The definition 2) implies that there exist two pairs $(u,v),\;
(u',v'),\;u\ne v,\;u'\ne v'$ for which the corresponding columns
of $X$ have the following form
$$
\x(u)=(\z,\w+\y),\;\x(v)=(\z,\y),\quad
\x(u')=(\z',\w+\y'),\;\x(v')=(\z',\y'),\leqno 3)
$$
where either $\z\ne\z'$ or $\y\ne\y'$.
It follows from 3) that
$$
\x(u)+\x(v')=\x(v)+\x(u').
$$
By virtue of the $(2,N)$-design definition, the last equality yields
$u=u',\;v=v'$, i.e., the both of inequalities $\z=\z',\;\y=\y'$ are true.

Lemma is proved.
\medskip

Define the number
$$
 l_n\eq\sum_{\x\in\{0,1\}^n}\mu^2(\x,\cdot)\ge t^2 2^{-n}.\leqno 4)
$$
Inequality 4) follows from 1) and the convex $\cup$ property of
the function $y=x^2$. 1)~and 2)~implies
$$
\sum_{\w}m(\w)=\sum_{\z}\mu^2(\z,\cdot)=l_n.
$$
Therefore, we can define an $(N-n)$-ary random variable $\xi
=(\xi_1,\xi_2,\dots,\xi_{N-n})\in\{-1,0,1\}^{N-n}$
with the distribution
$$
\Pr\{\xi=\w\}\eq\frac{m(\w)}{l_n}.\leqno 5)
$$
Let the set $W_n\eq\{\w:\;m(\w)=1\}$. For the Shannon entropy of
the random variable $\xi$, we have
$$
{\bf H}(\xi)\eq-\sum_{\w}\Pr\{\xi=\w\}\log\Pr\{\xi=\w\}\ge
$$
$$
\ge-\sum_{\w\in W_n}\Pr\{\xi=\w\}\log\Pr\{\xi=\w\}=
(1-t/l_n)\log l_n.
$$
Here we used 5) and Lemma. By virtue of 4),
$$
{\bf H}(\xi)\ge(1-2^n/t)(2\log t-n).\leqno 6)
$$

The subadditivity of the Shannon entropy gives
$$
{\bf H}(\xi)\le\sum_{i=1}^{N-n}{\bf H}(\xi_i).\leqno 7)
$$
For any $y=0,1$ and any $i=1,2,\dots,N-n$ define the set
$$
Y_i(y)\eq\{\y=(y_1,y_2,\dots,y_{N-n}):\;y_i=y\}
$$
and introduce
$$
\mu_i(\z,y)=\sum_{\y\in Y_i(y)}\mu(\z,y).
$$
We have $\mu_i(\z,0)+\mu_i(\z,1)=\mu(\z,\cdot)$ and, hence,
$$
\mu_i^2(\z,0)+\mu_i^2(\z,1)\ge\mu^2(\z,\cdot)/2.\leqno 8)
$$
By virtue of 2) and 5), the probability
$$
\Pr\{\xi_i=w\}=\sum_{\z}\sum_{y=0}^1\mu_i(\z,w+y)\mu_i(\z,y)/l_n,
\quad w\in\{-1,0,1\}.
$$
It follows, that $\Pr\{\xi_i=1\}=\Pr\{\xi_i=-1\}=(1-p)/2$,
and
$$
p\eq\Pr\{\xi_i=0\}=\frac{\sum_{\z}\sum_{y=0}^1\mu_i^2(\z,y)}{l_n}
\ge\frac12,\leqno 9)
$$
where we used inequality 8) and definition 4).

Inequality 9) implies that
$$
 {\bf H}(\xi_i)\le\max_{1/2\le p\le1}[-(1-p)\log\frac{1-p}{2}
-p\log p].\leqno 10)
$$
It is easy to check, that maximum in 10) is equal $3/2$ and
this maximum is achieved at $p=1/2$. On account
of 6) and 7), it follows
$$
(1-2^n/t)(2\log t -n)\le\frac32(N-n),\quad n=1,2,\dots,N-1.\leqno 11)
$$
Put $n\eq\lfloor\log(t/\log t)\rfloor$. Inequality 11) yields
$$
\log t\le\frac35\;\frac{N}{1-1/\log t}+\frac12\log\log t.
$$
Taking into account the definition of $R(2)$, it follows the
inequality $R(2)\le3/5$.

For $B_2$-codes, Theorem 2 is proved.

\subsection{Proof of Theorem 3}
\begin{center}
\end{center}

We say that the codeword $\x(u)$ is ``bad''
for code $X$, if it is not satisfied the definition of $B_s$-code.
This means, that there exist integers
$$
1\le i_1\le i_2\le\dots\le i_{s-1}\le t,\quad
1\le j_1\le j_2\le\dots\le i_s\le t,
$$
such that
$$
\x(u)+\x(i_1)+\x(i_2)+\cdots+\x(i_{s-1})=
\x(j_1)+\x(j_2)+\cdots+\x(j_s).
$$

Consider the random matrix (code)
$X=\|x_i(u)\|,\;i=1,2,\dots,N,\;u=1,2,\dots,t$, whose
components are independent identically distributed random
variables with distribution
$$
\Pr\{x_i(u)=0\}=Pr\{x_i(u)=1\}=1/2.\leqno 1)
$$
For the codeword $\x(u)$, denote by $P_s(N,t)$ the probability
to be ``bad''. Obviously, this probability does not depend
on $u=1,2,\dots,N$ and the average number of ``bad'' words does
not exceed $tP_s(N,t)$. It follows
\medskip

{\bf Lemma 1}. {\it If $P_s(N,t)<1/2$, then there exists $B_s$-code
of length $N$ and size} $t/2$.

The definition of the rate $R(s)$ and Lemma 1 yield

{\bf Lemma 2}. {\it For any} $s\ge2$
$$
R(s)\ge\sup\{R\;:\;
\varliminf_{N\to\infty}\frac{-\log P_s(N,\lfloor\exp\{RN\}\rfloor)}{N}>0\}
$$
\medskip

Let $m\le k\le s$ be integers. Denote by $E(m,k)$ the collection of
${k-1\choose m-1}$ ordered sequences
of integers $\e=(e_1,e_2,\dots,e_m)$ that satisfy the conditions
$$
1\le e_i\le k,\quad i=1,2,\dots,m,\quad \sum_{i=1}^m e_i=k.
$$
Let $\beta_1,\beta_2,\dots,\beta_s;\;\beta'_1,\beta'_2,\dots,\beta'_s$
be the collection of $2\cdot s$ independent random variables with the same
distribution 1). The following upper bound
on the probability $P_s(N,t)$ is true:
$$
P_s(N,t)\le K_s\sum_{1\le n\le m\le s}t^{n+m-1}Q_s^N(n,m),\leqno 2)
$$
$$
Q_s(n,m)\eq\max_{m\le k\le s}\max_{\e\in E(m,k)}
\max_{\e'\in E(n,k)}
\Pr\{\sum_{i=1}^m e_i\beta_i=\sum_{i=1}^n e'_i\beta'_i\},
$$
where the value $K_s$ depends only on $s$.
To prove 2), we used the theorem that the probability of a union of
events does not exceed the sum of their probabilities, and we also
took into account  that the number of all possible pairs
of the form $(A_{n-1},A_m)$ (or $A_n,A_{m-1}),\;1\le n\le m\le s$,
where $A_k,\;k=1,2,\dots s$ is an $k$-subset of the set $[t]$ and
$A_{n-1}\cap A_m=\emptyset$ (or $A_n\cap A_{m-1}=\emptyset$), is
not greater than $t^{n+m-1}$.

Inequality 2) implies that
$$
P_s(N,\lfloor\exp\{RN\}\rfloor)
\le K_s\sum_{1\le n\le m\le s}\exp\{-N[-\log Q_s(n,m)-(m+n-1)R]\}.\leqno 3)
$$

Lemma 2 and inequality 3) yield

{\bf Lemma 3}. {\it The $B_s$-code rate}
$$
R(s)\ge R_s\eq\min_{1\le n\le m\le s}-\frac{\log Q_s(n,m)}{(m+n-1)}.
\leqno 4)
$$

Our subsequent aim will be to prove that minimum in 4) is achieved
at $n=m=s$ and the minimal value $R_s=(2s-1)^{-1}{\tilde H_s}$.
It is sufficient to establish Theorem 3.
\medskip

We introduce the independent random variables
$\mu_i=\beta_i-\beta'_i,\;i=1,2,\dots,s$ with distribution
$$
\Pr\{\mu_i=0\}=1/2,\quad \Pr\{\mu_i=-1\}=\Pr\{\mu_i=1\}=1/4.\leqno 5)
$$

{\bf Lemma 4}. {\it For any $1\le n\le m\le s$
$$
Q_s(n,m)\le\sqrt{Q_s(n,n)}\cdot\sqrt{Q_s(m,m)},\leqno 6)
$$
where the sign of equality iff $n=m$ and}
$$
Q_s(m,m)\eq\max_{m\le k\le s}\max_{\e\in E(m,k)}
\Pr\{\sum_{i=1}^me_i\mu_i=0\}.\leqno 7)
$$

{\bf Proof of Lemma 4.}
Inequality 6) follows from the Cauchy inequality~\cite{hardy}:
$$
\sum p_i q_i\le\sqrt{\sum p_i^2}\cdot\sqrt{\sum q_i^2},
$$
if the Cauchy inequality is used to obtain an upper bound on the
probability of coincidence of two independent discrete random variables.
\medskip

{\bf Corollary}. {\it Number $R_s$, defined by $4)$, could be written in
the form}
$$
R_s=\min_{1\le n\le m\le s}-\log(Q_s(n,n)\,Q_s(m,m))^{1/2(m+n-1)}.\leqno 8)
$$
\medskip

{\bf Lemma 5}. {\it For any collection of integers
$e_1,e_2,\dots,e_m,\;e_i\ge1$, the probability
$$
\Pr\{\sum_{i=1}^me_i\mu_i=0\}\le\Pr\{\sum_{i=1}^m\mu_i=0\}
=\frac{(2m-1)!!}{(2m)!!},
$$
where the sign of equality iff $e_1=e_2=\cdots=e_m$}.

{\bf Proof of Lemma 5.} Consider the characteristic functions of
random variables $e_j\mu_j$ and~$\mu_j$:
$$
\chi_j(u)={\bf M}e^{iue_j\mu_j}=\frac12(1+\cos e_ju),\quad
\chi(u)={\bf M}e^{iu\mu_j}=\frac12(1+\cos u)=
\cos^2 (u/2),
$$
where we used 5). The inversion formula~\cite{feller} and
the inequality between the geometric mean and the
arithmetic mean~\cite{hardy} yield
$$
\Pr\{\sum_{i=1}^me_i\mu_i=0\}=
(2\pi)^{-1}\int_{-\pi}^{\pi}(\prod_{j=1}^m\chi_j^m(u))^{1/m}\,du\le
$$
$$
\le\frac{1}{m}\sum_{j=1}^m(2\pi)^{-1}\int_{-\pi}^{\pi}\chi_j^m(u)\,du
=(2\pi)^{-1}\int_{-\pi}^{\pi}\chi^m(u)\,du=
$$
$$
=(\pi)^{-1}\int_{-\pi/2}^{\pi/2}\cos^{2m}u\,du
=\Pr\{\sum_{i=1}^m\mu_i=0\}=\frac{(2m-1)!!}{(2m)!!}.
$$
The last equality follows from the well-known formula of
the calculus:
$$
(\pi)^{-1}\int_{-\pi/2}^{\pi/2}\cos^{2m}u\,du=\frac{(2m-1)!!}{(2m)!!}.
$$

Lemma 5 is proved.
\medskip

Lemma 5 and definition 7) imply that
$$
Q_s(m,m)=(2\pi)^{-1}\int_{-\pi}^{\pi}\chi^m(u)\,du=\frac{(2m-1)!!}{(2m)!!}.
\leqno 9)
$$
Formula 9) and the property of monotonicity of the generalized
mean~\cite{hardy} give
\smallskip

{\bf Lemma 6}. {\it Sequence $Q_s(m,m)^{1/m},\;m=1,2,\dots,s$,
increases monotonically with $m$}.
\smallskip

Put, for brevity, $a_m\eq Q_s(m,m)<1$. For $n\le m$,
Lemma 6 gives $a_n^{1/n}<a_m^{1/m}$, where the sign of
equality iff $n=m$.  In addition, for $n\le m$, the
following inequalities are true
$$
a_n\le a_m^{n/m}\le 1,\quad \frac{n+m}{n+m-1}\ge\frac{2m}{2m-1},
$$
where the sign of equality iff $n=m$. It follows that
for any $n\le m$,
$$
(a_n\;a_m)^{1/2(n+m-1)}\le(a_m^{1/2m})^{\frac{m+n}{m+n-1}}
\le(a_m^{1/2m})^{\frac{2m}{2m-1}}=a_m^{1/2m-1},\leqno 10)
$$
where the sign of equality iff $n=m$.
By virtue of 8) and 10), we have
$$
R_s=\min_{m\le s}-\frac{\log a_m}{2m-1}.
$$
For $m\le s$, by virtue of Lemma 6,
$$
a_m^{1/2m-1}=(a_m^{1/2m})^{\frac{2m}{2m-1}}
\le(a_m^{1/2m})^{\frac{2s}{2s-1}}\le a_s^{\frac{1}{2s-1}}.
$$
Hence
$$
R_s=-\frac{\log a_s}{2s-1}=\frac{{\tilde H}_s}{2s-1}.
$$
Theorem 3 is proved.

\newpage

\begin{center}
Arkadii G. D'yachkov,\qquad  "Lectures on Designing
Screening Experiments"
\end{center}
\medskip

\section{Universal Decoding for Random Design of\\
Screening Experiments}
\begin{center}
\end{center}

In this section, we consider the problem of  screening
experiment design (DSE) for the probabilistic model of
multiple access channel (MAC). We will discuss the random
design of screening experiments (random DSE) and present a
method of universal decoding (U-decoding) which does not
depend on transition probabilities of MAC.  The logarithmic
asymptotic behavior  of error probability for the symmetric model of
random DSE is obtained. Sect.~6 is based on the results of
paper~\cite{univ89} which completed the series of preceding
works~\cite{d79-1,d79-2,d81}.

\subsection{Statement of the problem, formulation and\\
discussion of results}
\begin{center}
\end{center}

We need to remind some notations from Section~1.
Let $2\le s<t$ be fixed integers and
$$
\e\eq(\la_1,\la_2,\dots,\la_s), \quad \la_i\in[t], \quad
1\le \la_1<\la_2<\dots<\la_s\le t,
$$
be an arbitrary $s$-subset of $[t]$. Introduce
${\Lambda}(s,t)$, the collection of all such subsets. Note
that the cardinality $|{\Lambda}(s,t)|={t\choose s}$.

Suppose that among $t$ {\it factors}, numbered by integers
from 1 to $t$, there are some $s<t$ unknown factors called
{\it significant} factors. Each $s$-collection of
significant factors is prescribed as an $s$-subset $\e\in
{\Lambda}(s,t)$.  The problem of {\em designing screening
experiments} (DSE) is to find all significant factors, i.e.
to identify an unknown subset~$\e$.  To look for $\e$, one
can carry out $N$ {\em experiments}. Each experiment is a
{\em group test} of a prescribed subset of $[t]$.  These
$N$ subsets are interpreted as $N$ rows of a binary
$N\times t$ matrix
$$
X=\|x_i(u)\|,\quad x_i(u)\in\{0;1\}\quad
i=1,2,\dots,N,\quad u=1,2,\dots,t,
$$
where the elements of the $i$-th row
$\x_{i}\eq(x_i(1),x_i(2),\dots,x_i(t))$
are defined as follows
$$
x_i(u)\eq\cases{1, & if the $u$-th factor is included into
the $i$-th test,\cr
0, & otherwise.\cr}
$$
The matrix $X$ is called a {\em code} or a {\em design of
experiments}. The detailed discussion of the DSE
problem is presented in the survey~\cite{m83}.
English translation of~\cite{m83} is included in
the book~\cite{aw87}.
\medskip

Let the symbol
$
\x(u)\eq(x_1(u),x_2(u),\dots,x_N(u))\in (0,1)^N,\;
u=1,2,\dots,t,
$
denote a column called a {\em codeword} of $X$.
The number of $1$'s
$w_u\eq\sum\limits_{i=1}^N\,x_i(u)$ is called
a {\em weight} of the codeword $\x(u)$.  We say that the
given $s$-subset
$\e=(e_1,e_2,\dots,e_s)$ called a {\it message} is {\em
encoded} into the non-ordered $s$-collection of codewords
$$
\x(\e)\eq(\x(\la_1),\x(\la_2),\dots,\x(\la_s)).
$$

Let $\z=(z_1,z_2,\dots,z_N)$ be a {\it test outcome}. To
describe the model of such a test outcome, we will use the
terminology of a {\it memoryless multiple-access channel}
(MAC)~\cite{ck81}, which has $s$
{\em inputs} and {\em one output}.
 Let all $s$ input alphabets of MAC are
identical and coincide with $\{0,1\}$.  Denote by $Z$ a
finite output alphabet. This MAC is prescribed by a
matrix of transition probabilities
$$
\|P(z\,|\,x_1,x_2,\dots,x_s)\|,\qquad z\in Z,
\quad x_k\in\{0,1\},\quad k=1,2,\dots,s.
$$
If the collection $\x(\e)$ was transmitted over MAC, then
$$
P_N(\z\,|\,\x(\e))\eq
\prod_{i=1}^N P(z_i\,|\,x_i(\la_1),x_i(\la_2),\dots,
x_i(\la_s))
$$
is the conditional probability to receive the word (outcome
of test) $\z\in Z^N$ at the output of MAC.

We will focus on the {\it symmetric model} of
DSE, which is identified as a
symmetric MAC. This means that for each fixed $z$, the
conditional probability $P(z|x_1^s)$, where
$x_1^s=(x_1,x_2,\dots,x_s)\in\{1,0\}^s$,
depends only on the cardinal number of $1$'s in the
sequence $x^s_1$, i.e. on the sum
$\sum\limits_{k=1}^s x_k$. Thus, the conditional
probability $P_N(\z|\x(\e))$ does not depend on the order
of MAC inputs.
\medskip

Let $i=1,2,\dots,N$ and
$x_i(\e)\eq(x_i(\la_1),x_i(\la_2),\dots,x_i(\la_s))\in
\{0,1\}^s$ be the $i$-th row of the $s$
collection~$\x(\la)$. Introduce the concept of {\it
composition} $C(\x(\e),\z)$ of a pair $(\x(\e),\z)$, i.e.
the collection of integers $\|n(x_1^s,z)\|$,
$x_1^s\in\{0,1\}^s$, $z\in Z$, where the element
$n(x_1^s,z)$ is the cardinal number of positions
$i=1,2,\dots,N$, in which $x_i(\la)=x_1^s$, $z_i=z$.  Using
this concept, the definition of MAC can be written in the
form
$$
P_N(\z|\x(\e))=
\prod_{x_1^s} \prod_{z} P(z|x_1^s)^{n(x_1^s,z)},\quad
\mbox{where}\quad \|n(x^s_1,z)\|\eq C(\x(\e),\z).
\eqno(1)
$$
Notice that
$$
\sum_{x_1^s} \sum_{z} n(x_1^s,z) = N.
$$

Let $1\le v\le k\le s$ and the symbol
$x^k_v\eq(x_v,x_{v+1},\dots,x_k)$ be a part of the sequence
$x^s_1$. For the codeword $\x(\la_k)$, we introduce
a {\em marginal} composition
$$
C(\x(\la_k))\eq\|n(x_k)\|,\quad x_k \in\{0,1\},
\quad k=1,2,\dots,s,
$$
where
$$
n(x_k)\eq\sum_{x^{k-1}_1} \sum_{x_{k+1}^s}
\sum_{z} n(x^s_1,z)\eqno(2)
$$
is the cardinal number of positions $i=1,2,\dots,N$,
in which $x_i(\la_k)=x_k$. We will also
use the similar symbols
to denote another marginal composition. For instance, the
marginal composition $\|n(x^s_{k+1},z)\|$,
$k=1,2,\dots,s-1$, is the composition with elements
$$
n(x^s_{k+1},z)\eq\sum_{x^k_1} n(x^s_1,z).
$$
\medskip

Let $\Q\eq(Q(0),Q(1))$, $0<Q(0)<1$, $Q(1)=1-Q(0)$,
be a fixed probability distribution
on $\{0,1\}$.  For any pair $(\x(\e),\z)$ we introduce an
{\it universal decoding} (U-decoding) $D_Q$, which does not
depend on transition probability of MAC:
$$
D_Q(\x(\e),\z)=D_Q(\|n(x^s_1,z)\|)\eq
\frac{\prod\limits_{x^s_1} \prod\limits_{z} n(x^s_1,z)!}
{\prod\limits_{k=1}^s \prod\limits_{x_k} Q(x_k)^{n(x_k)}},
\quad\mbox{if}\quad C(\x(\e),\z)=\|n(x^s_1,z)\|.
$$
An {\it average error
probability} of code $X$ and decoding $D_Q$
is defined as follows:
$$
P_Q(X)\eq{t\choose s}^{-1}\sum_{\e} \sum_{\z}
P_N(\z|\x(\e))\;[1-\phi_Q(\e,\z)],\eqno(3)
$$
where $\phi_Q(\e,\z)$ is the characteristic function of
the set
$$
{\cal A}^{\e}_Q\eq\{\z:\mbox{ for any }\e'\ne\e
\mbox{ the value }D_Q(\x(\e),\z)>D_Q(\x(\e'),\z)\}.
$$

Let the distribution $\Q$ be fixed. We will consider two
ensembles of codes with independent and identically
distributed codewords $\x(u)$, $u=1,2,\dots,t$.
\begin{enumerate}
\item
For a {\it completely randomized ensemble} (CRE), each
codeword $\x(u)$ is taken   with probability
$$
Q_N(\x(u))\eq\prod_{i=1}^N Q(x_i(u)).\quad
\eqno(4)
$$
\item
For a {\it constant-weight ensemble} (CWE), the
probability is
$$
Q_N(\x(u))\eq\left[
  \begin{array}{ll}
{N\choose\lfloor NQ(1)\rfloor}^{-1},
&\quad
\mbox{if the weight}\;
w_u=\sum\limits_{i=1}^N x_i(u)=
\lfloor NQ(1)\rfloor,\\
0,&\quad\mbox{otherwise.}\\
\end{array} \right.
\eqno(5)
$$
\end{enumerate}
Denote by $\overline{P_Q(X)}$ the  average error
probability over ensembles (4) and (5). This probability
is called a {\it random coding bound} for DSE~\cite{d81} .

Fix an arbitrary number $R>0$ called a {\it rate} and
consider the following asymptotic conditions
$$
N\to\infty,\quad t\to\infty,\quad \frac{\ln t}{N}\sim R,
\quad s=\mbox{const}.\eqno(6)
$$
Our aim is to investigate the
{\it logarithmic asymptotic behavior} of the random coding bound
for CRE and CWE under conditions (6).
\medskip

To formulate the results, we need the following notations.
Let
$$
\tau\eq\left\{\tau(x^s_1,z),\quad x^s_1\in\{0,1\}^s,
\;z\in Z,\;
P(z|x_1^s)=0\;\Rightarrow\;\tau(z|x_1^s)=0\right\}
$$
be an arbitrary probability distribution on the product
$\{0,1\}^s\cdot Z$, such that the conditional probability
$\tau(z|x_1^s)=0\,$ if $\,P(z|x_1^s)=0$. Consider
the functions
$$
{\cal H}(\Q,\tau)\eq
\sum\limits_{x_1^s\cdot z}\tau(x_1^s,z)
\ln\frac{\tau(x^s_1,z)}{
P(z|x_1^s)\cdot \prod\limits_{k=1}^s Q(x_k)},
\quad
I^{(k)}(\tau)\eq\sum\limits_{x^s_1\cdot z}\tau(x^s_1,z)
\ln\frac{\tau(x^k_1|x_{k+1}^s,z)}
{\prod\limits_{v=1}^k Q(x_v)},
\eqno(7)
$$
where $k=1,2,\dots,s$ and the standard symbols for
conditional probabilities are used. Let
$[a]^+ \eq\max\{0;a\}$.  The following main result
will be proved in Section~6.2.
\medskip

{\bf Theorem 1}. {\em If conditions $(6)$ are fulfilled,
then
$$
\overline{P_Q(X)}=\exp\{-N[E(R,\Q)+o(1)]\},
$$
where
$$
E(R,\Q)\eq\min\limits_{k=\overline{1,s}} E_k(R,\Q),
\eqno(8)
$$
$$
E_k(R,\Q)\eq\min\{{\cal H}(\Q,\tau)+[I^{(k)}(\tau)-kR]^+\}.
\eqno(9)
$$
For {\em CRE\/} $(4)$ the minimum in $(9)$ is taken over all
$\tau$, and for {\em CWE\/} $(5)$ the minimum in $(9)$ is
taken over distributions $\tau$, for which the marginal
probabilities on $x_k$ are fixed and coincide with $\Q$,
i.e., $\tau (x_k)=Q(x_k)$, $k=1,2,\dots,s$}.
\medskip

{\bf Remark 1.} One can easily understand that
Theorem 1 remains also true for the optimal maximum
likelihood decoding (ML-decoding), when the function (1) is
used as the decoding rule. Hence, this theorem completes our
preceding investigation~\cite{d79-1,d79-2,d81} of the error
probability  asymptotic behavior  for the symmetric model
of random~DSE.
\medskip

{\bf Remark 2.} Similar to the discrete memoryless
channel~\cite{dp82} one can show that Theorem~1 for CRE
remains true also, if (instead of CRE) we consider the
ensemble, when all $N$ rows of length $t$ are chosen
independently from the set of all binary
$t$--sequences containing $\lceil Q(0)t \rceil$ zeros and
$\lfloor Q(1)t \rfloor$ ones.
\medskip

Note that the functions (7) are non-negative and convex as
functions of $\tau$. Hence, using the standard
arguments~\cite{ck81}, we can easily obtain the following
properties of~(9) which are given in the form
of Propositions~1-3.
\bigskip

{\bf Proposition 1.} {\em Introduce the distribution
$$
\tau_Q\eq\left\{P(z|x_1^s)\cdot\prod_{i=1}^s Q(x_i),
\quad x^s_1\in\{0,1\}^s,\quad z\in Z\right\}.
$$
Then for $R\ge I^{(k)}(\tau_Q)/k$, the function
$E_k(R,\Q)=0$, and for $0<R<I^{(k)}(\tau_Q)/k$, the function
$E_k(R,\Q)$ is positive, convex and monotonically decreasing
with increasing~$R$}.
\medskip

{\bf Proposition 2.} {\em For each $k=1,2,\dots,s$ there
exists
the unique distribution $\tau_{cr}^{(k)}$, for which
$$
\min\left\{{\cal H}(\Q,\tau)+I^{(k)}(\tau)\right\}= {\cal
H}(\Q,\tau_{cr}^{(k)})+I^{(k)}(\tau_{cr}^{(k)}).\eqno(10)
$$
Conditions of minimization in $(10)$ for
{\em CRE\/} and {\em CWE\/} are
pointed out in the formulation of Theorem~$1$. In addition,
$$
0<I^{(k)}(\tau_{cr}^{(k)})\le I^{(k)}(\tau_Q).
$$}
\smallskip

It is evident that the extreme distribution
$\tau^{(k)}_{cr}$ for CRE may not coincide with the
extreme distribution $\tau^{(k)}_{cr}$ for CWE.
\medskip

{\bf Proposition 3.} {\em If $0<R\le
I^{(k)}(\tau_{cr}^{(k)})/k$, then
$$
E_k(R,\Q)={\cal H}(\Q,\tau_{cr}^{(k)})+I^{(k)}
(\tau_{cr}^{(k)})-kR.\eqno(11)
$$
If $I^{(k)}(\tau_{cr}^{(k)})/k \le R<I^{(k)}(\tau_Q)/k$,
then the minimum in $(9)$ is achieved at the unique
distribution $\tau_k$, for which
$$
R=\frac{I^{(k)}(\tau_{k})}{k}; \quad
E_k(R,\Q)={\cal H}(\Q,\tau_k).\eqno(12)
$$}

Some useful relations between the functions
$I^{(k)}(\tau), k=1,2,\dots,s$ are given by
\medskip

{\bf Proposition 4.} {\em Fix arbitrary $k=2,3,\dots,s $.
Let $\tau_k$ be the extreme distribution
satisfying~$(12)$. Then}
$$
\frac{I^{(k)}(\tau_k)}{k}\le\frac{I^{(k-1)}(\tau_k)}{k-1}
\le\dots\le I^{(1)}(\tau_k).\eqno(13)
$$
\smallskip

{\bf Proof.} By virtue of the MAC symmetry, the extreme
distribution
$$
\tau_k=\left\{\hat{\tau}(x^k_1,x^s_{k+1},z),\quad
x^k_1\in\{0,1\}^k,\; x^s_{k+1}\in \{0,1\}^{s-k},
\;z\in Z\right\}
$$
has the symmetry property as relative to components of
$x_1^k$ so to components of $x^s_{k+1}$. For instance, for any
$v=1,2,\dots,k$, probabilities $\hat
{\tau}(x_v,x^s_{v+1},z)$ coincide with the corresponding
probabilities $\hat {\tau}(x_1,x^s_{v+1},z)$ and
the marginal probabilities $\hat {\tau}(x_v)$ do not depend
on $v$. So
$$
\frac{I^{(v)}(\tau_k)}{v}=-\sum_{x_1}\hat
{\tau}(x_1)\ln
Q(x_1)-\frac{H_{\tau_k}(x^v_1|x^s_{v+1},z)}{v},\eqno(14) $$
where we use the standard notation of
the Shannon entropy of
distribution $\tau_k$. We have~\cite{ck81}
$$
H_{\tau_k}(x^v_1|x^s_{v+1},z)=\sum_{i=2}^{v+1}
H_{\tau_k}(x_1|x^s_i,z).
$$
Since
$$
H_{\tau_k}(x_1|x^s_i,z)\ge H_{\tau_k}(x_1|x^s_{i-1},z),
\quad i=3,4,\dots,v+1,
$$
then by virtue of (14) and the monotone property of the
arithmetic mean, we obtain the inequality (13).

Proposition 4 is proved.
\medskip

The behaviour of the random coding exponent (8) is
described by
\medskip

{\bf Proposition 5.} The following properties are true.
\begin{itemize}
\item
{\em If $R\ge I^{(s)}(\tau_Q)/s$, then
$E(R,{\Q})=0$, and if $0<R<I^s(\tau_Q)/s$,
then $E(R,{\Q})$ is positive and monotonically
decreasing with increasing $R$}.
\item
{\em There exists the interval
$0<R\le R_0,$ where $E(R,\Q)=E_1(R,\Q)$}.
\item
{\em There exists the interval
$R_1\le R\le I^{(s)}(\tau_Q)/s$, in which}
$E(R,\Q)=E_s(R,\Q)$.
\end{itemize}
\smallskip

{\bf Proof.} Using the definition (7), it is easy to verify
that for any fixed distribution $\tau$, the sequence
$I^{(k)}(\tau)$ is monotonically increasing with increasing
$k=1,2,\dots,s$.  Therefore, the sequence
$$
{\cal H}(\Q,\tau_{cr}^{(k)})+I^{(k)} (\tau_{cr}^{(k)}),
\quad k=1,2,\dots,s,
$$
defined by (10) is also
increasing.  The monotonic decreasing of the sequence
$I^{(k)}(\tau_Q)/k$,  $k=1,2,\dots,s$,
follows from~(13).\footnote[2]{This particular case
of (13) was established in~\cite{mm80}.}\quad
Hence, taking into account the
definition~(8) and Propositions~1 and~3, we
obtain all three assertions of Proposition~5.

Proposition 5 is proved.

Now we will consider the following question. How to solve
the extreme problem (9), i.e., how to evaluate the
exponent $E_k(R,\Q)$, $k=1,2,\dots,s\;$? Introduce the
function of parameter $\rho \ge 0$
$$
\begin{array}{ccl}
B_k(\rho,\Q)&\eq&\min \{ {\cal H}(\Q,\tau)+\rho
I^{(k)}(\tau)\}\\ &=&{\cal H}(\Q,\tau_k^{(\rho)})+\rho
I^{(k)}(\tau_k^{(\rho)}) \quad k=1,2,\dots,s,\\
\end{array}\eqno(15)
$$
where for CRE and CWE, the minimum in (15) is taken over the
distribution $\tau$, pointed out in the formulation of
Theorem~1, and $\tau_k^{(\rho)} $ is the extreme
distribution in~(15).
\medskip

{\bf Theorem 2.} {\em Let $k=1,2,\dots,s$. The function
$B_k(\rho,\Q)$, $\rho\ge0$, and the exponent
$E_k(R,\Q)$, $R>0$, satisfy the following assertions}.
\begin{enumerate}
\item
{\em  $B_k(0,\Q)=0$ and at $\rho >0$ the function
$B_k(\rho,\Q)$ is positive monotonically increasing and
concave}.
\item
{\em The derivative}\quad
$
\partial B_k(\rho,\Q)/\partial \rho=I^{(k)}
(\tau_k^{(\rho)}), \quad\rho\ge 0.
$
\item
$
E_k(R,\Q)=\max_{0\le\rho\le 1}
\left\{B_k(\rho,\Q)-k\rho R\right\}.
$
\item
$
I^{(k)}(\tau_Q)=\partial B_k(\rho,\Q)/\partial
\rho|_{\rho=0}
$,\quad
$
I^{(k)}(\tau_{cr}^{(k)})=\partial B_k
(\rho,\Q)/\partial \rho|_{\rho=1}
$.
\item
{\em Formulas $(11)$ and $(12)$ from
Proposition $3$ may be written as follows}:
\begin{enumerate}
\item
{\em if $0<R<I^{(k)}(\tau^{(k)}_{cr})/k$, then}
$$
E_k(R,\Q)=B_k(1,\Q)-kR;\eqno(11');
$$
\item
{\em if
$I^{k}(\tau^{(k)}_{cr})/k\le R\le I^{k}(\tau^{(k)}_{Q})/k$,
then the function $E_k(R,\Q)$ has the following parametric
form}
$$
\left[
\begin{array}{l}
R=k^{-1}\partial B_k(\rho,\Q)/\partial \rho,
\quad 0\le\rho\le 1,\\
E_k(R,\Q)=B_k(\rho,\Q)-\rho\partial B_k(\rho,\Q)/\partial
\rho.\\
\end{array}
\right.\eqno(12')
$$
\end{enumerate}
\item
{\em For {\em CWE,} the function
$$
B_k(\rho,\Q)=\max_{\Q_1,\Q_2} \{
G_k(\rho,\Q_1,\Q_2)-k(1+\rho)K(\Q,\Q_1)-
(s-k)K(\Q,\Q_2)\}\eqno(16)
$$
where $\Q_j=(Q_j(0),Q_j(1)),j=1,2$, are probability
distributions at $\{0,1\}$, and}
$$
\begin{array}{rcl}
G_k(\rho,\Q_1,\Q_2)&\eq&-\ln\left\{
\sum\limits_{x^s_{k+1}}
\prod\limits_{i=k+1}^s Q_2(x_i)\right.\\
&&\left.\cdot\sum_{z}\left[
\sum\limits_{x^k_1} \prod\limits_{i=1}^k
Q_1(x_i)P(z|x^s_1)^{1/(1+\rho)}
\right]^{1+\rho}\right\},\\
K(\Q,Q_j)&\eq&\
Q(0)\ln\frac{Q(0)}{Q_j(0)}+Q(1)\ln\frac{Q(1)}{Q_j(1)}
.\\
\end{array}
\eqno(17)
$$
\item
{\em For {\em CRE,} the function}
$
B_k(\rho,\Q)=G_k(\rho,\Q,\Q).
$
\end{enumerate}

{\bf Proof.} 1. The positivity and monotone increase of the
function $B_k(\rho,\Q)$ for $\rho >0$ are evident from
definition (15). If $\rho=0$, then the minimum in (15) is
achieved at $\tau=\tau_Q$ and this minimum equals zero, so
$B_k(0,\Q)=0$. To prove the concavity of $B_k(\rho,\Q)$ we
consider $\rho=\lambda\rho_1+(1-\lambda)\rho_2$, where
$\rho \ge 0$, and $0<\lambda<1$. We have
$$
{\cal H}(\Q,\tau)+\rho I^{(k)}(\tau)=
\lambda[{\cal H}(\Q,\tau)+\rho_1 I^{(k)}(\tau)]
+(1-\lambda)[{\cal H}(\Q,\tau)+\rho_2 I^{(k)}(\tau)].\eqno(18)
$$
Since the minimum of the sum of two functions is more than
or equal to the sum of their minimums, then from (18) and
definition (15) we obtain
$$
B_k(\rho,\Q)\ge\lambda B_k(\rho_1,\Q)+(1-\lambda)B_k(\rho_2,\Q).
$$

Statement~1 is proved.

To establish the other statements of Theorem~2, we can apply
the standard Lagrange method. In our case, functions (7) are
convex, and the Khun-Tucker theorem is applied. We omit the
detailed proofs, because the similar proofs were described
in~\cite{p82} for the more simple situation of
discrete memoryless channels.
\bigskip

Let
$$
B_s\eq\max_{\Q} B_1(1,\Q)=\max_{\Q} E(0,\Q),\eqno(19)
$$
where $E(0,\Q)\eq\lim\limits_{R\to 0} E(R,\Q)$ and the
second equality in (19) follows from Proposition~5.  Now
we consider the evaluation problem of $B_s$ for the
important particular case of~DSE.
\medskip

{\bf Example}.
({\em The disjunct channel model of DSE}, see Sect.~1).
This model
is the most interesting for applications~\cite{m83,d79-2}.
It is specified by the deterministic MAC, the output of
which $z$ is the Boolean sum of MAC inputs
$x_1,x_2,\dots,x_s$, i.e.
$$
z=\left[
\begin{array}{ll}
0,& \mbox{ if $x_1=x_2=\cdots=x_s=0$,}\\
1,& \mbox{ otherwise.}\\
\end{array}
\right.
$$
We conclude from (16) and (17) that for CWE
$$
\begin{array}{rl}
B_s=&\max\limits_{(\beta_1,\beta_2,Q)}
\left\{-\ln\left[1-2\beta_1(1-\beta_1)\beta_2^{s-1}
\right]
+2\left[Q\ln\frac{\beta_1}{Q}+
(1-Q)\ln\frac{1-\beta_1}{1-Q}\right]+\right.\\
&\left.+(s-1)\left[Q\ln\frac{\beta_2}{Q}+
(1-Q)\ln\frac{1-\beta_2}{1-Q}\right]\right\},\\
\end{array}
\eqno(20)
$$
where the maximum is taken over $(\beta_1,\beta_2,Q)$,
provided that
$$
0<\beta_j=Q_j(0)<1,\quad j=1,2,\quad
\mbox{and}\quad 0<Q=Q(0)<1.
$$
The following table gives the extreme parameters
$(\beta_1,\beta_2,Q)$ at which the maximum in (20) is
achieved:
\begin{center}
\begin{tabular}{|ccccccccc|}
\hline
$s$ & & $\beta_1$ & & $\beta_2$ & & $Q$ & & $B_s$\\
\hline
2 & & 0.62 & & 0.81 & & 0.69 & & 0.418\\
3 & & 0.69 & & 0.85 & & 0.78 & & 0.295\\
6 & & 0.82 & & 0.91 & & 0.89 & & 0.155\\
\hline
\end{tabular}
\end{center}
If $s\to\infty$, then
$$
\beta_1=1-\frac{2\ln 2}{s}+O(s^{-2}),\qquad
\beta_2=1-\frac{\ln 2}{s}+\frac{2(\ln2)^2}{s^2}+O(s^{-3}),
\eqno(21a)
$$
$$
Q=1-\frac{\ln 2}{s}+O(s^{-2}),\qquad
B_s=\frac{2(\ln 2)^2}{s}(1+o(1))=\frac{0.9609}{s}(1+o(1)).
\eqno(21b)
$$
With the help of Theorem~2 (statement~7) one
can check that for CRE the maximum in (19) is
achieved at $Q=(s/(1+s);1/(1+s))$, and
$$
B_s=-\ln\left[1-\frac{2s^s}{(1+s)^{1+s}}\right]
=\frac{2}{e\cdot s}(1+o(1))=
\frac{0.7358}{s}(1+o(1)).
$$
Let $N(s,t)$ be the {\em minimal possible length}
of the code $X$ with
{\em zero error probability} for the Boolean model of
DSE, and
$$
R^0_s\eq\limsup_{t\to\infty} \frac{\ln t}{N(s,t)}
$$
be the {\em zero error capacity}~\cite{m83}.
It is easy to understand that applying $(11')$,
we obtain the lower bound
$$
R^0_s\ge \max_Q \min_{k=1,2,\dots,s}
\frac{B_k(1,Q)}{s+k}.\eqno(22)
$$
Using Theorem 2, it is not
difficult to prove that for the Boolean model of DSE the
minimax in (22) is asymptotically achieved, when $k=1$ and
$Q$ satisfies (21). So the above mentioned asymptotic
formula for $B_s$ means that
$$
R^0_s\ge
\frac{0.9609}{s^2}(1+o(1))\quad\mbox{for CWE};\qquad
R^0_s\ge
\frac{0.7358}{s^2}(1+o(1))\quad\mbox{for CRE}.
\eqno(23)
$$
The bound (23) for CRE was obtained in~\cite{m75},
and the bound (23) for CWE
was obtained in~\cite{univ89}.
Note that the best known
upper bound~\cite{dr83} has the form
$$
R^0_s \le\frac{2\ln s}{s^2}(1+o(1)),\quad s\to\infty.
$$

\subsection{Proof of Theorem 1}
\begin{center}
\end{center}

Fix an arbitrary message $\e\in\Lambda(s,t)$
and an integer $k=1,2,\dots,s$. Let
$\e^{(s-k)}\subset\e$, $|\la^{(s-k)}|=s-k$, be an
arbitrary fixed $(s-k)$--subset of $\e$.
Consider the collection of messages
$$
\Lambda_k\eq\Lambda_k(\e^{(s-k)})
\eq\{\e'\in\Lambda(s,t)\;:\;\e'\cap\e=
\e^{(s-k)}\}.
$$
It is clear that $|\Lambda_k|={{t-s}\choose k}$.
For each $\e'\in \Lambda_k$ and $\z\in Z^N$, we
introduce the ensemble events
$$
X(\z,\e')\eq\{X\;:\;D_Q(\x(\e),\z)\le D_Q(\x(\e'),\z)\},
\quad
X^{(k)}(\z)\eq\bigcup_{\Lambda_k}\,X(\z,\e').
\eqno(24)
$$

Let
$\x_1^s\eq(\x_1,\x_2,\dots,\x_k,\x_{k+1},\dots,\x_s)\eq
(\x^k_1,\x^s_{k+1})$
be a fixed collection of $s$ binary columns
$\x_i\in\{0,1\}^N$. We define the probabilities
$$
\begin{array}{rcl}
q_k&\eq&\sum\limits_{\z}\,\sum\limits_{\x^s_1}
Q_N(\x^s_1)\,P_N(\z|\x^s_1)\,q_k(\x^s_1,\z),\\
q_k(\x^s_1,\z)&\eq&\Pr\,\{X^{(k)}(\z)|\x(\e)=\x^s_1\},\\
Q_N(\x^s_1)&\eq&\prod\limits_{i=1}^s Q_N(\x_i).
\end{array}
\eqno(25)
$$
Here and below, we will use the symbol $\Pr\{.\}$ to denote
the probability of an event in the ensemble.  Probabilities
$Q_N(\x_i)$ for CRE and CWE are calculated accordingly to
(4) and (5), respectively. Note that $q_k,k=1,2,\dots,s$,
is the average ensemble probability of the following event:
components $\e^{(s-k)}$ of the transmitted message were
decoded {\it correctly}, and the components
$\,\e\setminus\e^{(s-k)}\,$
were decoded {\it incorrectly}.  Hence,
 by virtue of the symmetry of MAC and definition
(3), we obtain
$$
\max_{k=1,2,\dots,s} q_k\le\overline{P_Q(X)}\le
\sum_{k=1}^s {s \choose k} q_k.\eqno(26)
$$
Under conditions (6), the following statements take place.
\medskip

{\bf Lemma 1.} {\em For each $k=1,2,\dots,s$}
$$
q_k\le\exp \{-N[E_k(R,\Q)+o(1)]\}.\eqno(27)
$$

{\bf Lemma 2.} {\em For fixed $k$, $k=1,2,\dots,s$, and
$0<R<I^k(\tau^{(k)}_{cr})/k$, the probability
$$
q_k\ge\exp\{-N[{\cal H}(\Q,\tau^{(k)}_{cr})+
I^k(\tau^{(k)}_{cr})
-kR+o(1)]\}.
\eqno(28)
$$}

{\bf Lemma 3.} {\em For each $k,\;k=1,2,\dots,s$
$$
\overline{P_Q(X)}\ge\exp \{-N[\hat {E}_k(r,\Q)+o(1)]\}.
\eqno(29)
$$
$$
\hat {E}_k(R,\Q)\eq\min {\cal H}(\Q,\tau),\eqno(30)
$$
where for {\em CRE\/} and {\em CWE\/} the minimum in
$(30)$ is taken over distributions $\tau$, which were
mentioned in the formulation of the Theorem~$1$, and such
that~$I^{(k)}(\tau)\le kR$}.
\medskip

With the help of arguments used in~\cite{ck81} for
investigating the sphere-packing exponent of a discrete
memoryless channel it is easy to understand that
$$
\hat E_k(R,\Q)=E_k(R,\Q)\quad \mbox{when }\quad
\frac{I^k(\tau^{(k)}_{cr})}{k}\le R\le\frac
{I^k(\tau_Q)}{k}.
$$
Therefore, the statement of Theorem~1 arises from
inequality (26), Lemmas~1-3, and Proposition~3
(see~\cite{d84}). To complete the proof of Theorem 1 we
need to establish the proofs of Lemmas~1-3.
\medskip

{\bf Proof of Lemma 1.} Denote by ${\cal N}(\Q)$ the set of
 all compositions $\|n(x^s_1,z)\|$ for which marginal
compositions $\|n(x_k)\|, k=1,2,\dots,s$ are the same and
$$
n(x_k)\eq\left\{
\begin{array}{ll}
\lceil NQ(0)\rceil,& \mbox{ if $x_k=0$,}\\
\lfloor NQ(1)\rfloor,& \mbox{ if $x_k=1$.}
\end{array}
\right.
$$
Let $1\le v\le k\le s$ be integers. In the case of CRE we
define
$$
\delta^k_v (\|n(x^s_1,z)\|)\eq
\prod_{i=v}^k \prod_{x_i} Q(x_i)^{n(x_i)}\eqno(31a)
$$
and in the case of CWE we define
$$
\delta^k_v (\|n(x^s_1,z)\|)\eq\left\{
\begin{array}{ll}
{N\choose\lfloor NQ(1)\rfloor}^{-(k-v+1)},&
\mbox{if}\quad \|n(x^s_1,z)\|\in{\cal N}(\Q),\\
0,&\mbox{otherwise.}
\end{array}
\right.\eqno(31b)
$$
Replace in (25) the sum over the pairs $(\z,\x^s_1)$ by
the sum over the compositions $\|n(x^s_1,z)\|$. Taking into
account (1), we have
$$
q_k=\sum_{\|n(x^s_1,z)\|} b(\|n(x^s_1,z)\|)\cdot
q^{(k)}(\|n(x^s_1,z)\|),\eqno(32)
$$
$$
b(\|n(x^s_1,z)\|)\eq N!\cdot\delta^s_1(\|n(x^s_1,z)\|)
\prod_{x^s_1} \prod_{z}
\frac{P(z|x^s_1)^{n(x^s_1,z)}}{n(x^s_1,z)!},\eqno(33)
$$
$$
q^{(k)}(\|n(x^s_1,z)\|)\eq\Pr\left\{X^{(k)}(\z)|C(\x(\e),\z)
=\|n(x^s_1,z)\|\right\}. \eqno(34)
$$

Fix any composition $\|n(x^s_1,z)\|$ and introduce for this
composition the set
$M_k=M_k(\|n(x^s_1,z)\|)$, $k=1,2,\dots,s$,
of compositions
$$
M_k\eq\left\{\|m(x^s_1,z)\|\,:\,\sum_{x^k_1}
m(x^s_1,z)=n(x^s_{k+1},z),\; D_Q(\|m(x^s_1,z)\|)\ge
D_Q(\|n(x^s_1,z)\|)\right\},
$$
where $D_Q$ is the considered
U-decoding. Then for
$\e'\in \Lambda_k=\Lambda_k(\e^{(s-k)})$ we have
$$
\Pr\left\{X(\z,\e')\,|\,C(\x(\e),\z)=\|n(x^s_1,z)\|\right\}
=\sum_{M_k} d_k(\|m(x^s_1,z)\|),
$$
where
$$
d_k(\|m(x^s_1,z)\|)\eq\prod_{x^s_{k+1}} \prod_{z}
\frac{n(x^s_{k+1},z)!}{\prod\limits_{x_1^k}m(x^s_1,z)!}\cdot
\delta^s_1(\|m(x^s_1,z)\|).\eqno(35)
$$
For the composition $\|m(x^s_1,z)\|\in M_k$ the inequality
$$
d_k(\|m(x^s_1,z)\|)\le d_k(\|n(x^s_1,z)\|)
$$
arises from the definition of U-decoding $D_Q$.
Therefore, applying the additive upper bound on the
probability of a union of events (34) we find that
$$
q^{(k)}(\|n(x^s_1,z)\|)\le\min
\left\{1\,;\,(N+1)^A\cdot
{{t-s}\choose k}\cdot d_k(\|n(x^s_1,z)\|)\right\},
$$
where $A\eq2^s\cdot|Z|$, and we use the fact that the
number of all compositions $\|m(x^s_1,z)\|$ does not exceed
$(N+1)^A$.  Hence, in virtue of (32),
$$
q_k\le(N+1)^A\cdot\max\left\{b(\|n(x^s_1,z)\|
\cdot\min\left[\;1;\;(N+1)^A{{t-s}\choose k}
d_k(\|n(x^s_1,z)\|)\right]\right\}.
\eqno(36)
$$
Note that for CRE the maximum in (36) is taken over all
compositions $\|n(x^s_1,z)\|$ and for CWE, by virtue of
(31) and (32), this maximum is taken only over those
compositions $\|n(x^s_1,z)\|$ which belong to ${\cal
N}(\Q)$.

Fix an arbitrary distribution
$\tau=\left\{\tau(x^s_1,z),\quad x^s_1\in\{0,1\}^s,\;z\in
Z\right\}$, which belongs to the set of distributions
pointed out in the formulation of Theorem 1. Suppose that
the elements of the composition $\|n(x^s_1,z)\|$ satisfy
the asymptotic equalities
$$
\|n(x^s_1,z)\|=N[\tau
(x^s_1,z)+o(1)], \qquad N\to\infty.\eqno(37)
$$
By virtue of
(33), (35), and (37), the Stirling formula yields
$$
b(\|n(x^s_1,z)\|)=\exp\left\{-N[{\cal
H}(\Q,\tau)+o(1)]\right\},\eqno(38)
$$
$$
d_k(\|n(x^s_1,z)\|)=
\exp\left\{-N[I^{(k)}(\tau)+o(1)]\right\},
\quad k=1,2,\dots,s,
\eqno(39)
$$
where the exponents in the right-hand sides are defined
by~(7). From (36)-(39) it follows the asymptotic bound (27).

Lemma 1 is proved.
\medskip

{\bf Proof of Lemma 2.} Fix an arbitrary composition
$\|n(x^s_1,z)\|$ and an integer $k$, $k=1,2,\dots,s$. For
each $\e'\in\Lambda_k=\Lambda_k(\la^{(s-k)}$
and $\z\in Z^N$ we introduce the
ensemble event
$$
\hat {X}(\z,\e')\eq X(\z,\e')\cap Y(\z,\e'),\quad
Y(\z,\e')\eq
\left\{X\,:\,C(\x(\e'),\z)=\|n(x^s_1,z)\|\right\},
$$
where $X(\z,\la')$ was defined by~(24). Note, that for
any $\la'\in\Lambda_k$ the conditional probability
$$
\Pr\left\{\hat {X}(\z,\e')\,|\,C(\x(\e'),\z)=
\|n(x^s_1,z)\|\right\}=
d_k(\|n(x^s_1,z)\|),
\eqno(40)
$$
where the notation of (35) is used.

Let $\e'\ne\e''\in \Lambda_k$.
Introduce $\e'_k\eq\e'\setminus\e$ and
$\e''_k\eq\e''\setminus\e$.
 Consider the collection $S=S(s,t,k,v)$
of all
pairs $(\e',\e'')$ for which
$|\e'_k\cap\e''_k|=k-v$, where
$v=1,2,\dots,k$. Note that the cardinality of this collection is
$$
|S|=
{t-s\choose {k-v}}{t-s-(k-v) \choose 2v}
{2v \choose v} < t^{k+v},\eqno(41)
$$
Further, for any pair $(\e',\e'')$ from $S$
the conditional probability
$$
\Pr\left\{\hat X(\z,\e')
\cap\hat X(\z,\e'')|C(\x(\e),\z)=
\|n(x^s_1,z)\|\right\}=
$$
$$
=\prod\limits_{z}\prod\limits_{x^s_{k+1}}
\frac{n(x^s_{k+1},z)!}{\prod\limits_{x^k_{v+1}}
n(x^s_{v+1},z)!} \delta^k_{v+1}(\|n(x^s_1,z)\|)
[d_v(\|n(x^s_1,z)\|)]^2= \eqno(42)
$$
$$
=d_v(\|n(x^s_1,z)\|)\cdot d_k(\|n(x^s_1,z)\|)
$$
where the notations of (31) and (35) are used. It is evident
that the conditional probability (34) is greater than or equal
to the conditional probability of the union of events
$\hat {X}(\z,\e')$, $\e'\in\Lambda_k$. Hence, applying the
standard lower bound
$$
\Pr\left\{\bigcup_i\theta_i\right\}\ge
\sum_i\Pr\{\theta_i\}-\sum_{i<j}
\Pr \{\theta_i\cap \theta_j\}
$$
and taking into account (40)-(42), we obtain
$$
q^{(k)}(\|n(x^s_1,z)\|)\ge {t-s \choose k}\cdot
d_k(\|n(x^s_1,z)\|)
-\sum\limits_{v=1}^k t^{k+v}
d_k(\|n(x^s_1,z)\|)\cdot d_v(\|n(x^s_1,z)\|).
\eqno(43)
$$

Consider the distribution $\tau^{(k)}_{cr}$ introduced in
Proposition~2. Restrict the summation in (32) by one and
only one composition $\|n(x^s_1,z)\|$, which satisfies the
asymptotic equality (37) when $\tau=\tau^{(k)}_{cr}$. By
substituting (38) and (39) in (43) and (32) we have
$$
q_k\ge\exp\{-N[{\cal H}(\Q,\tau^{(k)}_{cr})+I^{(k)}
(\tau^{(k)}_{cr})-R+o(1)]\},
$$
if for any $v,v=1,2,\dots,k$, the rate
$R<I^{(v)}(\tau^{(k)}_{cr})/v$. Applying Proposition~4 for
distribution $\tau^{(k)}_{cr}$, we obtain~(28).

Lemma 2 is proved.

{\bf Proof of Lemma 3.} Fix arbitrary
composition $\|n(x^s_1,z)\|$
and define on the product $\{0,1\}^s\cdot Z$
the probability distribution
$$
\tau=\left\{\tau(x^s_1,z)=\frac{n(x^s_1,z)}{N},
\quad x^s_1\in \{0,1\}^s,\;z\in Z\right\}.
$$
Let
$$
I_{\tau}(x^k_1,z|x^s_{k+1})\eq\sum_z \sum_{x^s_1}
\tau(x^s_1,z)\ln\frac{\tau(z|x^s_1)}{\tau(z|x^s_{k+1})},
\quad
K_p(\tau)\eq\sum_z \sum_{x^s_1} \tau(x^s_1,z)\ln
\frac{\tau(z|x^s_1)}{P(z|x^s_1)}
$$
be the information functions of Shannon and
Kullback~\cite{ck81}.  For the correctness of the
$K_p(\tau)$ definition we consider only such
compositions for which  $\tau(z|x^s_1)=0$ if~$P(z|x^s_1)=0$.

For a given composition $\|n(x^s_1,z)\|$, introduce the
corresponding marginal compositions $\|n(x^s_1)\|$ and
$\|n(x^s_{k+1})\|$. Fix an arbitrary $(N\times t)$ code $X$.
Denote by
$$
{\cal E}_X(\|n(x^s_1)\|)=\{\e\,:\,C(\x(\e))=\|n(x^s_1)\|\}
$$
--the set of messages $\e\in\Lambda(s,t)$ which are
encoded by $X$ into $s$--collection $\x(\e)$
with composition $(\|n(x^s_1)\|)$. Also introduce
$$
{\cal E}_X(\|n(x^s_{k+1})\|)=
\{\la^{(s-k)}\;:\;C(\x(\e^{(s-k)}))=\|n(x^s_{k+1})\|\}
$$
--the set of elements $\e^{(s-k)}\in\Lambda(s-k,t)$ which
are encoded by $X$ into $(s-k)$--collection
$\x(\e^{(s-k)})$ with composition $\|n(x^s_{k+1})\|$.
For each $k=1,2,\dots,s$ and any composition
$\|n(x^s_1,z)\|$, the error probability (3) satisfies the
inequality
$$
P_Q(X)\ge\exp\{-NK_p(\tau)\}\cdot\left[
\frac{|{\cal E}_X(\|n(x^s_1)\|)|}{(N+1)^A{t\choose s}}\right.-
$$
$$
-\frac{t^k {t \choose {s-k}}}{{t \choose s}}\;
\left.\frac{|{\cal E}_X(\|n(x^s_{k+1})\|)|}{{t\choose
{s-k}}}\cdot
\exp\{-N[kR-I_{\tau}(x^k_1,z|x^s_{k+1})]\}\right],
\eqno(44)
$$
where $R=\ln t/N,\quad A=2^s|Z|$. We omit the proof
of (44) and observe that it may be proved with the help of
arguments which were used in~\cite{dp82}. For comparison we
point also to the similar lower bounds which presented
in~\cite{d84,d88} for the error probability of an
individual code pair in MAC. Averaging over ensembles (4)
and (5), we have
$$
\frac{\overline{|{\cal E}_X(\|n(x^s_1)\|)|}}{{t\choose s}}=
\frac{N!}{\prod\limits_{x^s_1} n(x^s_1)!}
\cdot\delta^s_1(\|n(x^s_{1},z)\|),
\eqno(45a)
$$
$$
\frac{\overline{|{\cal E}_X(\|n(x^s_{k+1})\|)|}}{{t\choose{s-k}}}
=\frac{N!}{\prod\limits_{x^s_{k+1}} n(x^s_{k+1})!}
\cdot\delta^s_{k+1}(\|n(x^s_{1},z)\|),
\eqno(45b)
$$
where notations of (31) are used.

Fix an arbitrary distribution $\tau$ which satisfies the
conditions (depending on CRE or CWE) pointed out in the
formulation of Theorem~1. In virtue of (6), (31), and (45),
the averaging of inequality (44) yields
$$
\overline{P_Q(X)}\ge\{-N[{\cal H}(\Q,\tau)+o(1)]\}
\cdot[1-\exp\{-N[kR-I^{(k)}(\tau)+o(1)]\}],\eqno(46)
$$
$k=1,2,\dots,s$, where the exponents of the right-hand side
are defined by (7). From (46) it follows
the inequality (29).

Lemma 3 is proved.

\newpage

\centerline{Arkadii G. D'yachkov,\qquad "Lectures on
Designing Screening Experiments"}

\end{document}